\documentclass[useAMS,usenatbib]{mn2e}
\usepackage{myaasmacros}
\usepackage{graphicx}
\usepackage{ulem}

\def\ltsima{$\; \buildrel < \over \sim \;$}
\def\simlt{\lower.5ex\hbox{\ltsima}}   
\def\gtsima{$\; \buildrel > \over \sim \;$}
\def\simgt{\lower.5ex\hbox{\gtsima}}
\newcommand\bcite[1]{\citeauthor{#1} \citeyear{#1}}

\def\unabla{{\bf \nabla}}
\def\uv{{\bf v}}
\def\uvi{{\bf v}_i}
\def\uvj{{\bf v}_j}
\def\uvij{{\bf v}_{ij}}
\def\urij{{\bf r}_{ij}}
\def\ur{{\bf r}}
\def\uri{{\bf r}_{i}}
\def\urj{{\bf r}_{j}}

\def\mtij{m_j}
\def\uE{{\bf E}_0}
\def\uEi{{\bf E}_{0,i}}
\def\zero{{\bf 0}}
\def\uxij{{\bf x}_{ij}}
\def\unablax{{\bf \nabla}^x}

\def\wijtild{\overline{W}_{ij}}

\def\dwijtild{\unabla_i \overline{W}_{ij}}

\def\wij{W_{ij}}

\def\kijtild{\overline{K}_{ij} {\bf r}_{ij}}

\def\kij{\overline{K}_{ij}}

\def\hijtild{\overline{H}_{ij} {\bf r}_{ij}}

\def\hij{\overline{H}_{ij}}
\def\hi{H (|\urij|, h_i)}
\def\hj{H (|\urij|, h_j)}

\def\lijtild{\overline{L}_{ij} {\bf r}_{ij}}

\def\lij{\overline{L}_{ij}}

\def\phir{\eta}
\def\phiv{\phi}
\def\phiu{\zeta}
\def\dphir{\dot{\eta}}

\def\Mmr{{\bf R}}

\def\Mmv{{\bf V}}
\def\Sm{{\bf S}_{ij}}
\def\Id{{\bf I}}

\def\dx{x_{ij}}
\def\dy{y_{ij}}
\def\dz{z_{ij}}

\def\ud{\mathrm{d}}
\def\deriv#1#2{\frac{\ud #1}{\ud #2}}
\def\dderiv#1#2{\frac{\ud^2 #1}{\ud {#2}^2}}

\def\ppderiv#1#2{\frac{\partial^2 #1}{\partial {#2}^2}}
\def\ptderiv#1#2#3{\frac{\partial #1}{\partial #2 \partial #3}}
\def\ua{{\bf a}}
\def\uk{{\bf k}}
\def\uH{{\bf H}}
\def\uq{{\bf q}}
\def\ux{{\bf x}}

\def\dthr{\mathrm{d}^3{ r}}
\def\dthx{\mathrm{d}^3{ x}}
\def\dV{dV}

\def\MSPH{OSPH}
\def\Multiphase{Optimised}
\def\MULTIPHASE{OPTIMISED}
\def\SPHS{TSPH}

\newlength{\myfix}
\title[Resolving mixing in SPH]{Resolving mixing in Smoothed Particle
  Hydrodynamics}

\author[Read et al.]{J. I. Read$^{1,2}$\thanks{E-mail: justin.inglis.read@gmail.com},
  T. Hayfield$^3$ and O. Agertz$^1$\\
  $^1$Institute of Theoretical Physics, University of Z\"urich, Winterthurerstrasse 190, 8057 Z\"urich, Switzerland\\
  $^2$Department of Physics and Astronomy, University of Leicester, University Road, LE1 7RH Leicester, UK\\
  $^3$Department of Physics, ETH Z\"urich, Wolfgang-Pauli-Strasse 16, CH-8093 Z\"urich, Switzerland
  }

\begin{document}

\maketitle

\begin{abstract}
Standard formulations of smoothed particle hydrodynamics (SPH) are
unable to resolve mixing at fluid boundaries. We use an error and
stability analysis of the generalised SPH equations of motion to prove
that this is due to two distinct 
problems. The first is a leading order error in the momentum equation.
This should decrease 
with increasing neighbour number, but does not because 
numerical instabilities cause the kernel 
to be irregularly sampled. We identify two important instabilities:
the clumping instability and the banding
  instability, and we show that both are cured by a suitable choice
of kernel. The second problem is the local mixing instability
(LMI). This occurs as particles attempt 
to mix on the kernel scale, but are unable to due to entropy
conservation. The result is a pressure discontinuity at boundaries that
pushes fluids of different entropy apart. We cure the LMI by using a
weighted density estimate that ensures that pressures are single valued 
throughout the flow. This also gives a better volume estimate
for the particles, reducing errors in the continuity and momentum equations. 
We demonstrate mixing in our new \Multiphase\
Smoothed Particle Hydrodynamics (\MSPH) scheme using a Kelvin Helmholtz
instability (KHI) test with density contrast 1:2, and the `blob test'
-- a 1:10 density ratio gas sphere in a wind tunnel -- finding excellent
agreement between \MSPH\ and Eulerian codes.
\end{abstract}


\section{Introduction}\label{sec:introduction}

Smoothed Particle Hydrodynamics (SPH) was first introduced as a tool
for studying stellar structure (\bcite{1977MNRAS.181..375G};
\bcite{1977AJ.....82.1013L}), but has since found wide application in
all areas of theoretical astrophysics \citep{1992ARA&A..30..543M}, in
engineering \citep{1993JCoPh.109...67L}, and beyond
(e.g. \bcite{2008JCoPh.227.9195H}).

Although there are many varieties of SPH, the central idea is to
represent a fluid by discrete particles that move with the flow
(\bcite{1992ARA&A..30..543M}; \bcite{2005astro.ph..7472P}). Typically
these particles represent the fluid exactly, though in some variants
the fluid is advected on top of the particles (\bcite{1999Dilts};
\bcite{2003ApJ...595..564M}). The key advantages over
Eulerian schemes\footnote{This does not apply
  to Lagrangian moving mesh schemes that are  
  Galilean invariant
  \citep{2009arXiv0901.4107S}.}
are its Lagrangian nature that makes it Galilean invariant, and its
particle nature that makes it easy  
to couple to the fast multipole method for
gravity that scales as $O(N)$ (\bcite{2000ApJ...536L..39D};
\bcite{1987JCoPh..73..325G}). However, SPH has problems correctly
integrating fluid instabilities and mixing 
at boundaries (\bcite{1996PhDMorris}; \bcite{1999Dilts};
\bcite{2001MNRAS.323..743R}; \bcite{2003MNRAS.345..561M};
\bcite{2006astro.ph.10051A}). Several different reasons have been
suggested for this in the literature so far. \citet{1996PhDMorris} and
\citet{1999Dilts} argue that the problem owes to errors in the SPH
gradients that do not show good convergence for irregular particle
distributions. \citet{2007arXiv0709.2772P} argue that the problem owes to the
fact that entropies are discontinuous at boundaries, while the
densities are smooth. This gives spurious pressure blips at
boundaries that drive fluids of different entropy
apart. They find that adding thermal conductivity at boundaries to
smooth the entropies gives improved mixing in
SPH. \citet{2008MNRAS.387..427W} make a similar argument, phrasing the
problem in terms of an inability for SPH particles to mix and generate
entropy on the kernel scale. They find that adding a heat diffusion
term to model subgrid turbulence gives 
improved mixing in SPH. Finally, \citet{2001MNRAS.323..743R} suggest
that the problem lies in the SPH 
density estimate. They introduce a new temperature weighted density
estimate that is designed to give smoother pressures at flow
boundaries, thus combating the spurious boundary pressure
blip. 

In this paper, we perform an error and stability analysis of SPH in
its most general form to understand why mixing fails. In doing this,
we show that all of the above authors correctly identified one of two
distinct problems with mixing in SPH. The first is an $O(h^{-1})$
error in the momentum equation identified by \citet{1996PhDMorris} and
\citet{1999Dilts}. The second relates to entropy conservation
on the kernel scale, as addressed directly by
\citet{2007arXiv0709.2772P} and \citet{2008MNRAS.387..427W}, and
indirectly by \citet{2001MNRAS.323..743R}. Having identified the
problem, we present a new method -- \Multiphase\ Smoothed Particle
Hydrodynamics (\MSPH) -- that, given sufficient resolution, correctly
resolves multiphase fluid flow.

This paper is organised as follows. In \S\ref{sec:sph} and
\S\ref{sec:osph}, we briefly
review standard SPH schemes and introduce
our new \MSPH\ scheme. We show that there are two distinct problems with
mixing in SPH: the `$\uE$ error' in the momentum equation, and the
`local mixing instability' (LMI), and we show how both can be
cured. In \S\ref{sec:implement}, we present our implementation of \MSPH\ in the
{\tt GASOLINE} code \citep{2004NewA....9..137W}. In \S\ref{sec:kh}, we use a
Kelvin Helmholtz instability (KHI) test with density contrast 1:2 and 1:8 to
demonstrate mixing in \MSPH. We show the effect of turning on each of
the \MSPH\ improvements one at a time, arriving at a solution that is in
excellent agreement with the Eulerian code
{\tt RAMSES} \citep{2002A&A...385..337T}. In \S\ref{sec:sod}, we use
the standard Sod shock tube test  to demonstrate that \MSPH\ can successfully model shocks. In \S\ref{sec:blob}, we revisit 
the `blob test' introduced in \citet{2006astro.ph.10051A}, finding
excellent agreement between \MSPH\ and the Eulerian code {\tt FLASH}
\citep{2000ApJS..131..273F}. Finally, in \S\ref{sec:conclusions} we
present our conclusions.

\section{Smoothed Particle Hydrodynamics}\label{sec:sph}

In SPH, the fluid is represented by discrete particles
that move with the flow. The density of each particle is 
estimated by a weighted sum over its neighbours: 

\begin{equation}
\rho_i = \sum_j^N m_j W(|\urij|,h_i)
\label{eqn:sphcont}
\end{equation}
where $h_{i}$ and $m_j$ are the smoothing length and mass of particle
$i$ and $j$, respectively; we define $\urij = \uri - \urj$ and similarly for
other vectors; and $W$ is a symmetric kernel that obeys the
normalisation condition:
\begin{equation}
\int_{V} W(|\ur-\ur'|,h) \dthr' = 1
\label{eqn:normw}
\end{equation}
and the property:
\begin{equation}
\lim_{h\rightarrow 0} W(|\ur-\ur'|,h) = \delta(|\ur-\ur'|)
\end{equation}
In the limit $N\rightarrow\infty, h\rightarrow 0$ (and using
$m_j/\rho_j\rightarrow \dthr'$) equation \ref{eqn:sphcont} recovers the
continuum flow density.

The equations of motion for SPH are then derived by discretising the Euler equations -- the continuity, momentum and energy equations: 

\begin{equation}
\frac{d\rho}{dt} = -\rho\unabla\cdot\uv
\label{eqn:euler1}
\end{equation}
\begin{equation}
\frac{d\uv}{dt}  = -\frac{\unabla P}{\rho}
\label{eqn:euler2}
\end{equation}
\begin{equation}
\frac{du}{dt} = -\frac{P}{\rho}\unabla\cdot\uv
\label{eqn:euler3}
\end{equation}
where $\rho, v$ and $u$ are the density, velocity and internal energy
per unit mass of the flow, respectively.

The Euler equations can be derived from the
Lagrangian for hydrodynamics (e.g. \bcite{bennett}):

\begin{equation}
L = \int \left(\frac{1}{2}\rho v^2 - \rho u\right) \dV
\label{eqn:lhydro}
\end{equation}
and in many modern derivations of the
equations of motion for SPH, equation \ref{eqn:lhydro} is discretised,
rather than equations \ref{eqn:euler1}-\ref{eqn:euler3}. 

Replacing the volume element $\dV$ with the volume per SPH particle
$m/\rho$, we obtain (\bcite{2005astro.ph..7472P}): 

\begin{equation}
L = \sum_j m_j \left(\frac{1}{2}\uv_j^2 - u_j\right)
\label{eqn:lhydrosph}
\end{equation}
and the standard SPH equations of motion then follow from the
Euler-Lagrange equations:

\begin{equation}
\frac{d\rho_i}{dt} = \sum_j^N \mtij \uvij \cdot \unabla_i \wij
\label{eqn:sphconti}
\end{equation}
\begin{equation}
\frac{d\uv_i}{dt} = -\sum_j^N \mtij \left[\frac{P_i}{\rho_i^2} +
  \frac{P_j}{\rho_j^2}\right] \unabla_i \wij
\label{eqn:sphmomenti}
\end{equation}
\begin{equation}
\frac{du_i}{dt} = \frac{P_i}{\rho_i^2}\sum_j^N \mtij \uvij \cdot
\unabla_i \wij
\label{eqn:sphenergyi}
\end{equation}
where $\wij = W(|\urij|,h_i)$.

Note that equation \ref{eqn:sphconti} is {\it automatically}
satisfied by time derivative of the SPH density estimate (equation
\ref{eqn:sphcont}). For this reason, equation \ref{eqn:sphcont} is
often referred to as the {\it integral form} of the continuity
equation.

The above system of equations are closed by the equation of state: 

\begin{equation}
P_i = u_i \left(\gamma - 1\right) \rho_i
\label{eqn:state}
\end{equation}

This standard approach to deriving the SPH equations of motion has the
advantage that the resulting 
equations are {\it coherent}\footnote{Also called {\it
    consistent} \citep{2007Oger}.} by construction -- that is they are 
consistent with a Lagrangian. This gives very good conservation 
properties for the flow. It is also straightforward to calculate the
necessary correction terms that arise if the smoothing lengths are a
function of space and time $h = h(\ur,t)$ (see
e.g. \bcite{nelsonpapagradh94}; \bcite{2005astro.ph..7472P}). We do
not include these correction terms in this paper.

However, this standard derivation leads to a scheme that cannot
correctly model fluid mixing processes (see
\S\ref{sec:introduction}), which motivates us to move to a 
more general derivation. Discretising each of the Euler equations
separately leads to a free function for each equation: $\phir$,
$\phiv$ and $\phiu$, as well as a different smoothing kernel for
each. This is the approach we take next in \S\ref{sec:osph}. In
\S\ref{sec:error} and \S\ref{sec:minimise}, we will then use an error and
stability analysis of these more general equations of motion to constrain
the new functions $\phir, \phiv$ and $\phiu$ and our new kernels. By choosing these new free functions and kernels such that they minimise the integration error, we will arrive at a new scheme that can, with sufficient
resolution, correctly resolve multiphase fluid flow. 

\section{\MULTIPHASE\ Smoothed Particle Hydrodynamics}\label{sec:osph} 

In the previous section, we presented a standard derivation of the SPH equations of motion. However, this standard derivation leads to a scheme that cannot correctly model fluid mixing processes (see \S\ref{sec:introduction}). In this section, we move to a more general derivation of the SPH equations of motion. We show that, in general, we have a free function for each of the Euler equations: $\phir$, $\phiv$ and $\phiu$, as well as a different smoothing kernel for each. There is also a freedom in the energy equation in the choice of integration variable (energy or entropy; \S\ref{sec:engent}). In \S\ref{sec:error} and \S\ref{sec:minimise}, we will then use an error and stability analysis of these more general equations of motion to constrain the new functions $\phir, \phiv$ and $\phiu$ and our new kernels. By choosing these new free functions and kernels such that they minimise the integration error, we will arrive at a new scheme that can, with sufficient resolution, correctly resolve multiphase fluid flow. 

\subsection{A general derivation of SPH}\label{sec:generalsph}

In general, we have some freedom in how we discretise the Euler equations (equations \ref{eqn:euler1}-\ref{eqn:euler3}) to obtain the equations of motion for SPH \citep[see e.g.][]{1992ARA&A..30..543M,2005astro.ph..7472P,2009NewAR..53...78R}. The gradients in the Euler equations can be expanded to include a new free function for each equation: $\phir$, $\phiv$ and $\phiu$:

\begin{equation}
\frac{d\rho}{dt} = \phir\left[\uv\cdot
  \unabla\left(\frac{\rho}{\phir}\right) - \unabla \cdot
  \left(\frac{\rho\uv}{\phir}\right)\right] 
\end{equation}
\begin{equation}
\frac{d\uv}{dt}  =
-\left[\frac{P\phiv}{\rho^2}\unabla\left(\frac{\rho}{\phiv}\right) +
  \frac{1}{\phiv}\unabla\left(\frac{P\phiv}{\rho}\right)\right] 
\end{equation}
\begin{equation}
\frac{du}{dt} = \frac{P}{\rho^2}\phiu\left[\uv\cdot
  \unabla\left(\frac{\rho}{\phiu}\right) - \unabla \cdot
  \left(\frac{\rho\uv}{\phiu}\right)\right] 
\end{equation}
In the continuum form, above, $\phir, \phiv$ and $\phiu$ cancel. But in the
discrete SPH form, they remain giving a useful additional freedom
\citep{2005astro.ph..7472P}: 
\begin{equation}
\frac{d\rho_i}{dt} = \sum_j^N \mtij \frac{\phir_i}{\phir_j}\uvij \cdot
\hijtild 
\label{eqn:cont}
\end{equation}
\begin{equation}
\frac{d\uv_i}{dt} = -\sum_j^N \mtij
\left[\frac{P_i}{\rho_i^2}\frac{\phiv_i}{\phiv_j} +
  \frac{P_j}{\rho_j^2}\frac{\phiv_j}{\phiv_i}\right] \kijtild 
\label{eqn:moment}
\end{equation}
\begin{equation}
\frac{du_i}{dt} = \frac{P_i}{\rho_i^2}\sum_j^N \mtij
\frac{\phiu_i}{\phiu_j}\uvij \cdot \lijtild 
\label{eqn:energy}
\end{equation}
where $\hij = \left[\hi + \hj \right]/2$, $\kij$ and $\lij$ are
symmetrised smoothing kernels -- one for each Euler equation. Standard
SPH (SPH from here on) is a special case of the above with 
$\phir = \phiv = \phiu = 1$ and $\hijtild 
= \kijtild = \lijtild = \dwijtild$. 

Equation \ref{eqn:cont} casts the continuity equation in differential
form. This is problematic since, in this case, the particles no longer
represent the fluid exactly. Instead they represent 
a moving mesh on which the Euler equations are solved. This leads to
the danger that high density regions will contain few particles
leading to large errors \citep{2003ApJ...595..564M}. For this reason,
we use instead a generalised integral form for the continuity
equation: 

\begin{equation}
\rho_i = \sum_j^N m_j \frac{\phir_i}{\phir_j} \wijtild
\label{eqn:sphcontgen}
\end{equation}
which, taking the time derivative, gives: 

\begin{equation}
\frac{d\rho_i}{dt} = \sum_j^N \mtij \frac{\phir_i}{\phir_j}\uvij \cdot
\unabla_i \wijtild + \epsilon
\end{equation}
where:
\begin{equation}
\epsilon = \sum_j^N \mtij\left(\frac{\dphir_i}{\phir_i} -
  \frac{\dphir_j}{\phir_j}\right)\frac{\phir_i}{\phir_j}\wijtild
\label{eqn:epsilerror}
\end{equation}
and $\dphir = \frac{d\eta}{dt}$.

This reduces to the continuity equation (equation \ref{eqn:cont})
under the kernel constraint: $\hijtild = \unabla_i\wijtild$, and for
$\epsilon = 0$. The latter can be satisfied by construction if
$\phir_i = \phir_j$ (as is the case for SPH), or if
$\dphir=0$. However, in the continuum limit ($N\rightarrow \infty$, $h\rightarrow 0$), $\epsilon \rightarrow 0$ and so
$\epsilon$ will vanish with increasing resolution. For this reason, equation
\ref{eqn:sphcontgen} gives a valid approximation to the continuity
equation for any choice of $\phir$, with $\epsilon$ simply
contributing an additional error term.

\subsection{Energy versus entropy forms of SPH}\label{sec:engent}

A final freedom in the equations motion for SPH comes from the energy
equation. Equation \ref{eqn:energy} is the standard energy form
of SPH, but there is also an entropy form
(\bcite{1991ApJ...378..637G}; \bcite{2002MNRAS.333..649S}). Instead of
the internal energy, $u$, we evolve a function $A(s)$ -- the {\it
  entropy function} -- that is a monotonic function of the entropy $s$
defined by the equation of state:

\begin{equation}
P_i = A_i(s) \rho_i^{\gamma}
\label{eqn:entropystate}
\end{equation}
Away from shocks and in the absence of thermal sources or sinks, $A_i$ is a constant of motion. Thus, taking the time derivative of equation \ref{eqn:entropystate} and substituting for equation \ref{eqn:state}, we recover: 

\begin{equation}
\frac{du_i}{dt} = \frac{P_i}{\rho_i^2}\frac{d\rho_i}{dt}
\label{eqn:thermconst}
\end{equation}
by construction. Schemes that obey equation \ref{eqn:thermconst} are
called {\it thermodynamically consistent}.

In practice, we find -- for the tests presented in this paper -- that
the energy and entropy forms of SPH give near-identical results,
provided that equation \ref{eqn:thermconst} is
satisfied (for adiabatic flow). We use the
thermodynamically consistent energy form throughout this paper. This
gives us the constraints: $\phiu=\phir$ and $\lijtild = \hijtild =
\unabla_i\wijtild$, which we apply from here on. We also use $\kijtild =
\unabla_i\wijtild$, as in standard SPH. This is not a formal requirement, but ensures that coherence is recovered in the limit of constant density. 

\subsection{Errors: choosing the free functions}\label{sec:error}

In this section, we perform an error analysis of the generalised equations for SPH (equations \ref{eqn:moment}, \ref{eqn:energy} and \ref{eqn:sphcontgen}) derived in \S\ref{sec:generalsph}. We will then choose our free functions $\phir$, $\phiv$ and $\phiu$ so that these errors are minimised. 

\subsubsection{Error analysis} 

We assume that the pressure and velocity of the flow are smooth. In this case, we can Taylor expand to give: 

\begin{equation}
P_j \simeq 
P_i + h \uxij \cdot \unabla_i P_i + O(h^2)
\label{eqn:ptaylor}
\end{equation}
and 
\begin{equation}
\uvj \simeq \uvi +
 h (\uxij \cdot \unabla_i)\uvi + O(h^2)
\label{eqn:vtaylor}
\end{equation} 
where $\ux_{ij} = \ur_{ij}/h$, and we have assumed a constant smoothing length $h$. 
 
Substituting equations \ref{eqn:ptaylor} and \ref{eqn:vtaylor} into the continuity and momentum equations gives:
\begin{equation}
\frac{d\rho_i}{dt} \simeq -\rho_i \left(\Mmr_i \unabla_i\right) \cdot \uvi
+ \epsilon + O(h) 
\label{eqn:conterr}
\end{equation}
and
\begin{equation}
\frac{d\uv_i}{dt}  \simeq  -\frac{P_i}{h \rho_i} \uEi  - \frac{\left(\Mmv_i \unabla_i\right) P_i}{\rho_i}+ O(h) 
\label{eqn:momenterr}
\end{equation}
where $\uEi$ is a dimensionless error vector given by:

\begin{equation}
\uEi = \sum_j^N \frac{\mtij}{\rho_j} \left[g_{ij} + g^{-1}_{ij} \right] \unablax_i \wijtild
\label{eqn:e0int}
\end{equation}
and $\Mmr_i$ and $\Mmv_i$ are dimensionless error matrices given by:

\begin{equation}
\Mmr_i =  \sum_j^N \frac{\mtij}{\rho_j} f_{ij} \Sm \qquad ;
\qquad \Mmv_i =  \sum_j^N \frac{\mtij}{\rho_j} g^{-1}_{ij} \Sm
\end{equation}
with:
\begin{equation}
\Sm =  \frac{1}{x} \frac{\partial
  \wijtild}{\partial x} \left(\begin{array}{ccc}
\dx^2 & \dx \dy & \dx \dz \\
 \dy  \dx &  \dy^2 &  \dy \dz \\
 \dz \dx &  \dz \dy &  \dz^2 \end{array} \right)
\end{equation}
where $\unablax_i = h \unabla_i$; $\uxij = (\dx, \dy, \dz)$; $x=|\uxij|$; $f_{ij} = \frac{\rho_j}{\rho_i}\frac{\phir_i}{\phir_j}$; and $g_{ij} = \frac{\rho_j}{\rho_i}\frac{\phiv_i}{\phiv_j}$. 

The accuracy of the continuity equation (\ref{eqn:conterr}) is given by the extent to which $\epsilon = 0$ (see equation \ref{eqn:epsilerror}) and $\Mmr_i = \Id$, the identity matrix. The accuracy of the momentum equation (\ref{eqn:momenterr}) is given by the extent to which $\uEi = \zero$ and $\Mmv_i = \Id$. (The energy equation behaves similarly to the continuity equation with $\epsilon = 0$.)

\subsubsection{Minimising errors: the continuity equation} 

Let us consider how accurately equation \ref{eqn:conterr} approximates its Euler equation equivalent (equation \ref{eqn:euler1}). First, consider standard SPH where $\phir = 1$ and $\epsilon = 0$ by construction. Typically in the literature, the error is calculated only in the continuum limit ($N\rightarrow\infty$; $h\rightarrow 0$; see e.g. \cite{2005astro.ph..7472P}). In this case, the sums become integrals, and (using $m_j/\rho_j\rightarrow \dthx'$) we obtain terms like: 

\begin{equation}
\lim_{N\rightarrow\infty} \Mmr_{33}(\ux) = \int_{V} \dthx'
f(\ux,\ux') \frac{(z-z')^2}{|\ux-\ux'|} \frac{\partial W}{\partial x}
\label{eqn:normcond1}
\end{equation}
and
\begin{equation}
\lim_{N\rightarrow\infty} \Mmr_{12}(\ux) = \int_{V} \dthx'
f(\ux,\ux') \frac{(x-x')(y-y')}{|\ux-\ux'|} \frac{\partial W}{\partial
  x}
\label{eqn:normcond2}
\end{equation}
where the notation $_{33}$ refers to element $[3,3]$ in the matrix
$\Mmr$.

If we assume smooth densities, then we can Taylor expand $f$ also to obtain: 

\begin{equation}
f = \frac{\rho(\ur')}{\rho(\ur)} \simeq 1
+ h \frac{(\ux - \ux')}{\rho}\cdot\unabla\rho + O(h^2)
\end{equation}
and we see that, by symmetry of $W$, $\Mmr = \Id$ to $O(h^2)$. In fact, Taylor expanding to an order higher than above, it is straightforward to show that the whole continuity equation is accurate to $O(h^2)$ in the limit $N\rightarrow\infty$ (see e.g. \bcite{2005astro.ph..7472P}). A similar argument applies to the other SPH equations of motion and leads to the conclusion that SPH is accurate to $O(h^2)$. However -- and this is a key point -- this formal calculation is {\it only valid for smoothly distributed particles in the limit $N\rightarrow\infty$}. In practical situations, where we have a finite number of particles within the kernel and these are not perfectly smoothly distributed, the leading order errors in the continuity equation appear at $O(0)$ and are contained within the matrix $\Mmr$. We will quote orders of error from here on in this finite particle limit. 

We can think of each term of $\Mmr$ as a finite sum approximation to a dimensionless integral that should be either 0 (for the off diagonal terms), or 1 (for the diagonal terms). For smooth particle distributions, this approximation is a good one since $f_{ij}\simeq 1$, while $m_j/\rho_j$ gives a good estimate of the volume of each particle within the kernel. However, if the particles are distributed irregularly on the kernel scale -- for example at a sharp density step -- then $f_{ij}$ can grow arbitrarily large, while $m_j/\rho_j$ becomes a poor volume estimate. We will demonstrate this in \S\ref{sec:kh}. 

We can improve matters by choosing $\phir = \rho$, which fixes $f=1$ always. However, the integral form of the continuity equation then becomes:

\begin{equation}
\rho_i = \sum_j^N \frac{\rho_i}{\rho_j} m_j \wijtild
\label{eqn:sphcontiter}
\end{equation}
which must be solved iteratively and is not guaranteed to
converge. Worse still, $\epsilon$ is now no longer zero and
contributes an additional error. 

\citet{2001MNRAS.323..743R} present an interesting solution to this dilemma. If the pressures are approximately constant across the kernel ($P_i \simeq P_j$) then, for the energy form of SPH (see equation \ref{eqn:state} and \S\ref{sec:engent}), $\frac{\rho_i}{\rho_j} \simeq \frac{u_j}{u_i}$ and equation \ref{eqn:sphcontiter} is well approximated by the integral continuity equation:

\begin{equation}
\rho_i = \sum_j^N \frac{u_j}{u_i} m_j \wijtild
\label{eqn:sphcontrteng}
\end{equation}
This can be solved without the need for iteration. 

The above suggests that we use $\phir = 1/u$. Thermodynamic consistency then requires that we set $\phiu = \phir = 1/u$ (see \S\ref{sec:engent}). 

There may be some advantage, however, to using the {\it entropy} form of SPH. In this case, the equation of state is given by equation \ref{eqn:entropystate}. For approximately constant pressure across the kernel, we now have that $\frac{\rho_i}{\rho_j} = \left(\frac{A_j}{A_i}\right)^{\frac{1}{\gamma}}$, and the integral continuity equation becomes: 

\begin{equation}
\rho_i = \sum_j^N \left(\frac{A_j}{A_i}\right)^{\frac{1}{\gamma}} m_j \wijtild
\label{eqn:sphcontrtent}
\end{equation}
This has the advantage that, in the absence of shocks or thermal sources/sinks, $\dot{A}=0$ and so the error term $\epsilon = 0$ by construction (see equation \ref{eqn:epsilerror}). In practice, however, we find no appreciable difference between the energy and entropy forms of SPH for the tests presented in this paper. This suggests that $\epsilon$ is not a significant source of error. 

Equations \ref{eqn:sphcontrteng} and \ref{eqn:sphcontrtent} retain the desirable integral form for the density, while giving significantly improved error properties. They also have a second important advantage that we discuss in \S\ref{sec:localmix}. We refer to equation \ref{eqn:sphcontrteng} as the `RT' density estimator for the energy form of SPH; and equation \ref{eqn:sphcontrtent} as the RT density estimator in the entropy form. 

\begin{center}
\begin{figure*}
Cubic Spline (CS) kernel:\\
\includegraphics[height=0.33\textwidth]{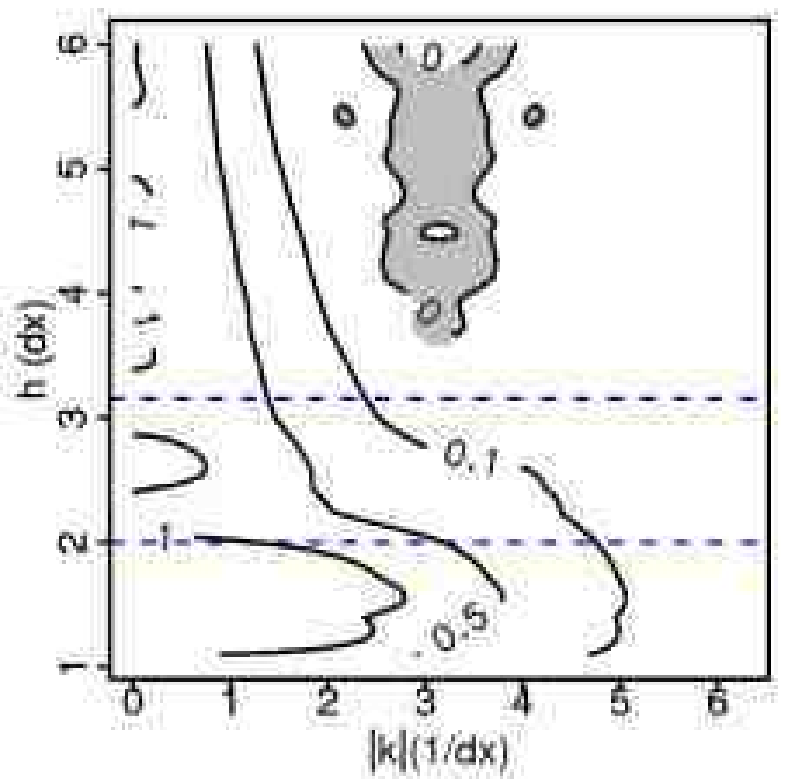}\hspace{\myfix}
\includegraphics[height=0.33\textwidth]{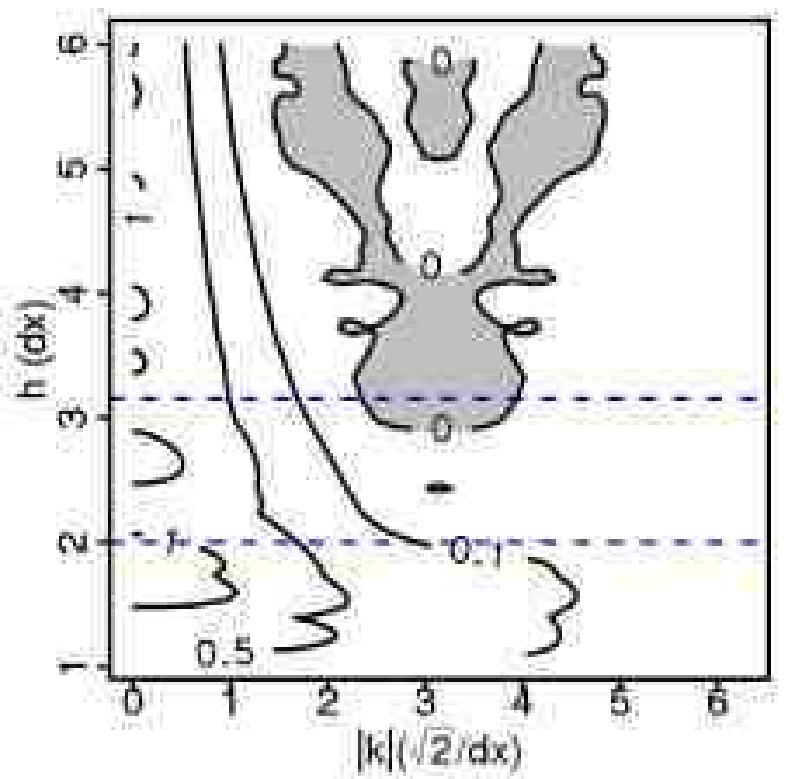}\hspace{\myfix}
\includegraphics[height=0.33\textwidth]{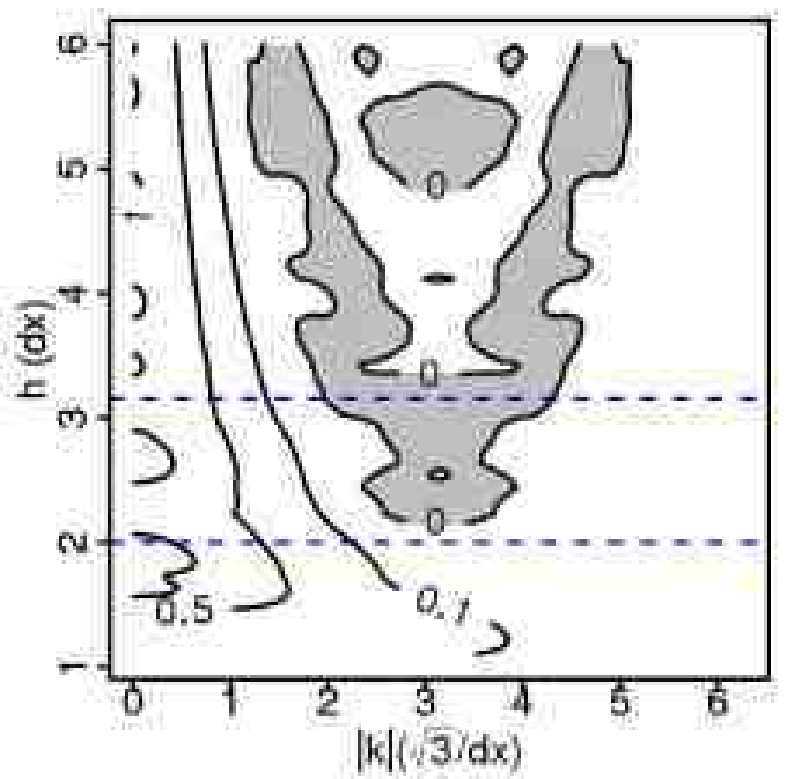}\\
\includegraphics[height=0.33\textwidth]{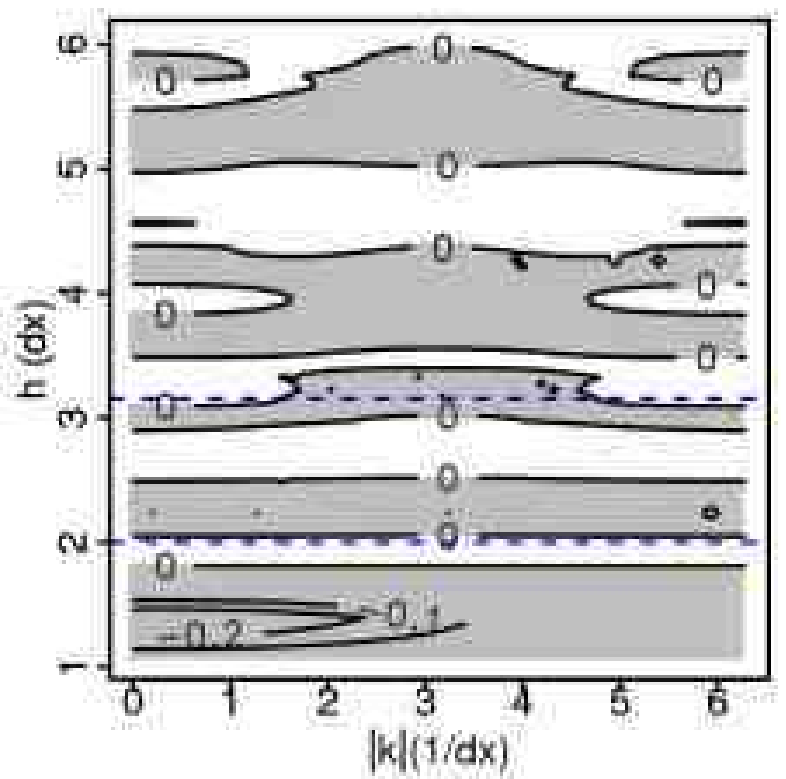}\hspace{\myfix}
\includegraphics[height=0.33\textwidth]{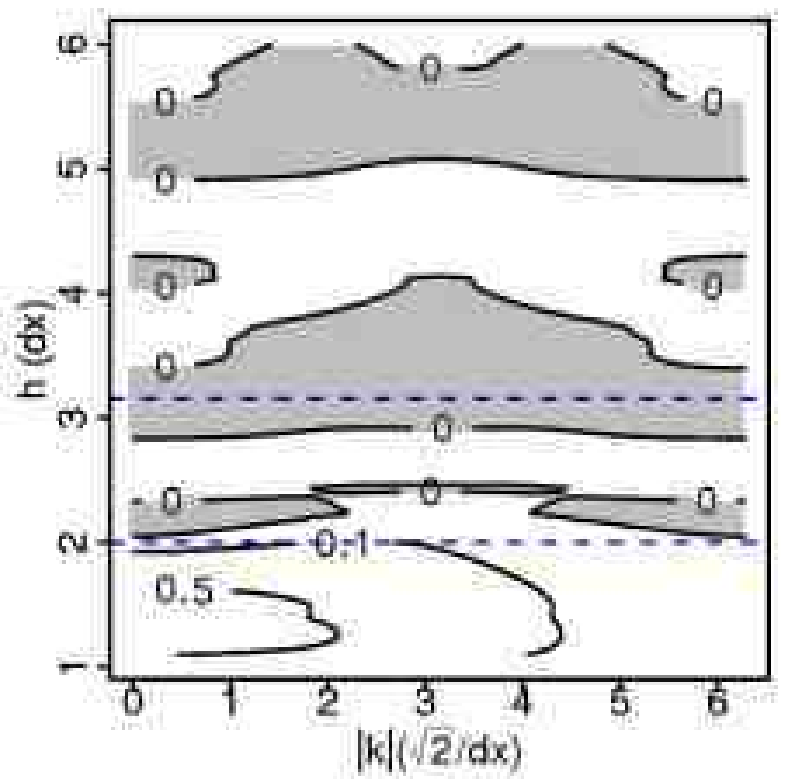}\hspace{\myfix}
\includegraphics[height=0.33\textwidth]{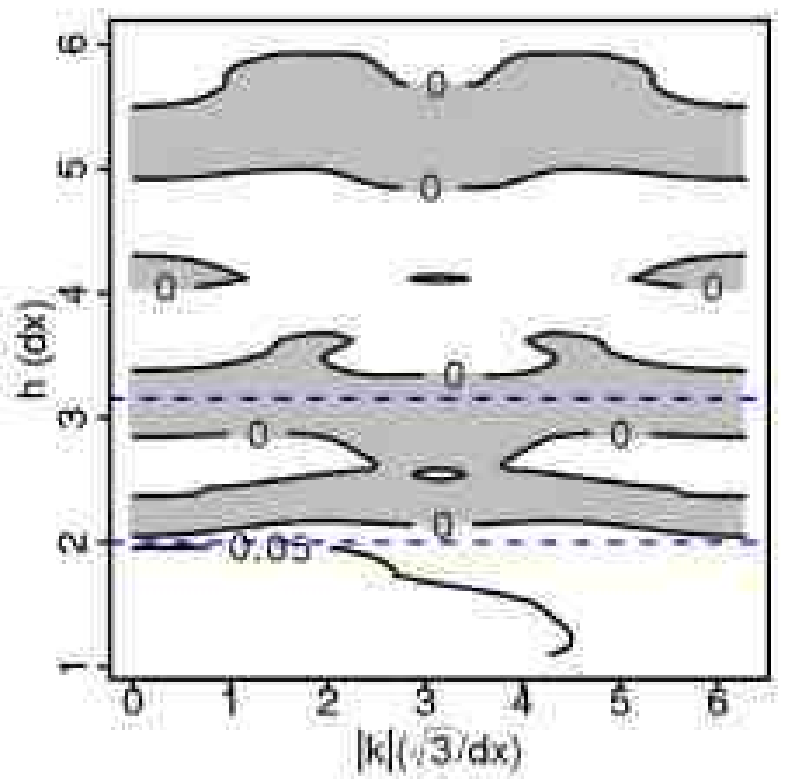}\\
\includegraphics[height=0.33\textwidth]{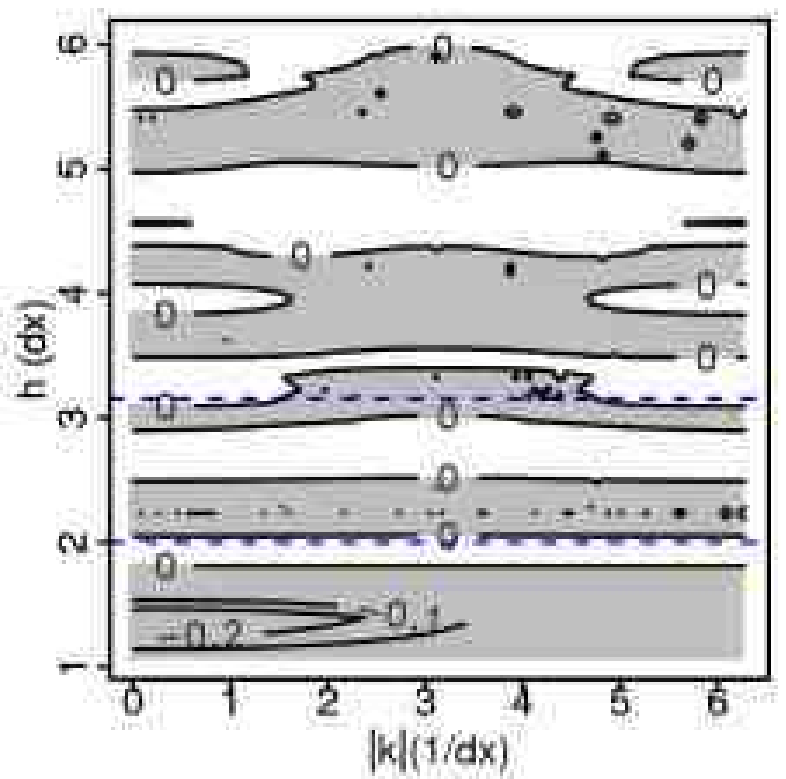}\hspace{\myfix}
\includegraphics[height=0.33\textwidth]{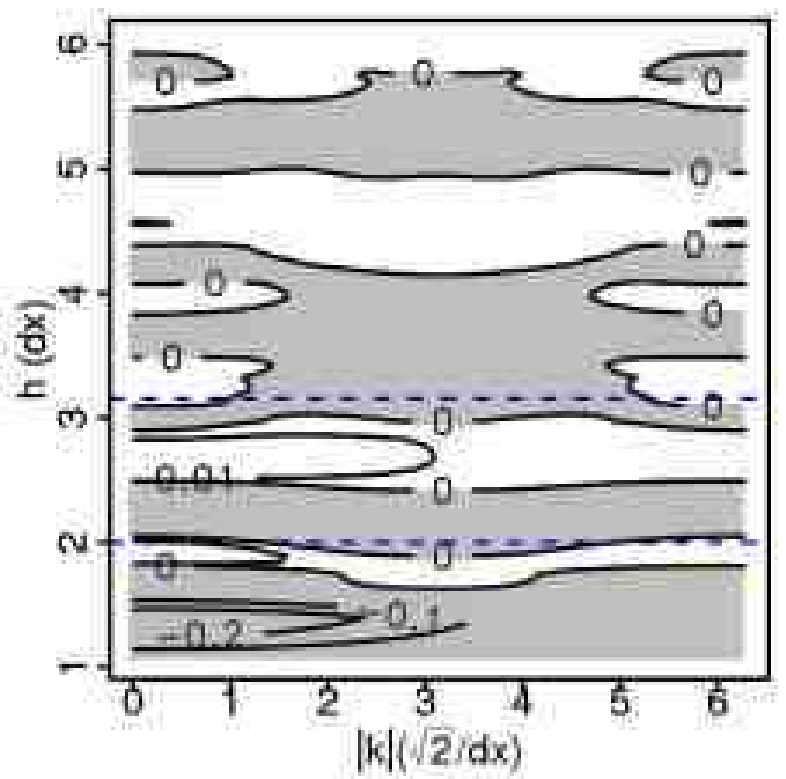}\hspace{\myfix}
\includegraphics[height=0.33\textwidth]{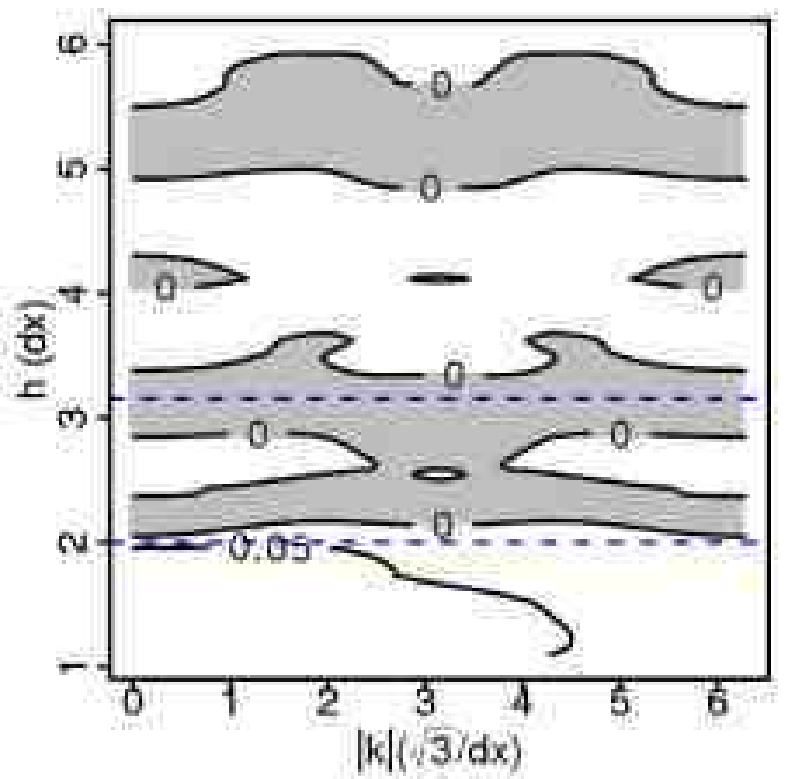}\\

\caption{Stability plots for the Cubic Spline (CS) kernel (equation
  \ref{eqn:cubicspline}) in \MSPH. The plots show contours of the frequency
  $\omega^2/k^2/c_s^2$ of plane waves impacting a regular lattice of
  particles, as a function of the wavenumber $k$ and the smoothing
  length $h$, in units of the inter-particle spacing $dx = 1$. From
  left to right the plots show 
  $(k_x,k_y,k_z) = k(1,0,0), k(1,1,0)$ and $k(1,1,1)$. The three rows
  show the longitudinal wave and the two transverse
  waves for each of these orientations. Also marked by the blue dashed
  lines are the $h$ that corresponds to 32 neighbours (bottom line)
  and 128 neighbours (top line). \MSPH\ is unstable if $\omega^2 <
  0$ (grey regions).}
\label{fig:stab}
\end{figure*}
\end{center}

\begin{center}
\begin{figure*}
Core Triangle (CT) kernel:\\
\includegraphics[height=0.33\textwidth]{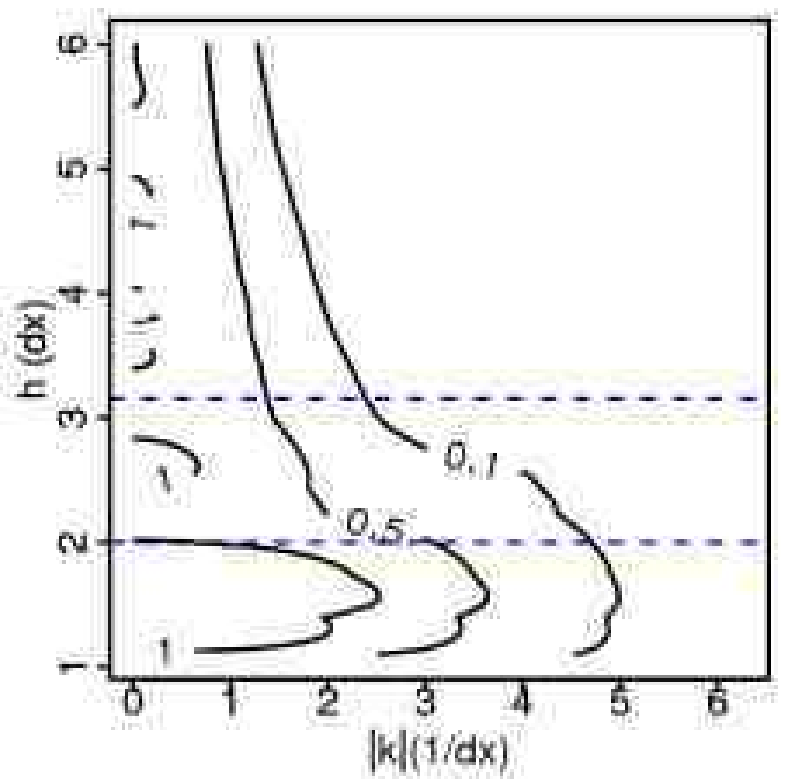}\hspace{\myfix}
\includegraphics[height=0.33\textwidth]{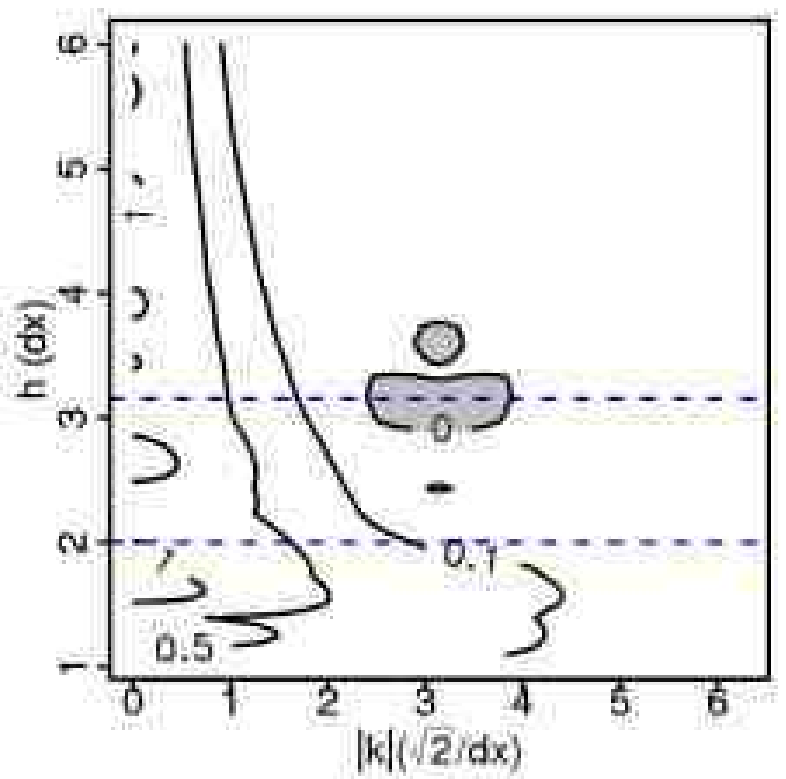}\hspace{\myfix}
\includegraphics[height=0.33\textwidth]{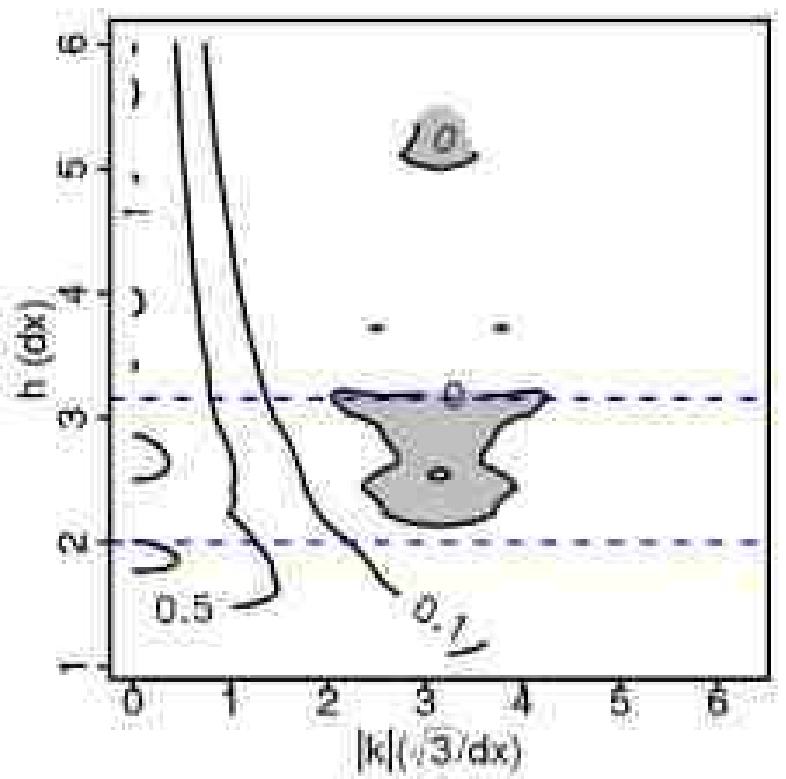}\\
\includegraphics[height=0.33\textwidth]{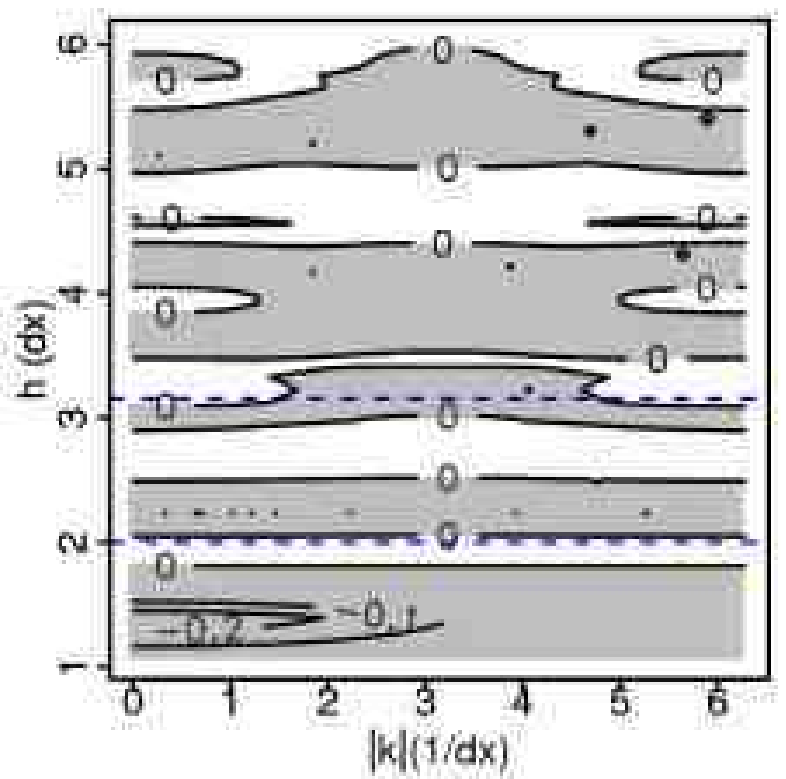}\hspace{\myfix}
\includegraphics[height=0.33\textwidth]{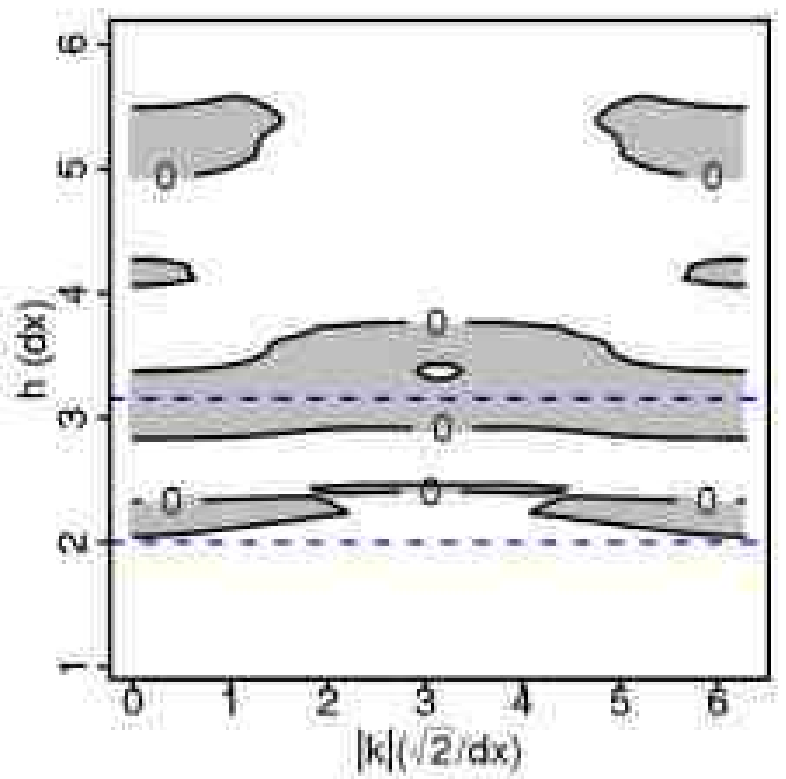}\hspace{\myfix}
\includegraphics[height=0.33\textwidth]{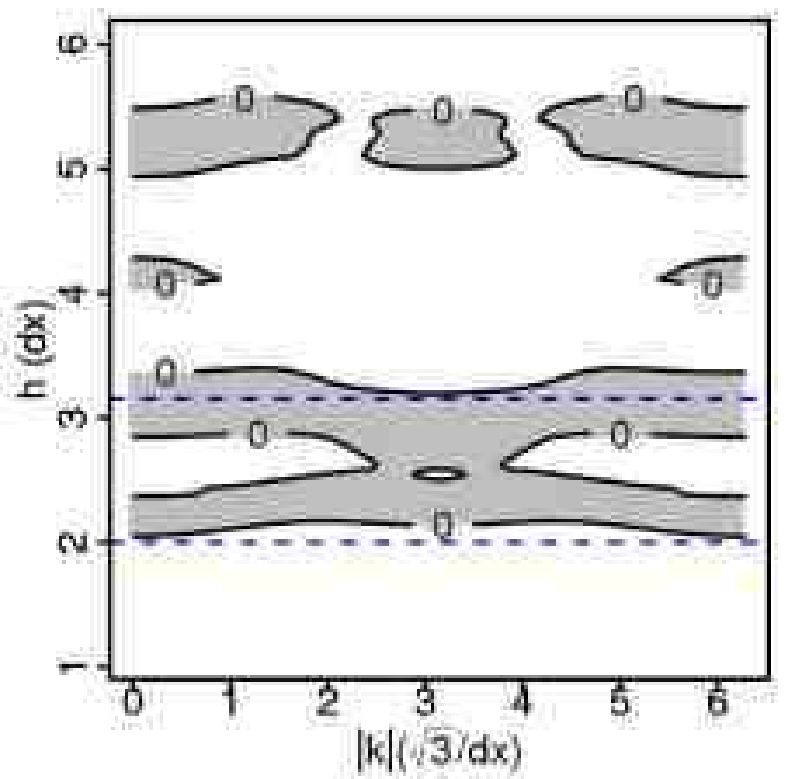}\\
\includegraphics[height=0.33\textwidth]{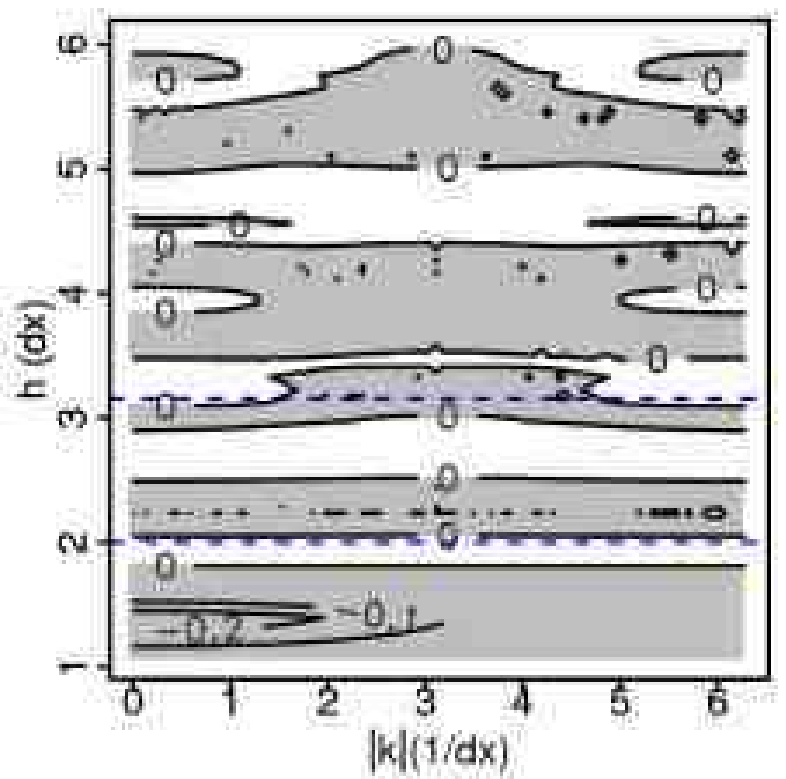}\hspace{\myfix}
\includegraphics[height=0.33\textwidth]{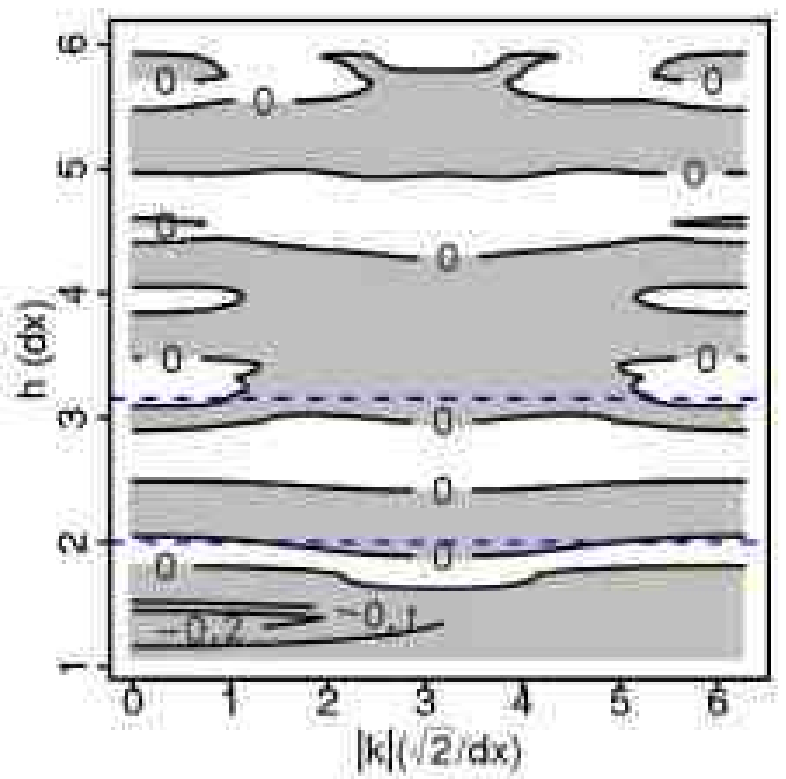}\hspace{\myfix}
\includegraphics[height=0.33\textwidth]{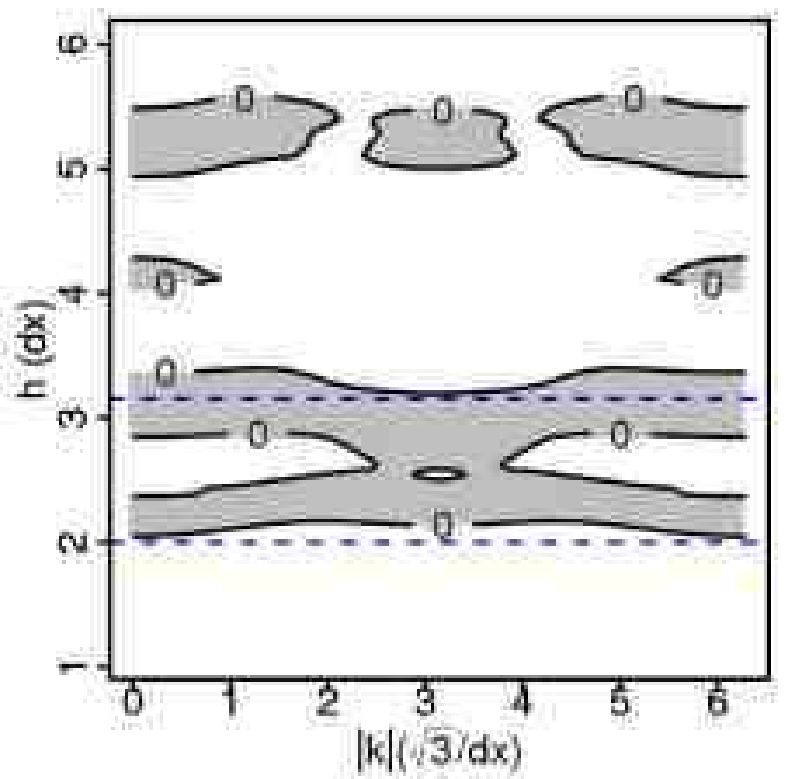}\\

\caption{As Figure \ref{fig:stab}, but for the Core Triangle (CT)
  kernel.}
\label{fig:stabct}
\end{figure*}
\end{center}

\begin{center}
\begin{figure*}
High Order Core Triangle HOCT4 kernel:\\
\includegraphics[height=0.33\textwidth]{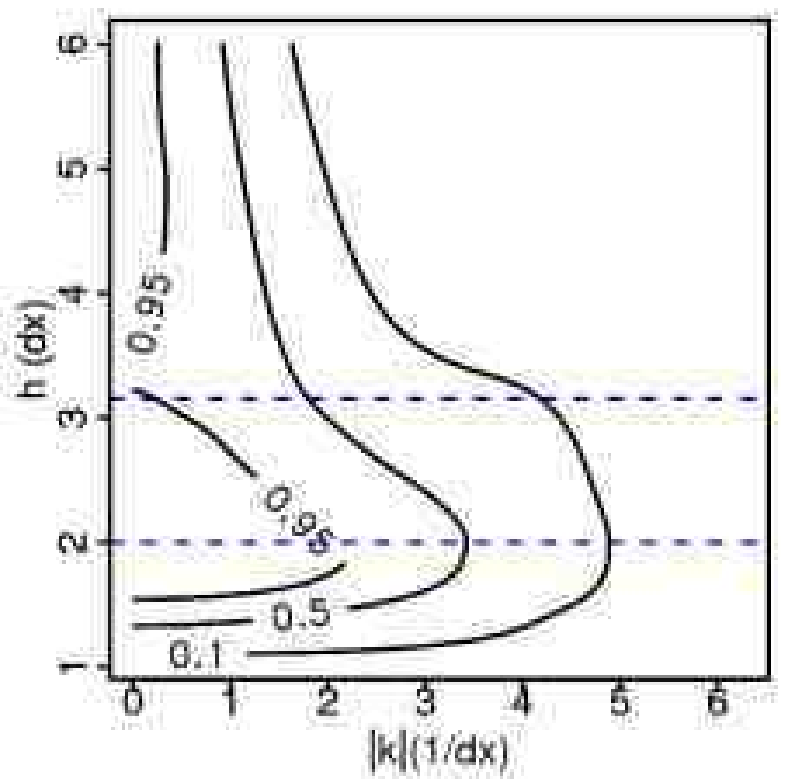}\hspace{\myfix}
\includegraphics[height=0.33\textwidth]{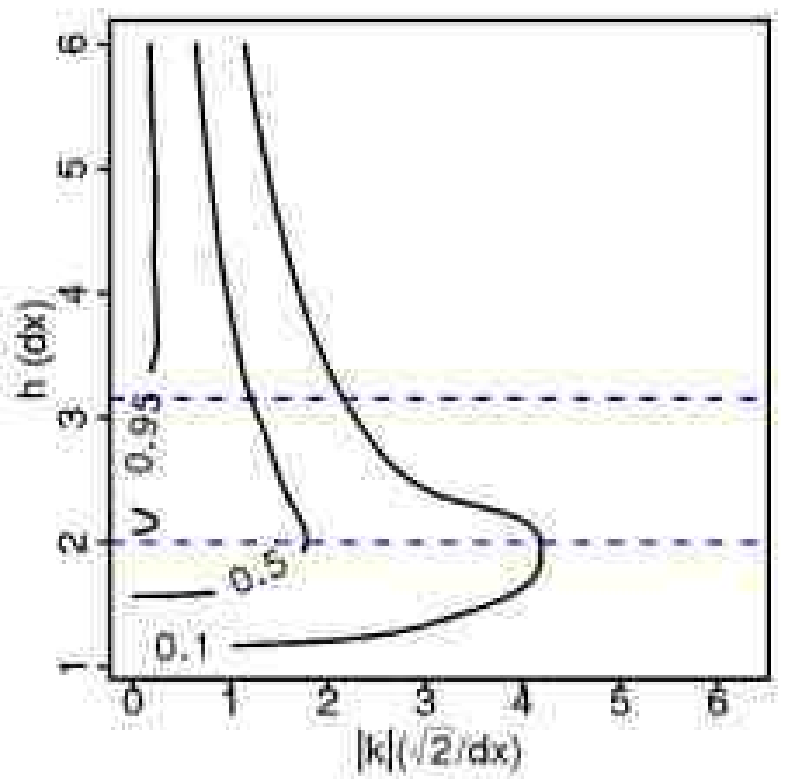}\hspace{\myfix}
\includegraphics[height=0.33\textwidth]{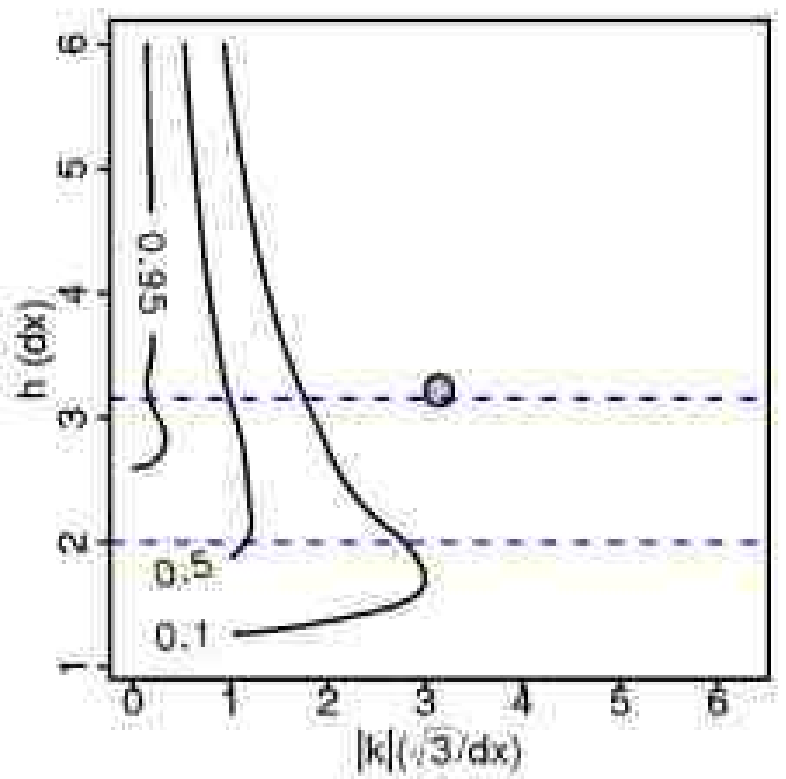}\\
\includegraphics[height=0.33\textwidth]{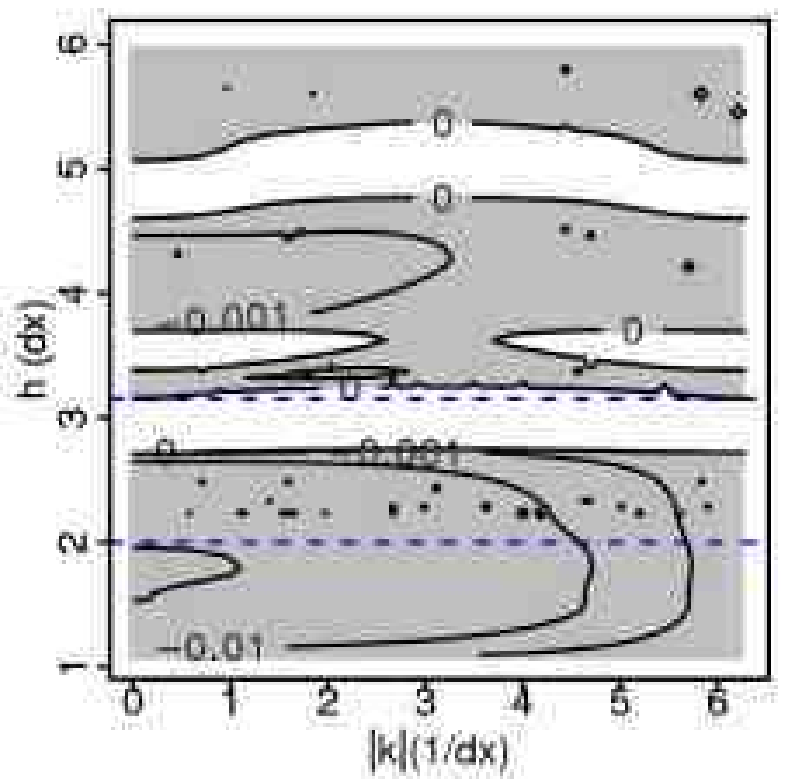}\hspace{\myfix}
\includegraphics[height=0.33\textwidth]{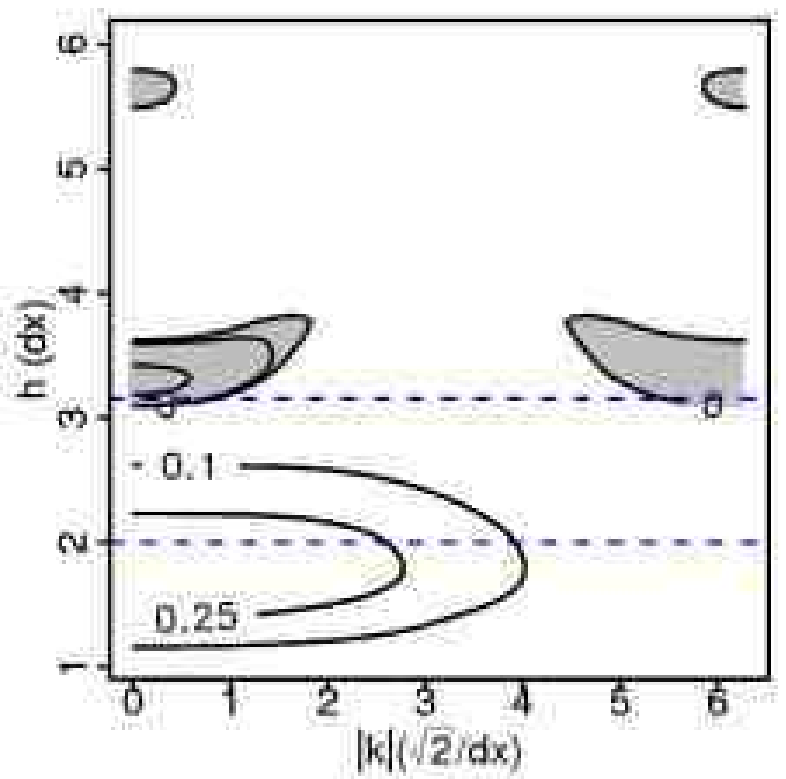}\hspace{\myfix}
\includegraphics[height=0.33\textwidth]{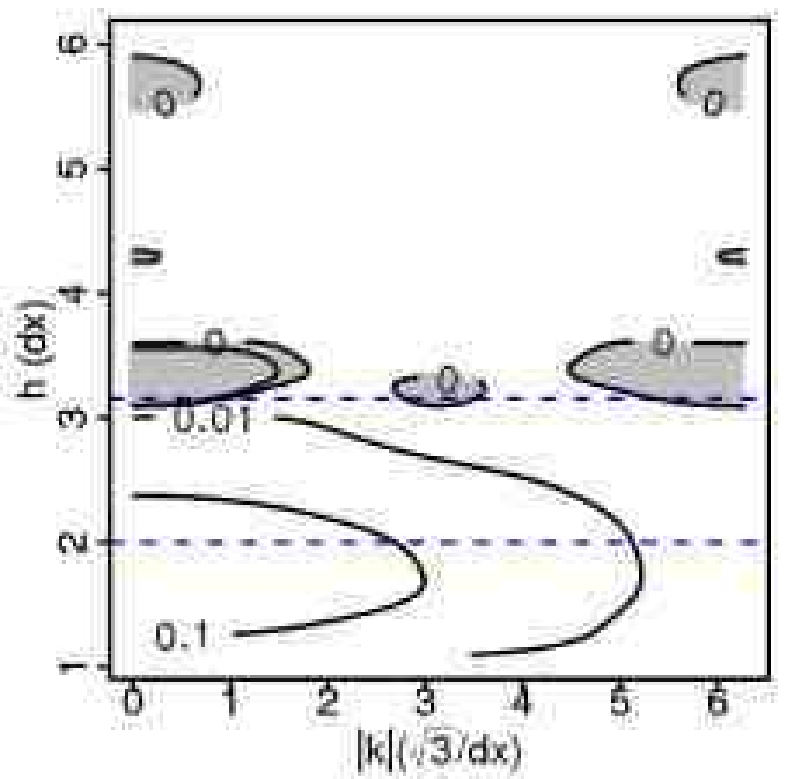}\\
\includegraphics[height=0.33\textwidth]{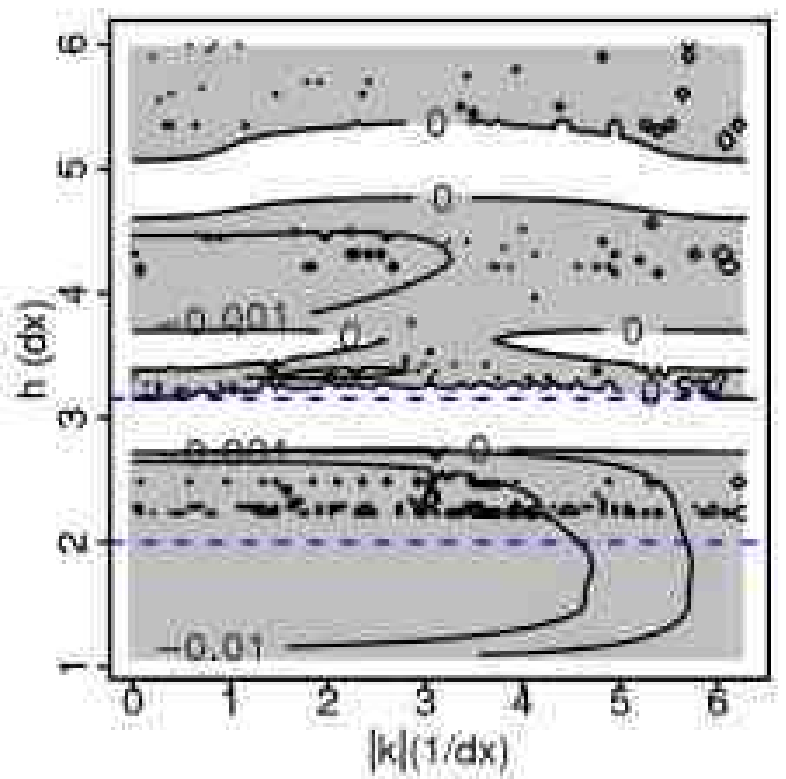}\hspace{\myfix}
\includegraphics[height=0.33\textwidth]{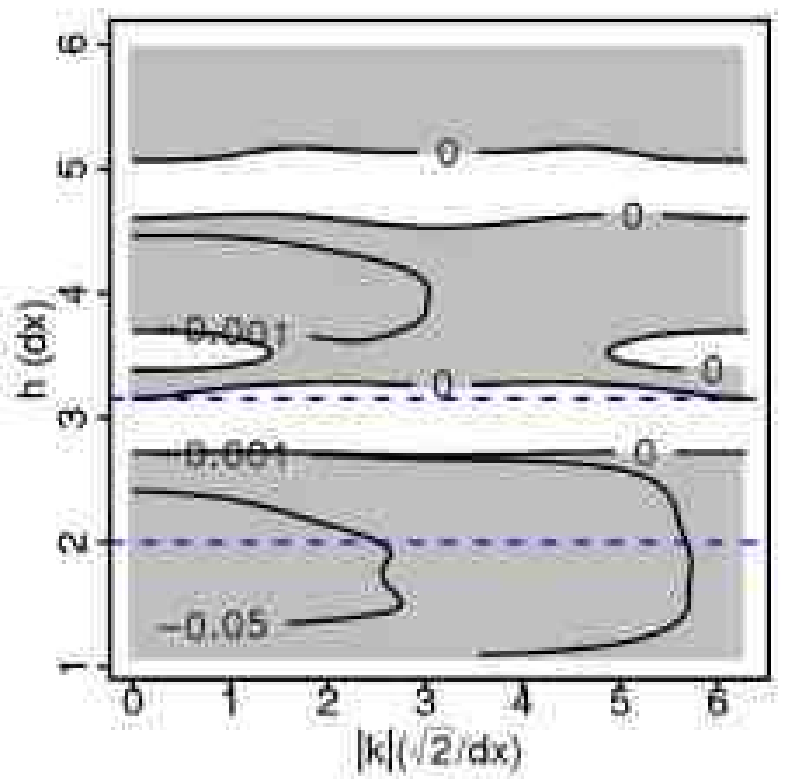}\hspace{\myfix}
\includegraphics[height=0.33\textwidth]{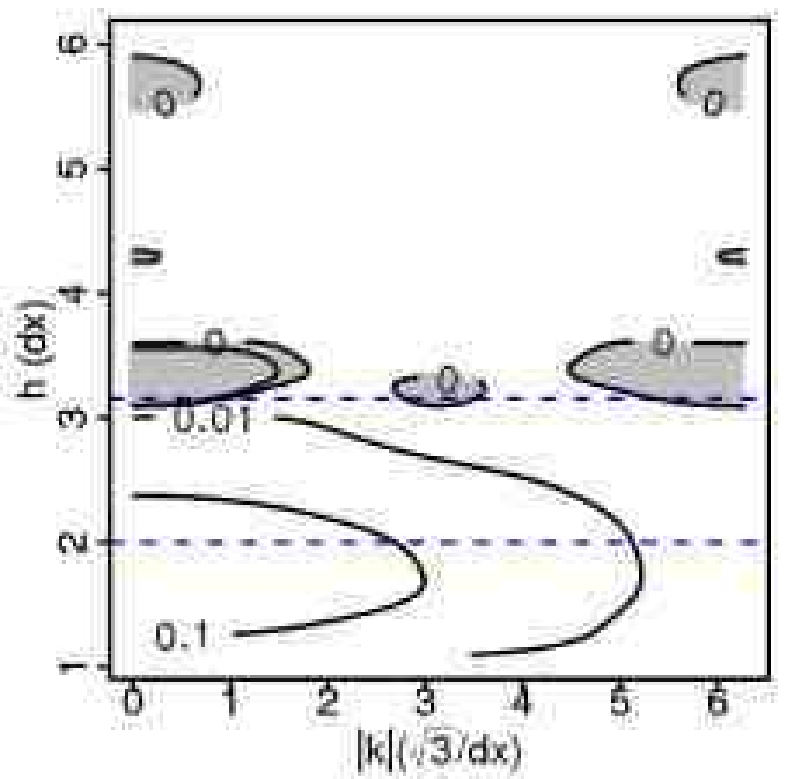}\\
\caption{As Figure \ref{fig:stab}, but for the High Order Core
  Triangle (HOCT) kernel with $n_k = 4$.}
\label{fig:stabhoct}
\end{figure*}
\end{center}
\vspace{-20mm}

\subsubsection{Minimising errors: the momentum equation} 

The momentum equation (equation \ref{eqn:momenterr}) is more problematic than the continuity equation. Its accuracy is governed not only by the extent to which $\Mmv_i = \Id$, but primarily by the leading $\uEi$ term that should vanish. 

First, consider the situation in standard SPH where $\phiv = 1$ and $g_{ij} = \rho_j / \rho_i$. As for the continuity equation, in the continuum limit ($N\rightarrow \infty; h\rightarrow 0$), $\uE = \zero + O(h^2)$ since $\unablax_i \wijtild$ is antisymmetric. However, this analysis is only relevant if the particles are smoothly distributed on the kernel scale. For irregularly distributed particles, $g_{ij}$ can grow arbitrarily large, while $m_j/\rho_j$ is not guaranteed to be a good volume estimate. In such situations, $\uE$ contributes a significant error. Worse still, moving to higher resolution is not guaranteed to help. In order for the SPH integration to converge as $h \rightarrow 0$, we require that $\uEi$ shrinks faster than $\frac{P_i}{h \rho_i}$. This requires some care in making sure that $h$ does not shrink too fast as the number of particles is increased.  

A density step is an extreme example of an irregular particle distribution, and this suggests that the $\uE$ error is at least in part responsible for SPH's failure to correctly model mixing processes between different fluid phases. We demonstrate this in \S\ref{sec:kh}.

There are three key problems with ensuring that $\frac{P_i}{h \rho_i} \uEi$ will shrink with increasing resolution. The first is the function $g_{ij}$. In SPH, this is the ratio $\rho_j/\rho_i$ which is large when there are large density gradients. We can significantly improve on this we choose our free function $\phiv = \rho$. In this case, we have $g_{ij} = g_{ij}^{-1} = 1$ by construction, and $g_{ij}$ no longer contributes to the $\uE$ error even for large density gradients across the kernel\footnote{It is
  interesting to note that other work in the literature has also found that $\phiv =
  \rho$ is the preferred choice if density gradients are large
  (\bcite{2007Oger}; \bcite{2005astro.ph..7472P};
  \bcite{2001MNRAS.323..743R}; \bcite{1999Dilts}). But we could not
  find a detailed proof similar to that presented here. Interestingly,
  \citet{2003MNRAS.345..561M} find empirically that $\phiv =
  \rho^{3/2}$ gives the best performance for multiphase flow in their
  tests. Our analytic results here suggest that this is not the
  optimal choice, though perhaps the inclusion of cooling and/or other
  physics makes a difference.}. The second problem relates to kernel scale smoothness. If particles clump or band on the kernel scale, then we will have poor kernel sampling and $\uE$ will not approach its integral limit even at very high resolution. Ensuring that this does not happen means ensuring that our \MSPH\ scheme is {\it stable} to perturbations. We discuss this next in \S\ref{sec:minimise}. The third and final problem is the volume estimate of each particle $m_j/\rho_j$. This will be poor if the particles are irregularly distributed on the kernel scale (for example at a density step) leading to a large $\uE$ error. We discuss this further in \S\ref{sec:localmix}.

The choices $\phiu = \phir = 1/u$; $\phiv =
  \rho$ and the kernel constraints $\hijtild = \lijtild = \kijtild = 
  \unabla_i\wijtild$ are the first important ingredients in our \MSPH\
  scheme. These choices mean that we are no longer coherent\footnotemark, but this
  only introduces tolerable $O(h^2)$ errors in the energy conservation 
\citep{1989ApJS...70..419H}.

\footnotetext{Note that it is possible to construct pseudo-coherent versions of \MSPH\ using $\phiu = \phir = \phiv = 1/u$ for the energy form, or $\phiu = \phir = \phiv = 1/A^{\frac{1}{\gamma}}$ for the entropy form. Introducing `grad $h$' terms as in \citet{nelsonpapagradh94}, such schemes can then be made to conserve energy exactly in the limit of constant timesteps. However, they are only truly coherent up to the approximation that the $\epsilon$ error in the continuity equation is small (see equation \ref{eqn:epsilerror}). Nonetheless, it would be interesting to explore such schemes in future work.}

\subsection{Stability: the choice of kernel function}\label{sec:minimise}

In \S\ref{sec:error}, we used an error analysis of the generalised SPH equations of motion to show that the dominant source of error in SPH is in the momentum equation -- the $\uE$ error. We showed that choosing the free functions $\phiu = \phir = 1/u$; $\phiv = \rho$ and the kernel constraints $\hijtild = \lijtild = \kijtild = \unabla_i\wijtild$ should minimise both this error and errors in the continuity equation, and we called these choices \MSPH. 

In \MSPH, provided the particles are regularly distributed on the kernel scale, we can make $\uE$ arbitrarily small simply by increasing the neighbour number. However, if the particles are irregularly distributed, $\uE$ can shrink very slowly with increasing resolution. In this section, we show that for large neighbour number the cubic spline kernel typically used in SPH calculations is unstable to both particle clumping (\S\ref{sec:clumpinginst}) and particle banding (\ref{sec:bandinginst}), and we derive a new class of kernels that are stable to both even for large neighbour number. In \S\ref{sec:kh}, we show that these new kernels give significantly improved performance at no additional computational cost.

\subsubsection{The clumping instability}\label{sec:clumpinginst}

The clumping instability\footnote{Also called the tensile instability.} can be derived from a linear 3D stability analysis of the \MSPH\ equations of motion. Following \bcite{1996PASA...13...97M} and \bcite{1996PhDMorris}, we imagine a lattice of equal masses $m$ of equal separation $(\Delta x_{0},\Delta y_{0},\Delta z_{0})$ with initial density $\rho_0$ and pressure $P_0$. We perturb these with a linear wave of the form:

\begin{equation}
\ux_i = \ux_{0,i} + \ua \exp[i(\uk\cdot\ux_{0,i} - \omega t)]
\label{eqn:a}
\end{equation}
\begin{equation}
\rho_{i} = \rho_0 + D \exp[i(\uk\cdot\ux_{0,i} - \omega t)]
\label{eqn:pert2}
\end{equation}
\begin{equation}
P_{i} = P_0 + c_s^2 D \exp[i(\uk\cdot\ux_{0,i} - \omega t)]
\label{eqn:pert3}
\end{equation}
and similar for particle $j$, where $c_s^2 = \frac{\partial
  P}{\partial \rho} = \gamma P_0/\rho_0$ is the sound speed assuming
  an adiabatic equation of state ($\gamma = 1$ gives an isothermal
  equation of state), $\ua = (X,Y,Z)$ is the amplitude of the
  perturbation and $\uk = (k_x,k_y,k_z)$ is the wave vector.

To simplify the analysis, we assume that we have a lattice symmetry
such that for every displacement vector $\ux_{0,ij} = \ux_{0,i} - \ux_{0,j}$ to a
  neighbour, there is also one at $-\ux_{0,ij}$. Then, plugging
  equations \ref{eqn:a},  
  \ref{eqn:pert2} and \ref{eqn:pert3} in to
\ref{eqn:moment}, discarding 
terms higher than first order, and connecting $D$ to $X,Y,Z$ through
the continuity equation\footnote{We use here the full \MSPH\ continuity
  equation in the entropy form (equation \ref{eqn:cont} with
  $\phir=1/A^{1/\gamma}$). However, for plane waves on a constant
  density lattice $A_j/A_i = 1$ and so this is identical to 
  the SPH continuity equation with $\phir=1$.} ($D = m\sum_j
(1-e^{i\phi_{ji}})\left[\frac{\partial \wijtild}{\partial x_i}X +
  \frac{\partial \wijtild}{\partial y_i}Y+\frac{\partial
  \wijtild}{\partial z_i}Z\right]$), we obtain the 3D \MSPH\ dispersion
relation\footnote{This is actually identical to the SPH dispersion
  relation derived under the same assumptions in 3D
  \citep{1996PhDMorris}.}:

\begin{eqnarray}
  \omega^2\ua & = & \left[\frac{2mP_0}{\rho_0^2}\sum_j \uH
    (\overline{W}_{0,ij}) (1-\cos\uk\cdot\ux_{0,ij})
    +\right.\nonumber \\ 
 & &  \left. (\gamma - 2)\frac{m^2 P_0}{\rho_0^3}(\uq_i \wedge \uq_i)\right]\cdot\ua
\label{eqn:disp}
\end{eqnarray}
where $\uq_i\wedge\uq_i$ is the outer product of $\uq_i$, $\uH(\overline{W})$ is the Hessian\footnote{Recall that the outer product of two vectors is a matrix, while the Hessian is a square matrix of second order partial derivatives.} of $\overline{W}$:

\begin{equation}
    H_{aa} = \ppderiv{\overline{W}}{x_a} = \dderiv{\overline{W}(r)}{r}
    \frac{{x_a}^2}{r^2} 
    + 
    \deriv{\overline{W}(r)}{r} \frac{1}{r}\left(1 - \frac{{x_a}^2}{r^2} \right)
\label{eqn:haa}
\end{equation}
\begin{equation}
    H_{ab} = \ptderiv{\overline{W}}{x_a}{x_b} =
    \dderiv{\overline{W}(r)}{r} \frac{x_a 
    x_b}{r^2} - 
    \deriv{\overline{W}(r)}{r} \frac{x_a x_b}{r^3}
\label{eqn:hbb}
\end{equation}
and $\uq_i$ is given by:

\begin{equation} \label{eqn:q}
  \uq_i = \sum_j \sin \uk\cdot\ux_{0,ij} \unabla_i \overline{W}_{0,ij}
\end{equation}
Our scheme is stable if $\omega^2 \ge 0$. It is also desirable for
the numerical sound speed to equal the true sound speed: $\omega^2/k^2
= c_s^2$. 

In SPH it is typical to use the cubic spline (CS) kernel given by:

\begin{equation}
W = \frac{8}{\pi h^3}\left\{\begin{array}{lr}
 1 - 6x^2 + 6x^3 & 0 < x \le \frac{1}{2}\\
2(1-x)^3 & \frac{1}{2} < x \le 1\\ 
0 & \mathrm{otherwise} \end{array}\right.
\label{eqn:cubicspline}
\end{equation}
where $x=r/h$ is the distance from the centre of the kernel in units
of the smoothing length.

In Figure \ref{fig:stab}, we show contours of
$\omega^2 / k^2 / c_s^2$ as a function of wavenumber $k$
and the smoothing length $h$ in units of the
inter-particle spacing $dx=1$, for the CS kernel. We assume an adiabatic
equation of state with $\gamma = 5/3$. From left to right the
plots show $(k_x,k_y,k_z) = k(1,0,0), k(1,1,0)$ and
$k(1,1,1)$. The three rows show the longitudinal wave and the two
transverse waves for these orientations. 

From Figure \ref{fig:stab}, it is clear that the CS kernel in 3D is
unstable to longitudinal waves for $h \simgt 2$, and very unstable to
transverse waves. The unstable longitudinal waves drive the clumping, or
tensile instability that causes particles to clump on the kernel scale
(\bcite{1981A&A....97..373S}; \bcite{1992MNRAS.257...11T};
\bcite{1994MmSAI..65.1013H};  \bcite{1996PASA...13...97M};
\bcite{2000Monaghan}).

\begin{center}
\begin{figure*}
SPH-CS-128\\
\includegraphics[height=0.23\textwidth]{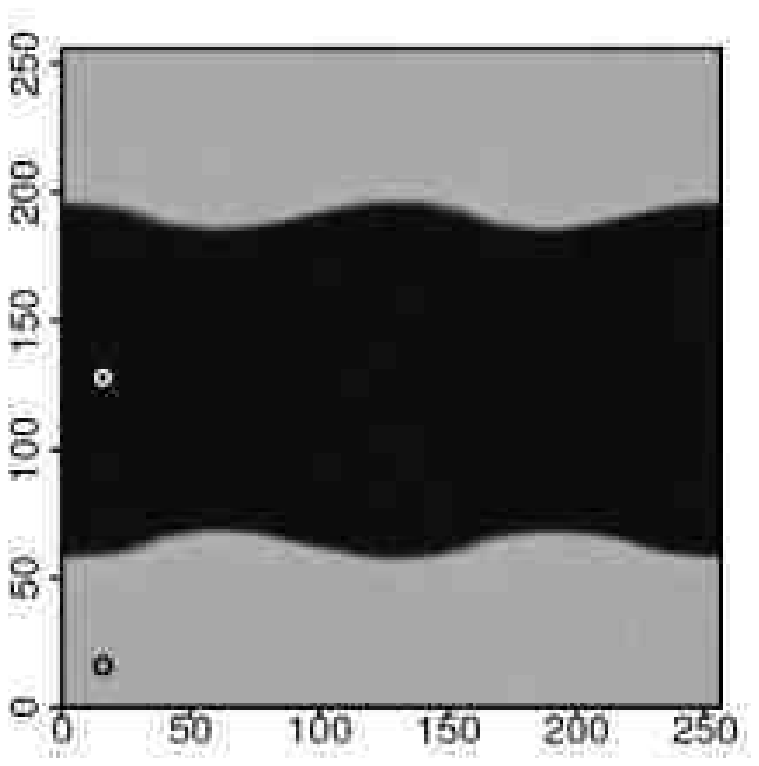}\hspace{\myfix}
\includegraphics[height=0.23\textwidth]{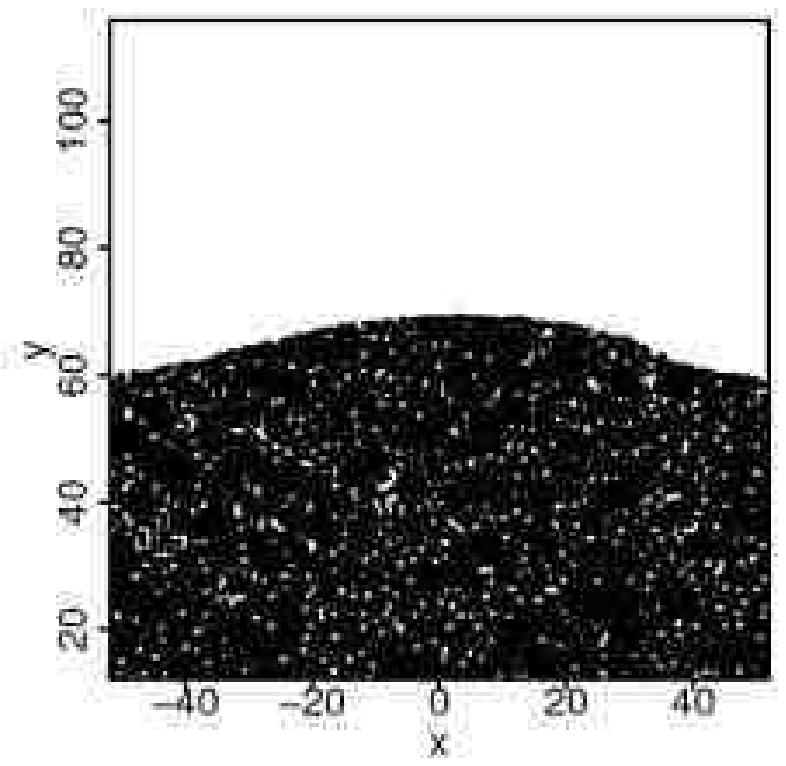}\hspace{\myfix}
\includegraphics[height=0.23\textwidth]{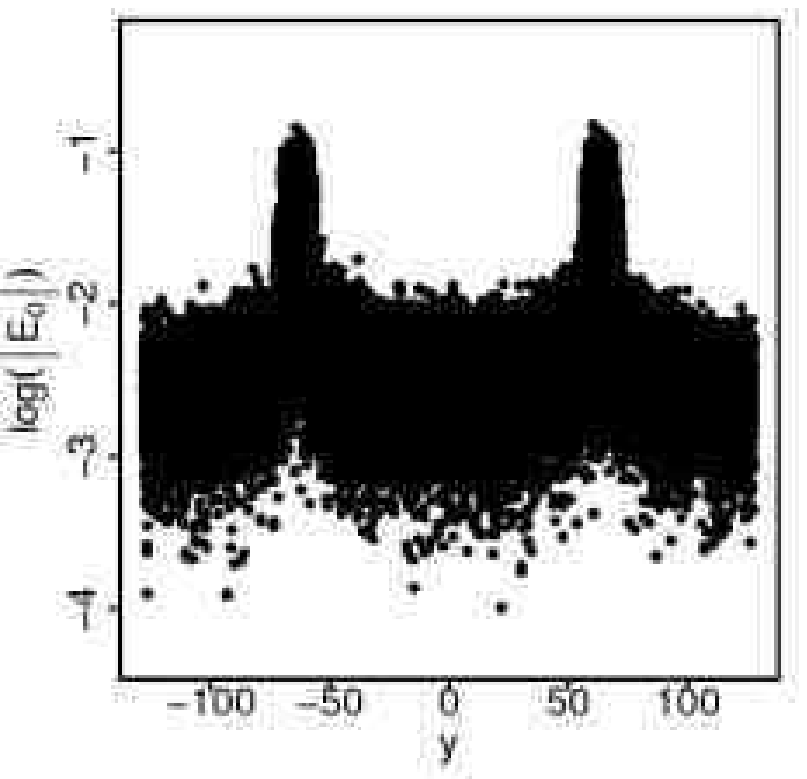}\hspace{\myfix}
\includegraphics[height=0.23\textwidth]{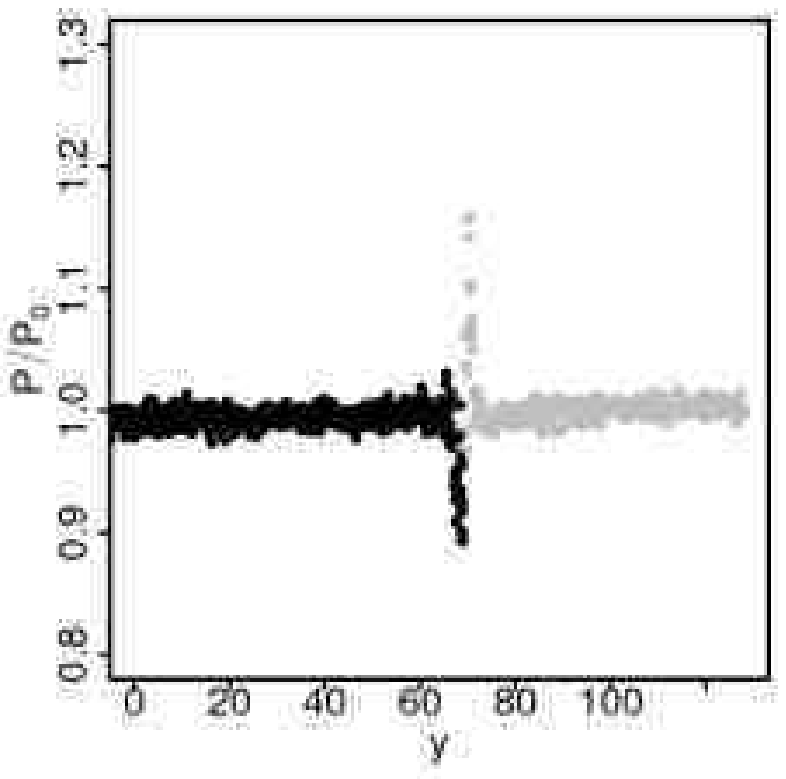}\\
\SPHS-CS-128\\
\includegraphics[height=0.23\textwidth]{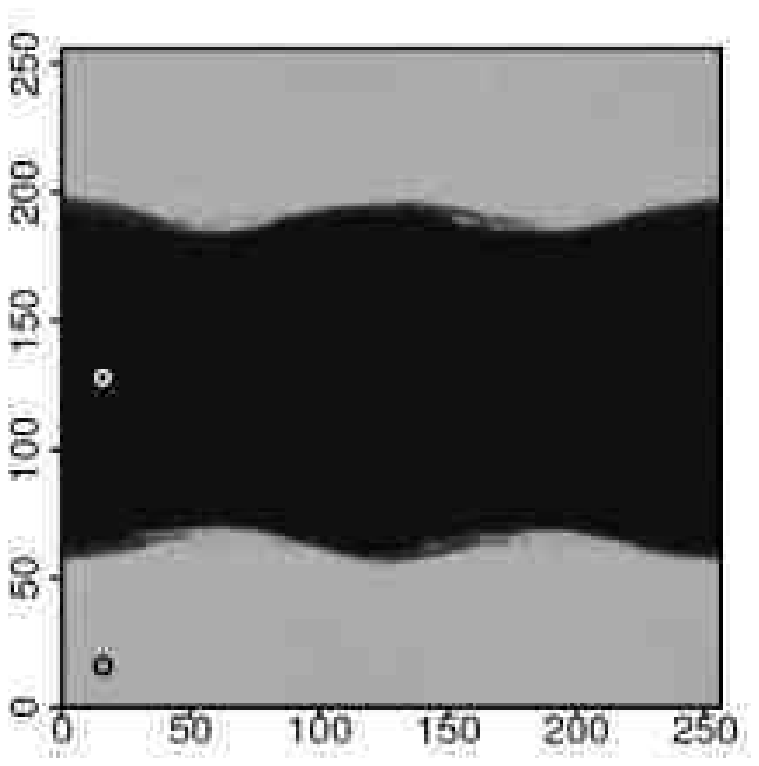}\hspace{\myfix}
\includegraphics[height=0.23\textwidth]{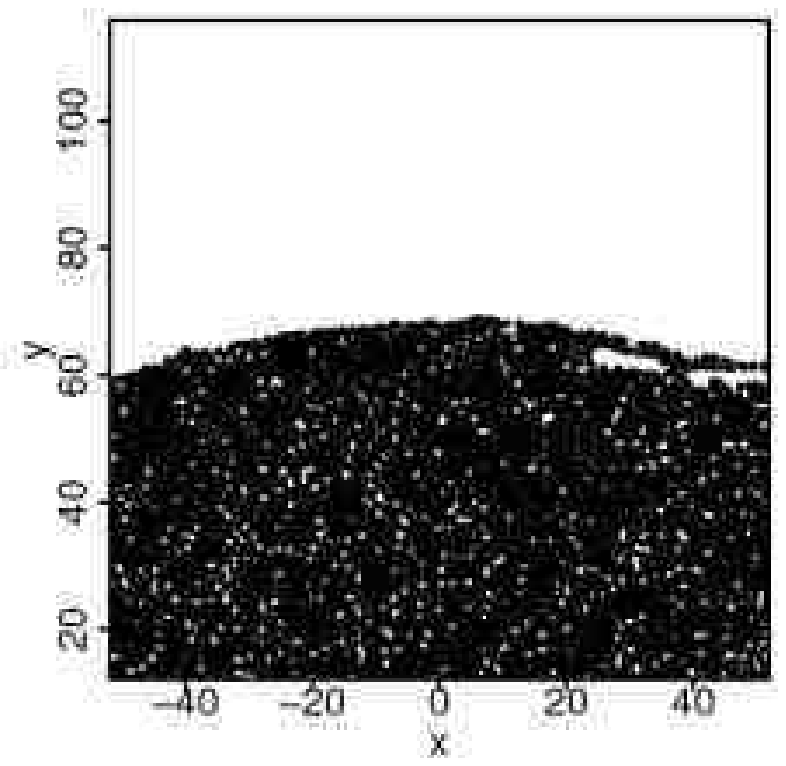}\hspace{\myfix}
\includegraphics[height=0.23\textwidth]{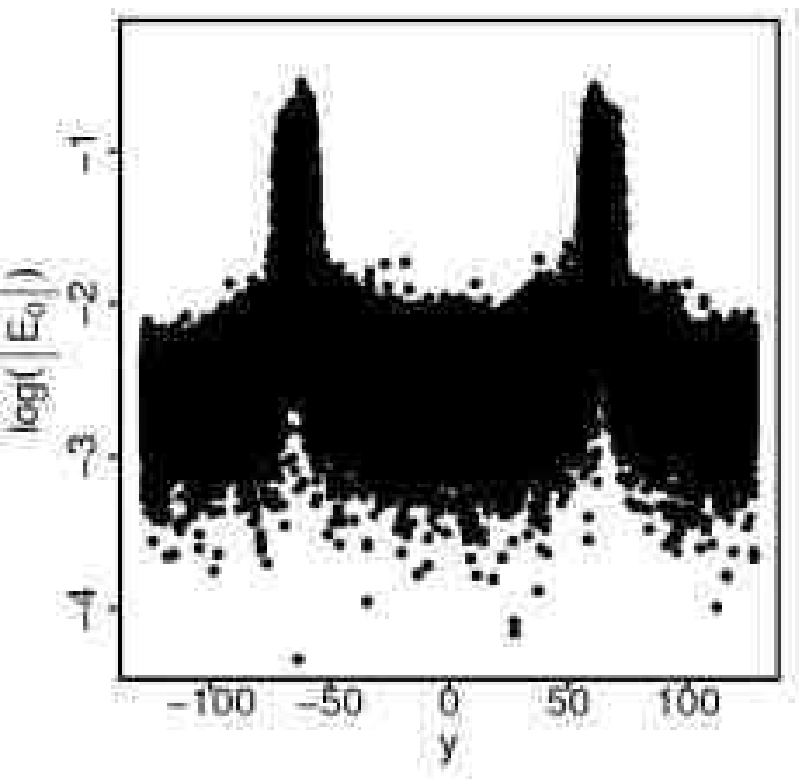}\hspace{\myfix}
\includegraphics[height=0.23\textwidth]{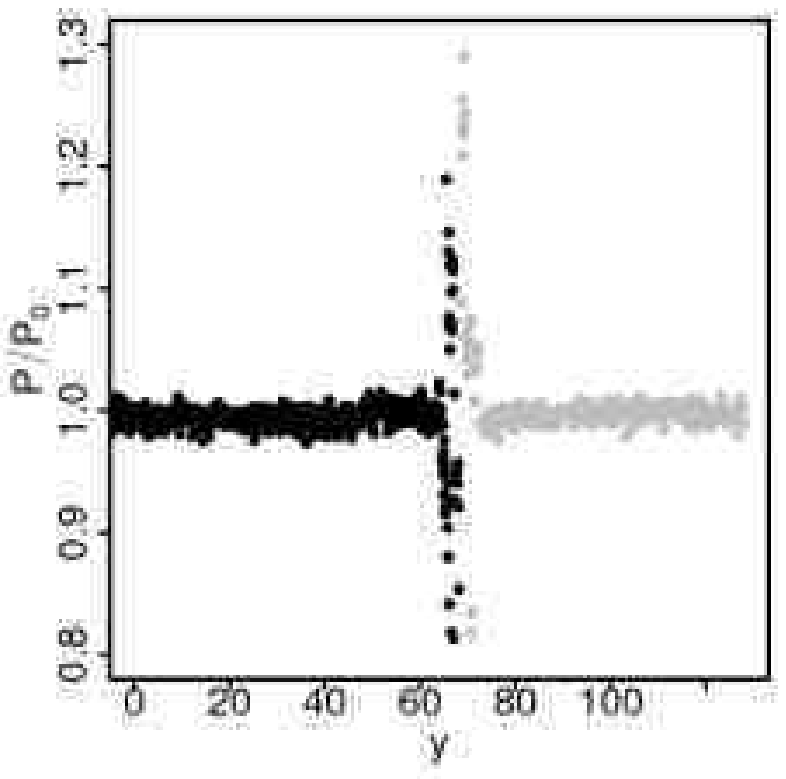}\\
\SPHS-CT-128\\
\includegraphics[height=0.23\textwidth]{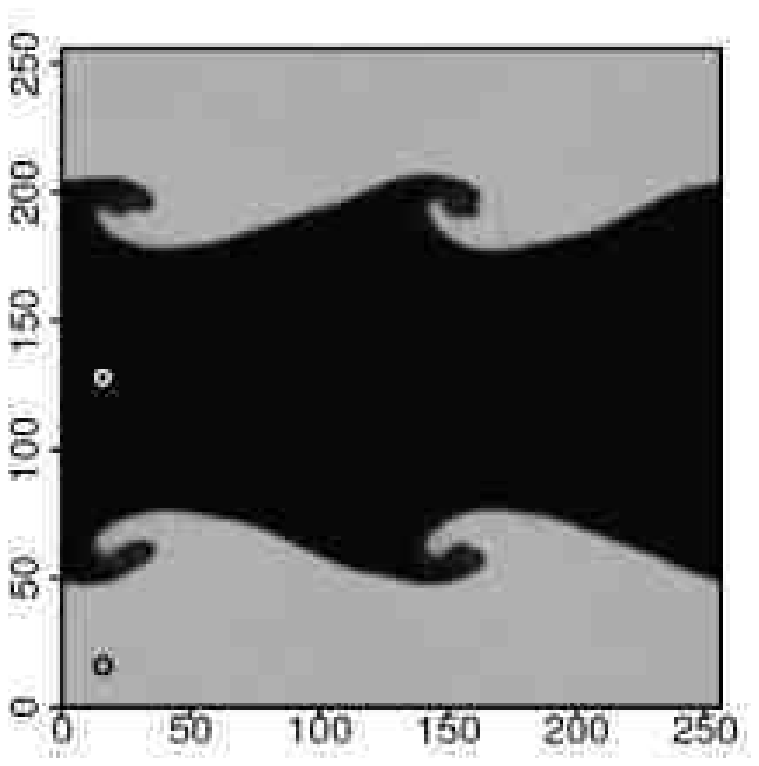}\hspace{\myfix}
\includegraphics[height=0.23\textwidth]{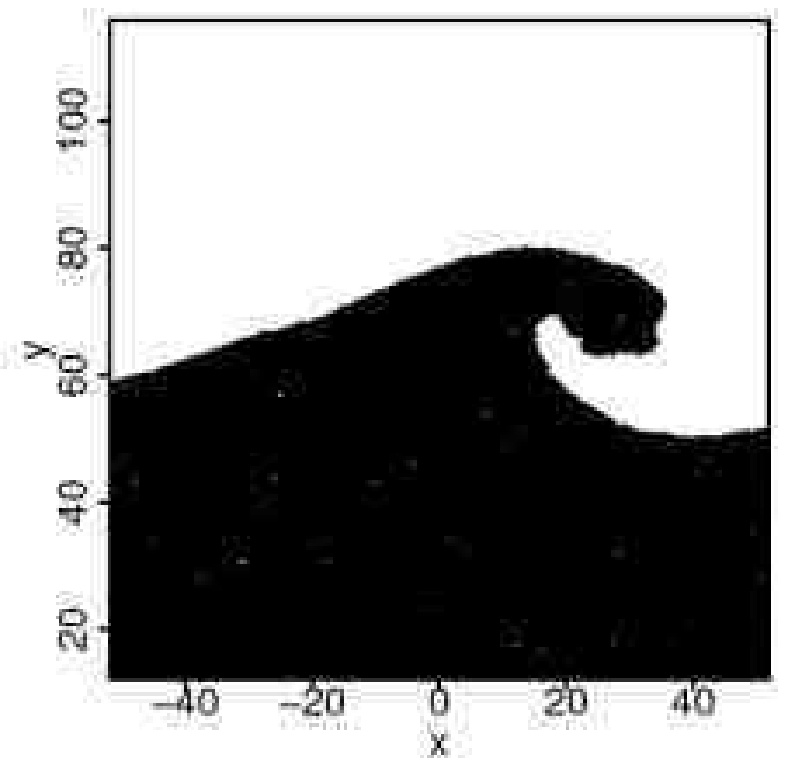}\hspace{\myfix}
\includegraphics[height=0.23\textwidth]{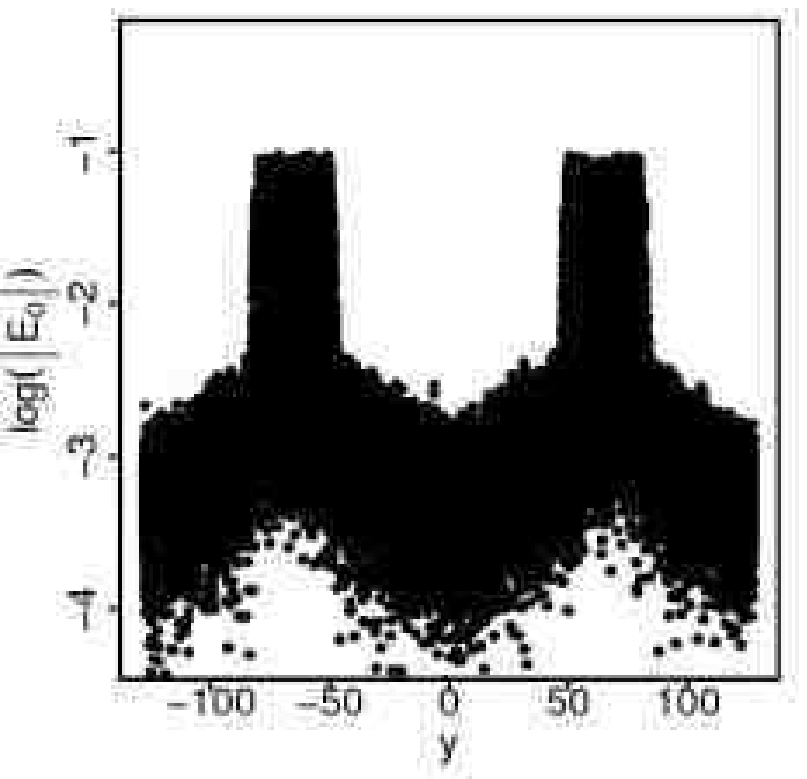}\hspace{\myfix}
\includegraphics[height=0.23\textwidth]{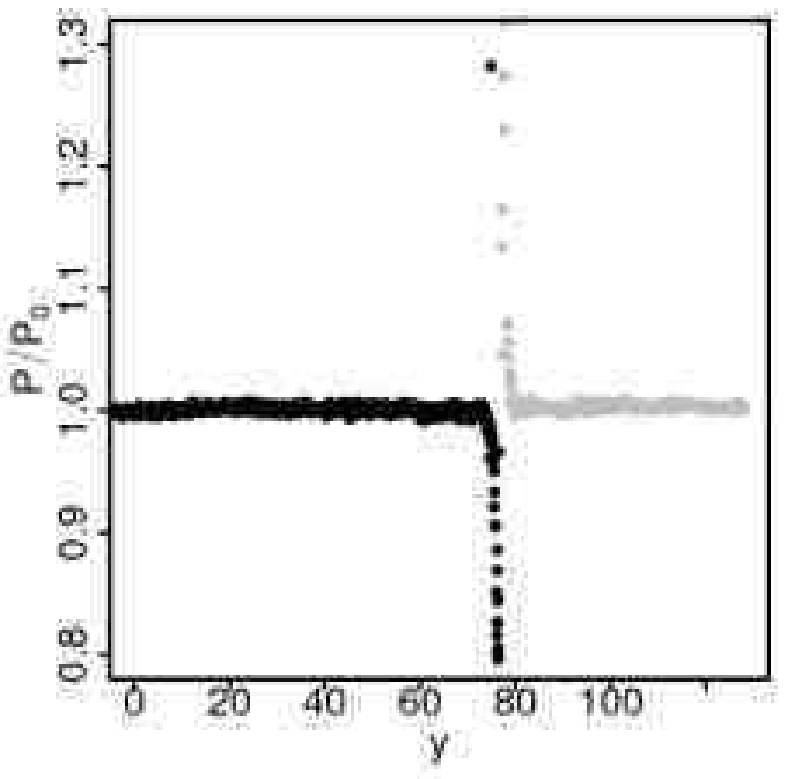}\\
\SPHS-HOCT4-442\\
\includegraphics[height=0.23\textwidth]{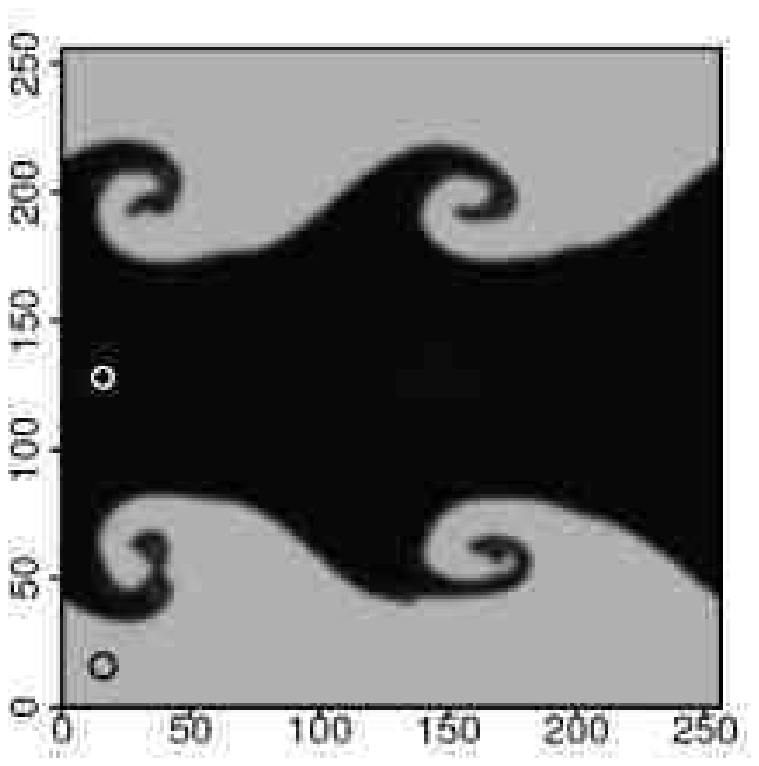}\hspace{\myfix}
\includegraphics[height=0.23\textwidth]{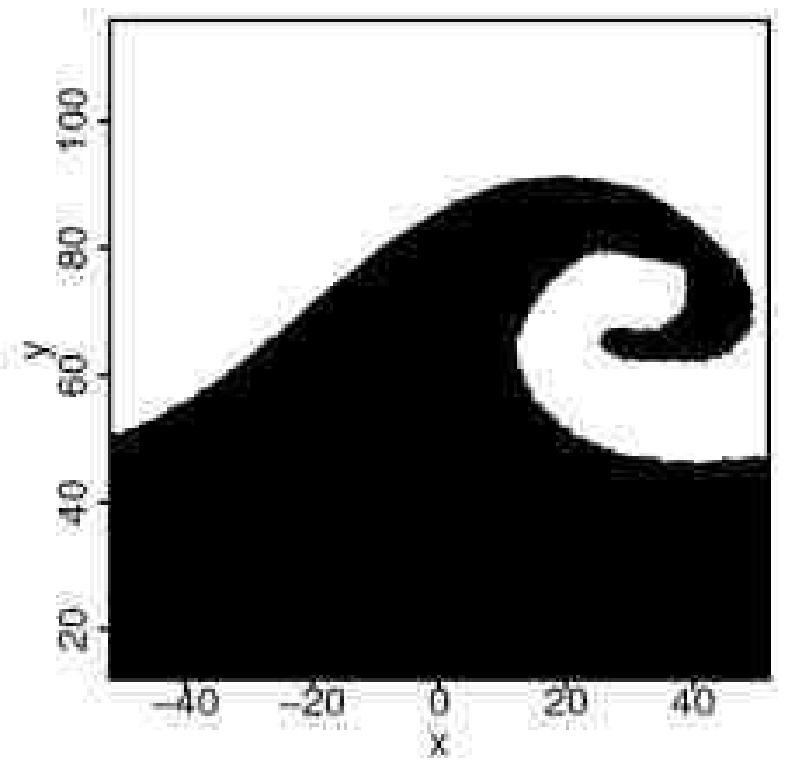}\hspace{\myfix}
\includegraphics[height=0.23\textwidth]{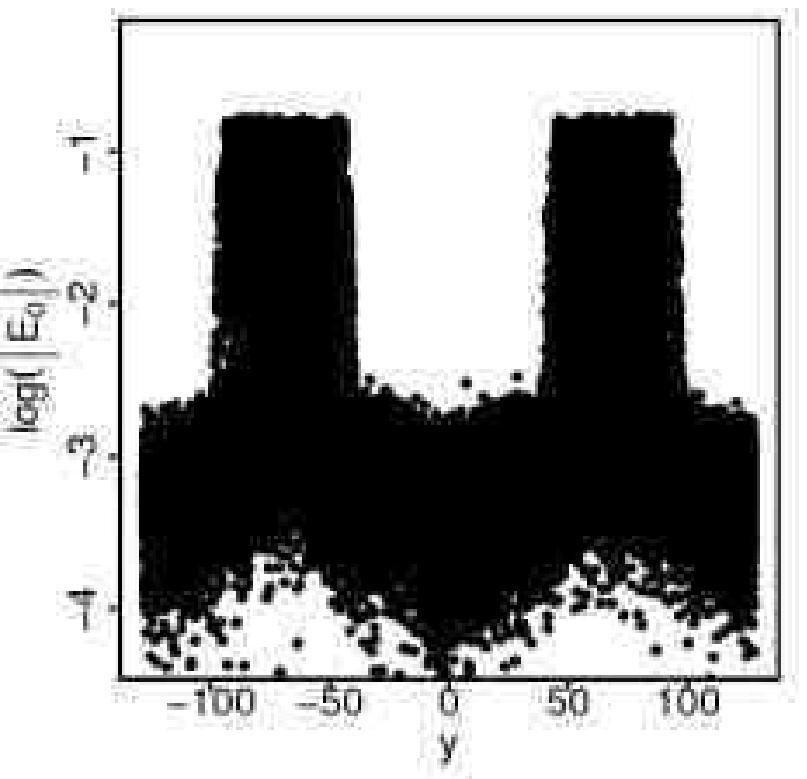}\hspace{\myfix}
\includegraphics[height=0.23\textwidth]{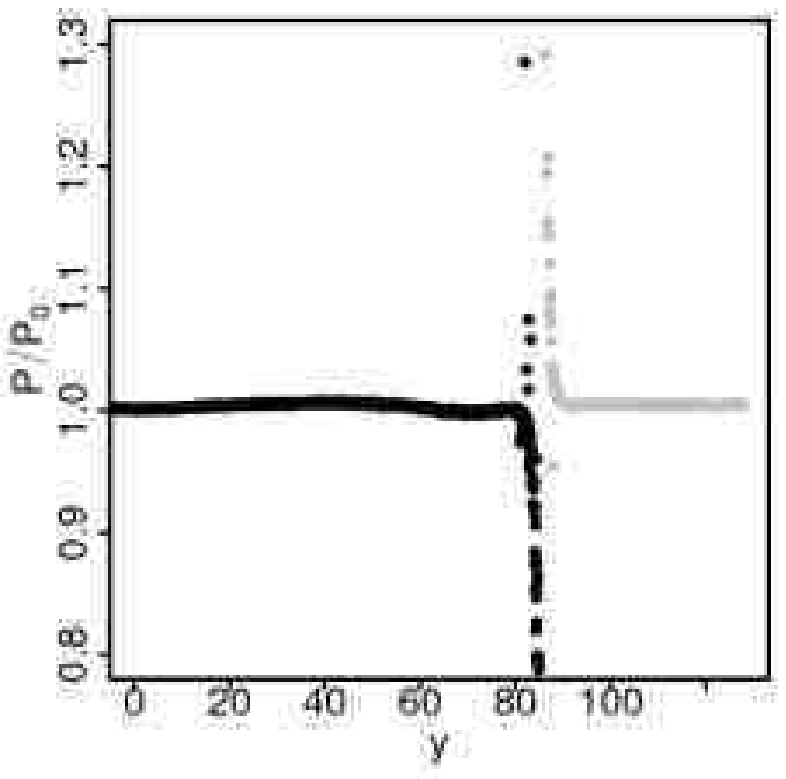}\\
\MSPH-HOCT4-442\\
\includegraphics[height=0.23\textwidth]{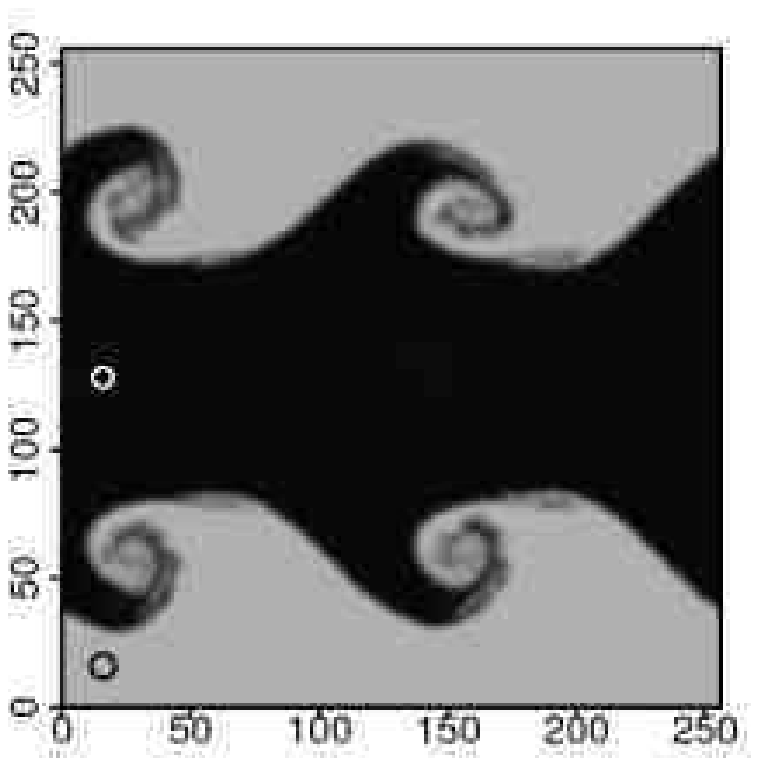}\hspace{\myfix}
\includegraphics[height=0.23\textwidth]{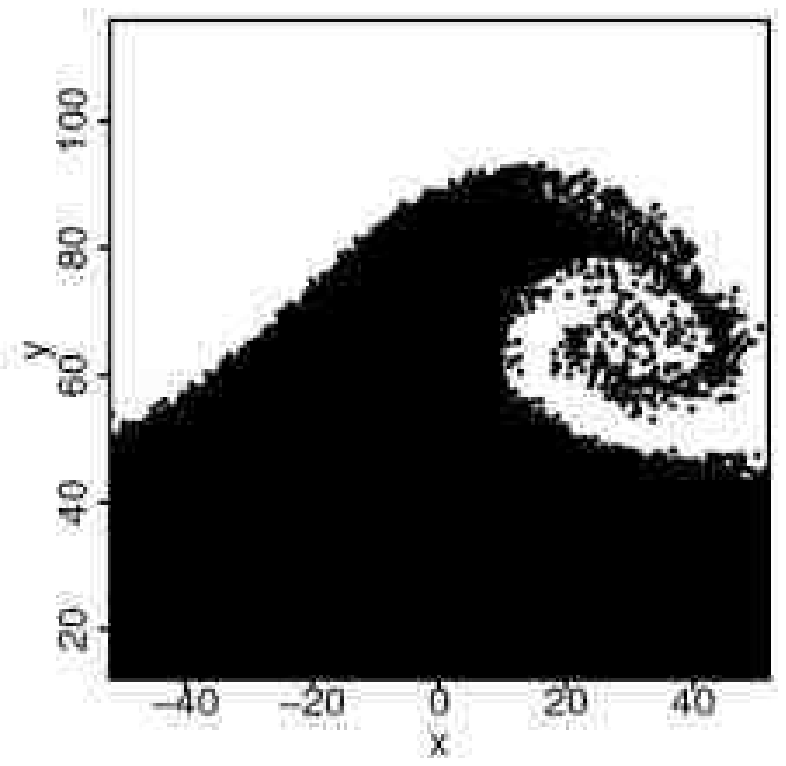}\hspace{\myfix}
\includegraphics[height=0.23\textwidth]{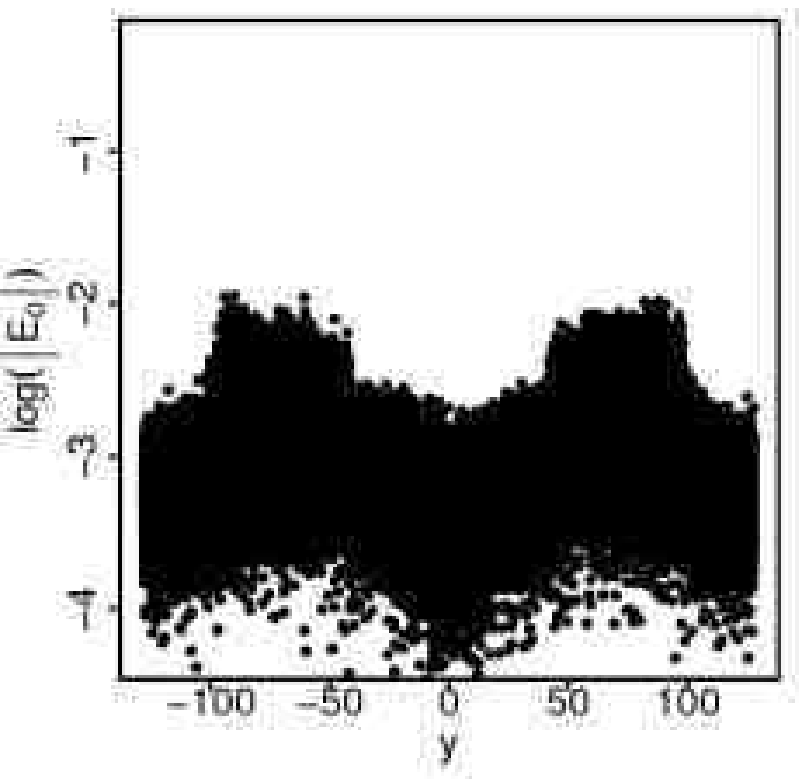}\hspace{\myfix}
\includegraphics[height=0.23\textwidth]{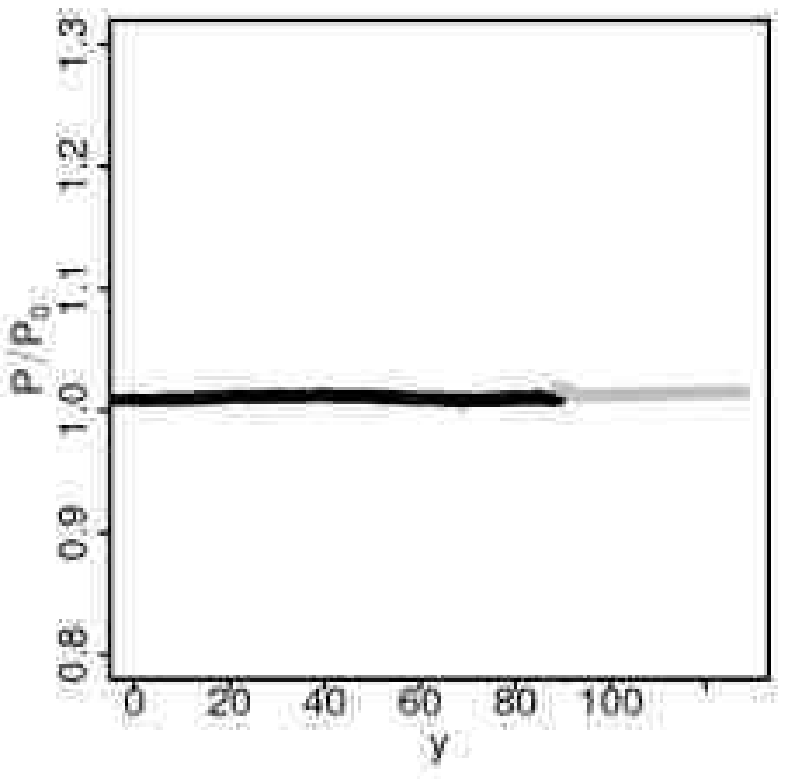}\\
\caption{A Kelvin-Helmholtz instability (density ratio $R_\rho = 2$)
  at $\tau_\mathrm{KH}=1$ modelled with SPH, \SPHS\ and \MSPH\ using CS,
  CT and HOCT4 
  kernels (see equations \ref{eqn:cubicspline}, \ref{eqn:wkern} and
  \ref{eqn:hoctkern}). From left to right the plots show, in a slice of
  width $dx=1$ about the z-axis: density
  contours; a zoom-in on the particle distribution around one of
  the rolls; the magnitude of the $|\uE|$ error
  (see equation \ref{eqn:e0int}) as a function of $y$; and the
  pressure in a slice of width $dx=1$ about the x-axis, as a function of $y$. The circles on the density contour plots mark the size of the smoothing kernel, $h$.}
\label{fig:clump}
\end{figure*}
\end{center}

\begin{center} 
\begin{figure*}
\includegraphics[height=0.24\textwidth]{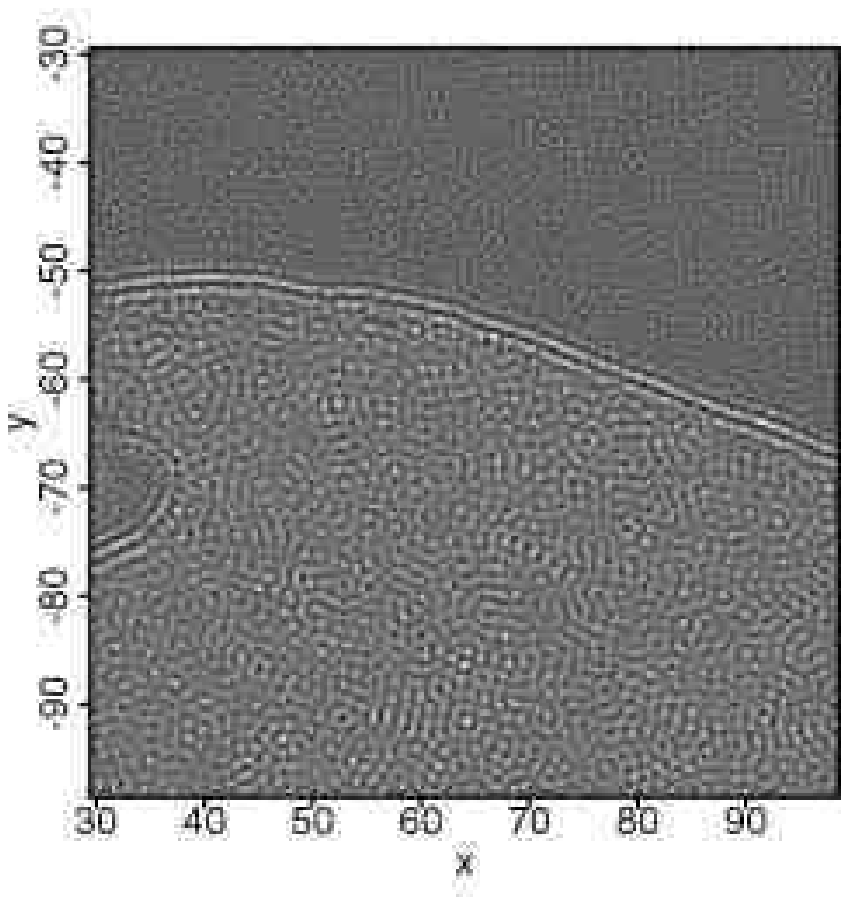}\hspace{\myfix}
\includegraphics[height=0.24\textwidth]{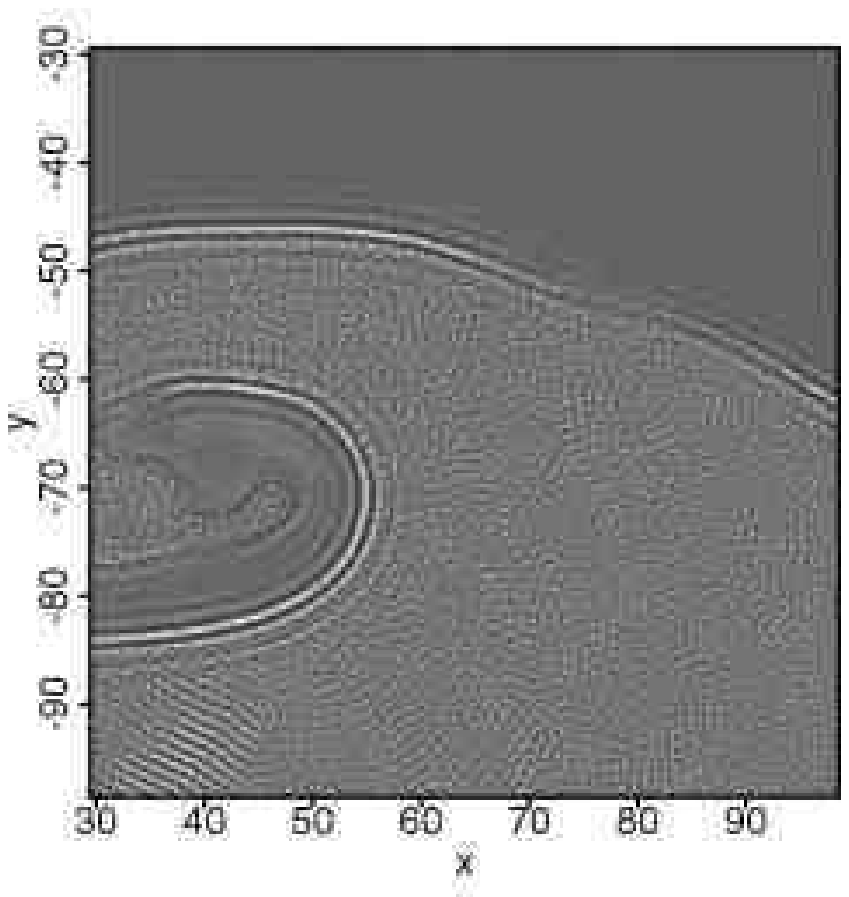}\hspace{\myfix}
\includegraphics[height=0.24\textwidth]{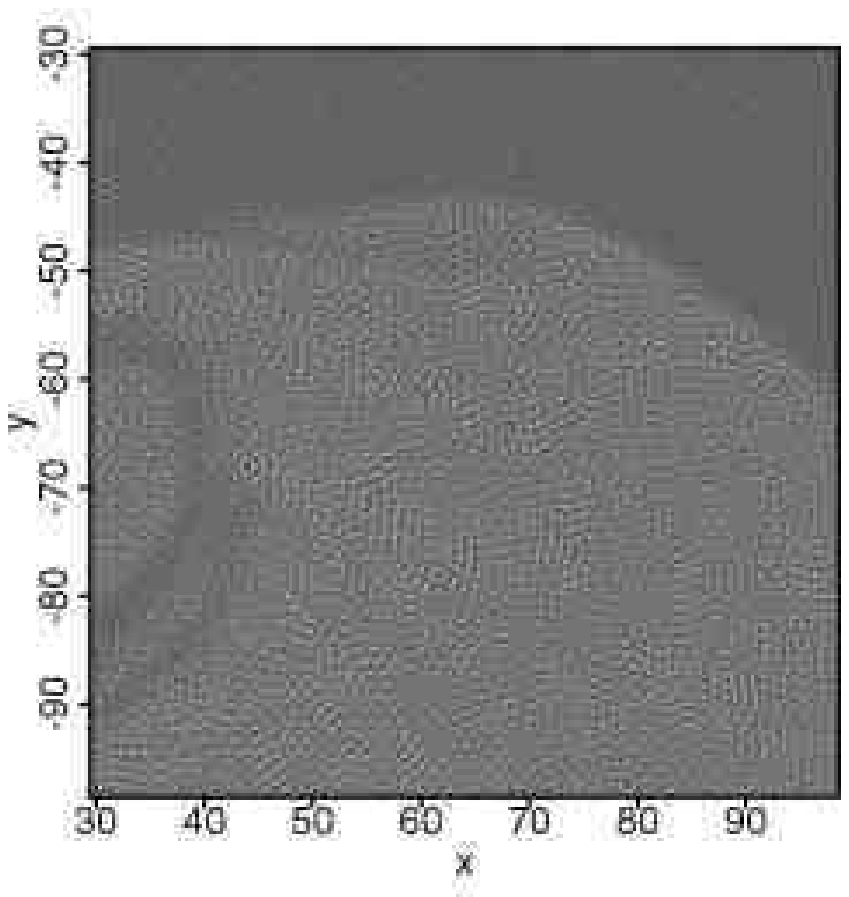}\hspace{\myfix}
\caption{A zoom in on the boundary for the KH test in (from left to
  right): \SPHS-CT-128, \SPHS-HOCT4-442 and \MSPH-HOCT4-442
  at $\tau_\mathrm{KH}=1$. The plots are vertical projections of all
  the points within a $64\times64\times z$ cuboid. Each point is plotted as a
  solid black circle with some transparency in order for the gaps
  to stand out.}
\label{fig:banding}
\end{figure*}
\end{center}

\begin{center} 
\begin{figure*}
\SPHS-CT-128\\
\includegraphics[height=0.23\textwidth]{osph_kp_energy_1tkh_ctr.eps}\hspace{\myfix}
\includegraphics[height=0.23\textwidth]{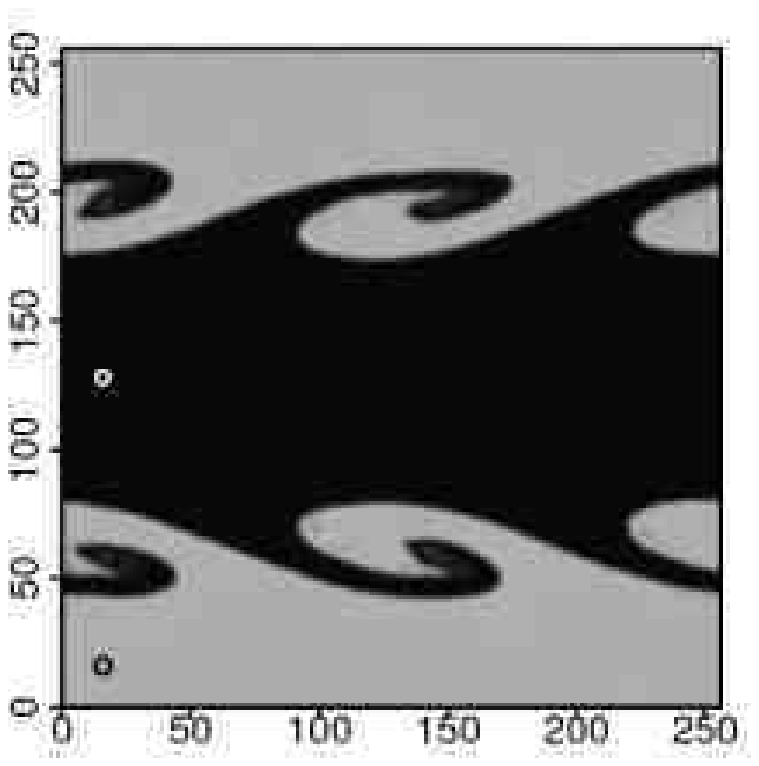}\hspace{\myfix}
\includegraphics[height=0.23\textwidth]{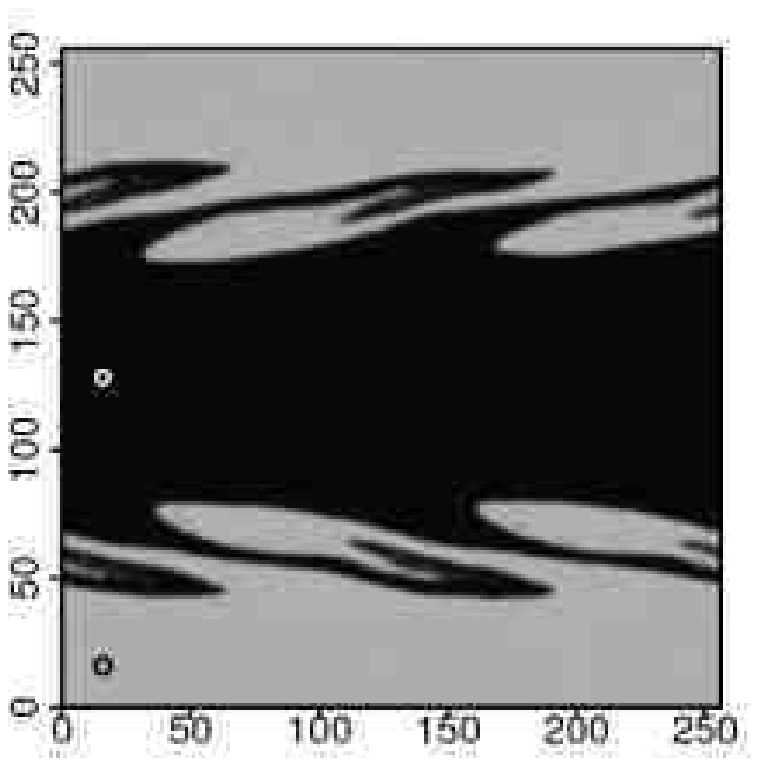}\\
\SPHS-HOCT4-442\\
\includegraphics[height=0.23\textwidth]{osph_hoct4_442_1tkh_ctr.eps}\hspace{\myfix}
\includegraphics[height=0.23\textwidth]{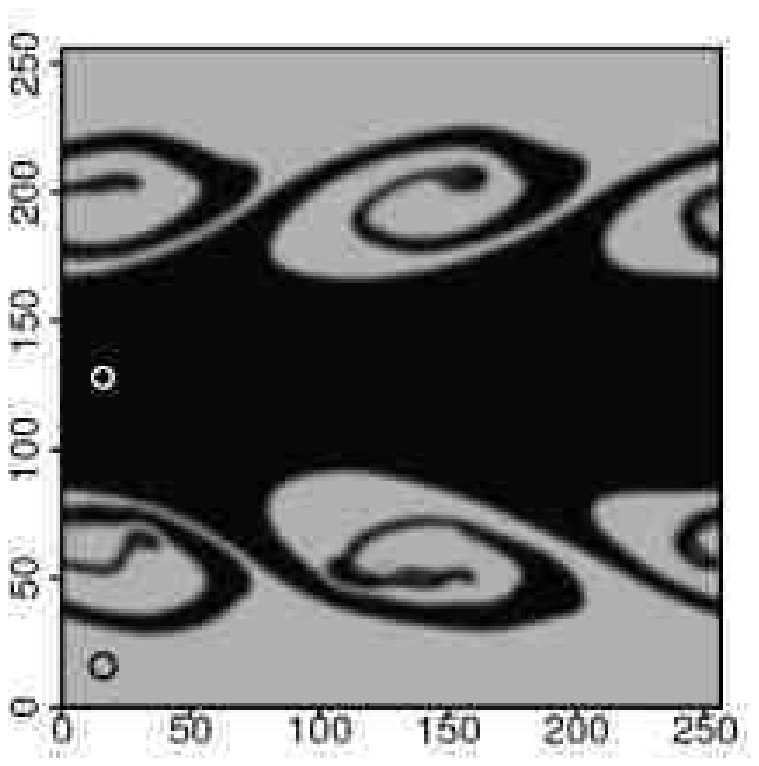}\hspace{\myfix}
\includegraphics[height=0.23\textwidth]{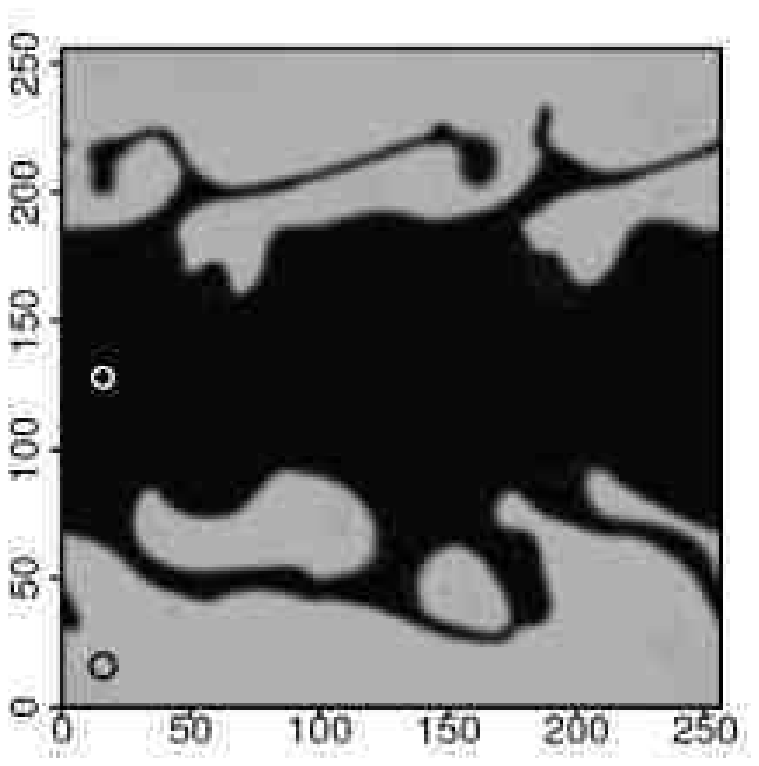}\\
\MSPH-HOCT4-442\\
\includegraphics[height=0.23\textwidth]{osph_rt_hoct4_442_1tkh_ctr.eps}\hspace{\myfix}
\includegraphics[height=0.23\textwidth]{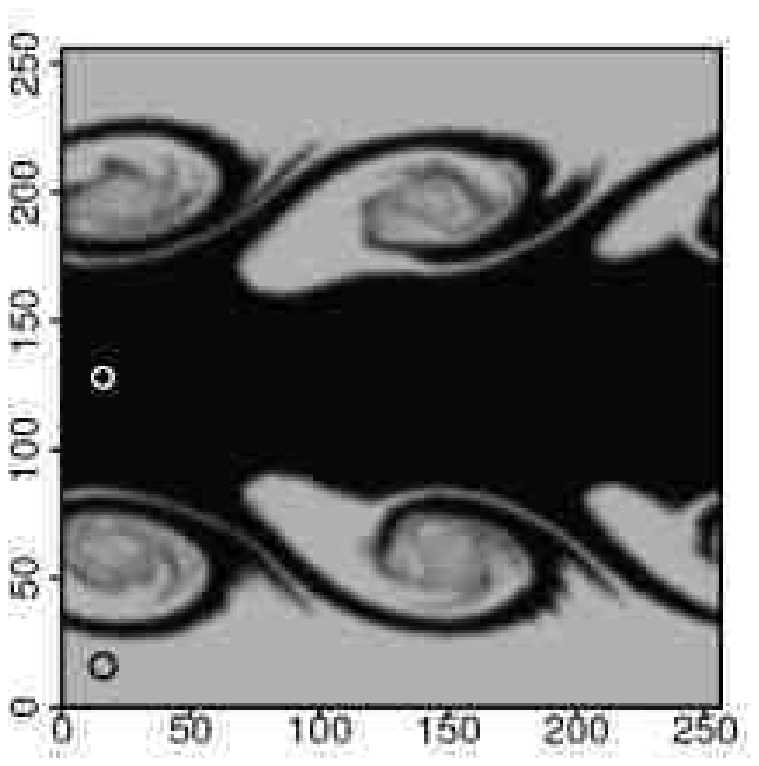}\hspace{\myfix}
\includegraphics[height=0.23\textwidth]{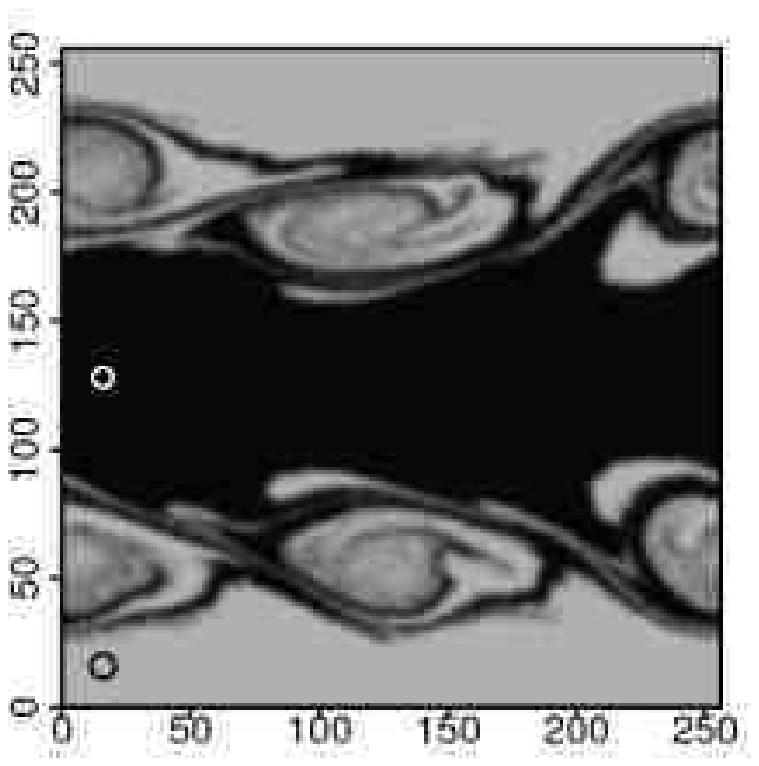}\\
\MSPH-HOCT4-442; 1:8 density ratio\\
\includegraphics[height=0.23\textwidth]{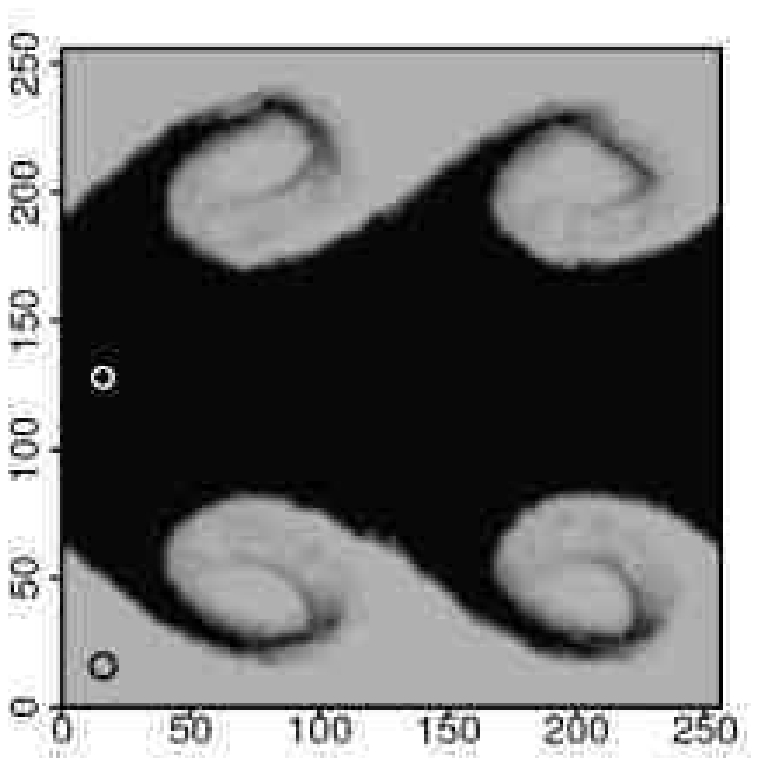}\hspace{\myfix}
\includegraphics[height=0.23\textwidth]{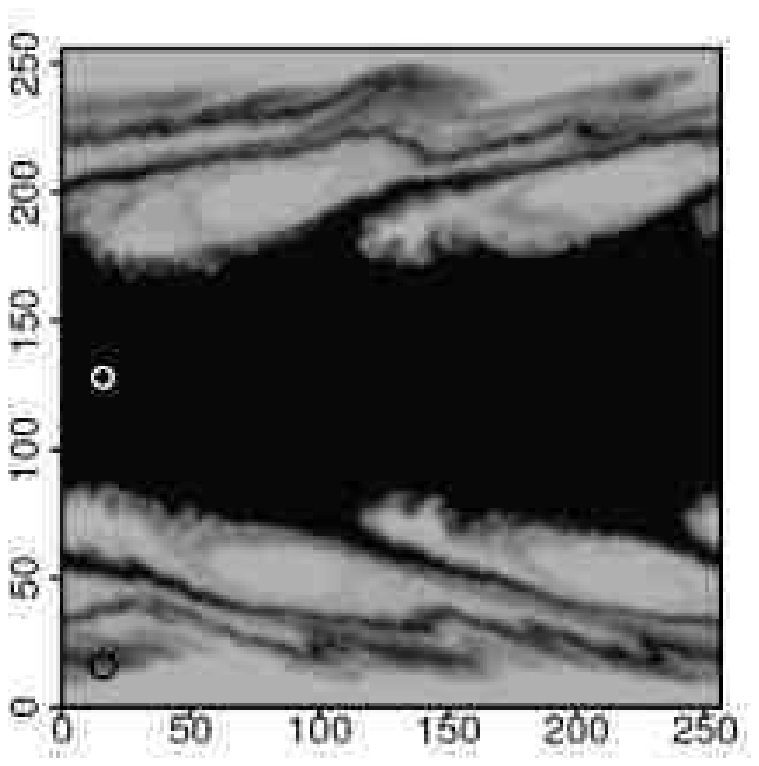}\hspace{\myfix}
\includegraphics[height=0.23\textwidth]{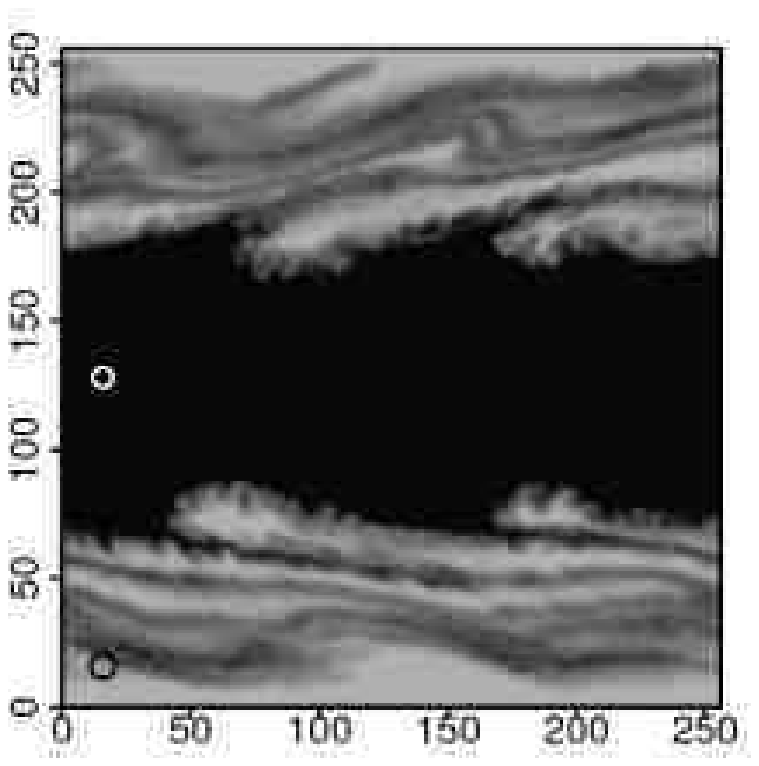}\\

{\tt RAMSES}; $256\times256$ cells, no refinement, LLF Riemann solver\\
\includegraphics[height=0.23\textwidth]{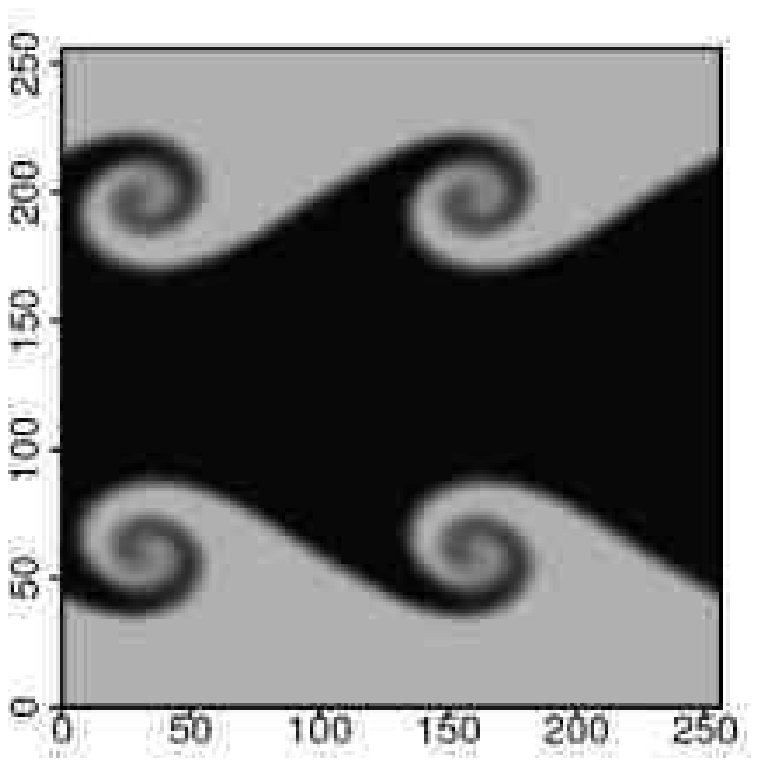}\hspace{\myfix}
\includegraphics[height=0.23\textwidth]{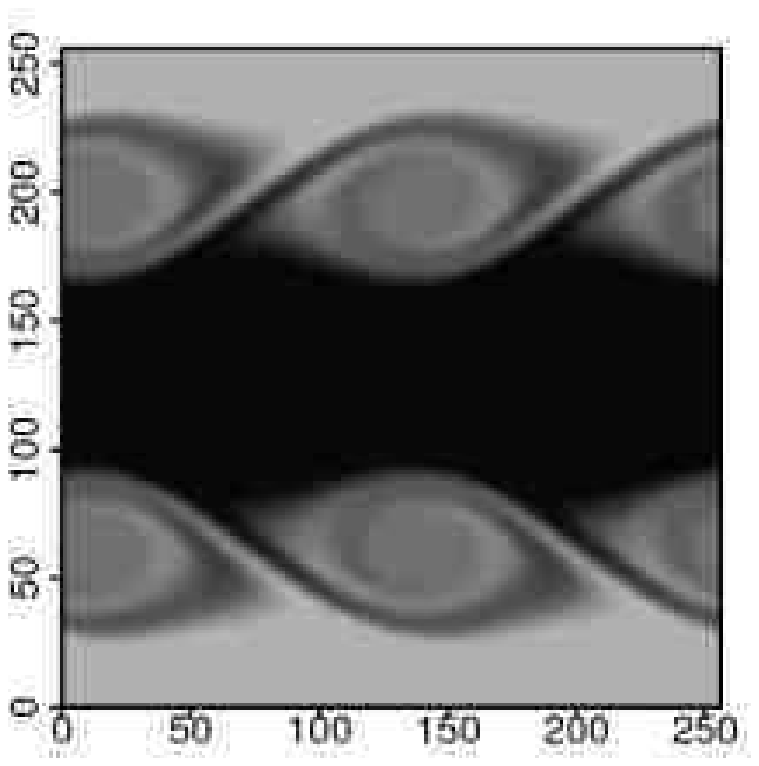}\hspace{\myfix}
\includegraphics[height=0.23\textwidth]{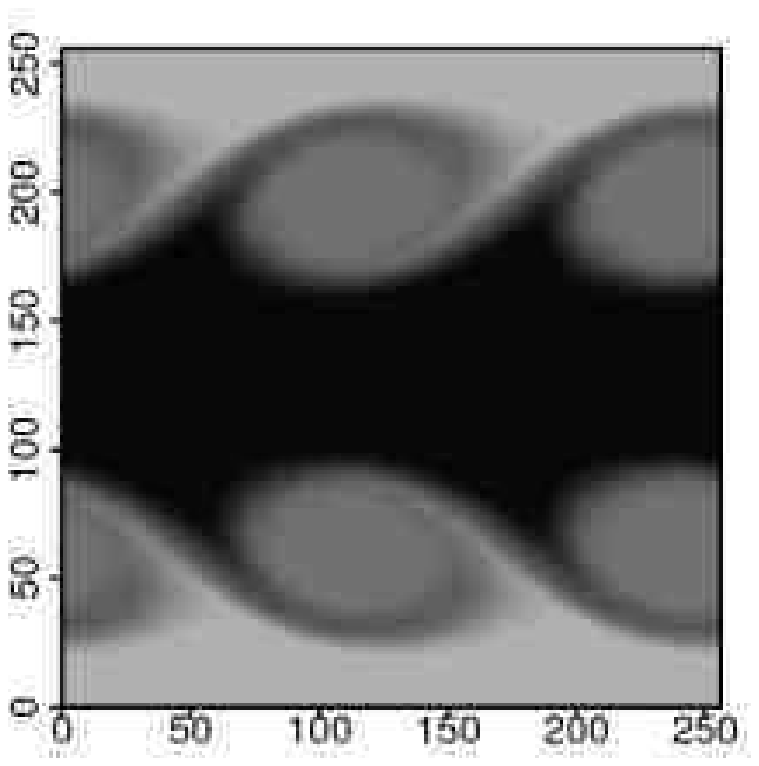}\\
\caption{Long term evolution of the KH instability in \SPHS\ and
  \MSPH\ versus the Eulerian code {\tt RAMSES}. From left to right, the panels show
  density contours in a slice of width $dx=1$ about the z-axis at
  times $\tau_\mathrm{KH}=1,2$ and 3.}
\label{fig:longterm}
\end{figure*}
\end{center}

\begin{center}
\begin{figure*}
%SPH-CS-32\\
%\includegraphics[height=0.23\textwidth]{sph_sod_32_rho.eps}\hspace{\myfix}
%\includegraphics[height=0.23\textwidth]{sph_sod_32_vel.eps}\hspace{\myfix}
%\includegraphics[height=0.23\textwidth]{sph_sod_32_pres.eps}\hspace{\myfix}
%\includegraphics[height=0.23\textwidth]{sph_sod_32_temp.eps}\\
SPH-CS-442\\
\includegraphics[height=0.23\textwidth]{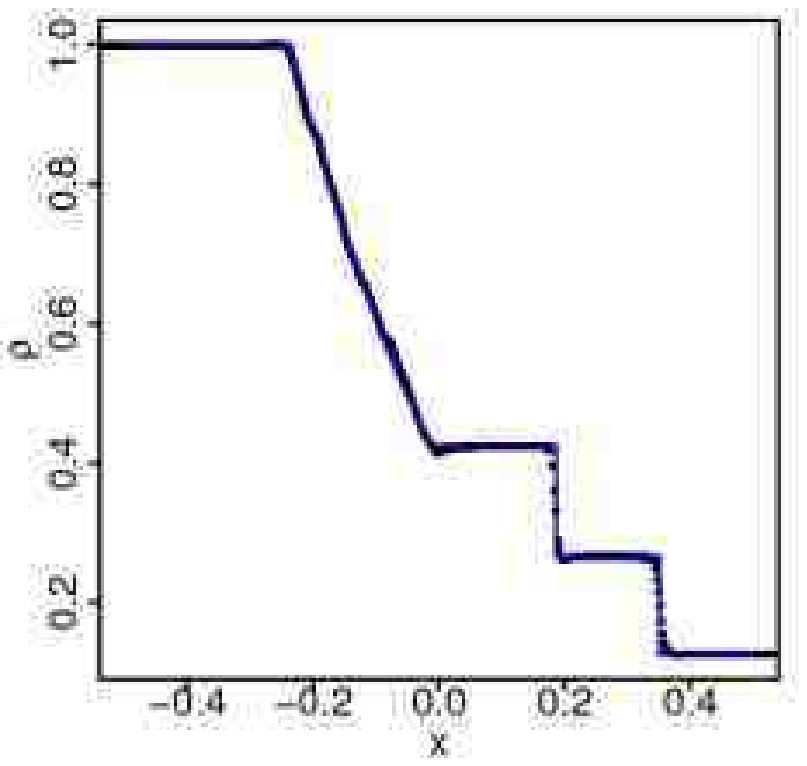}\hspace{\myfix}
\includegraphics[height=0.23\textwidth]{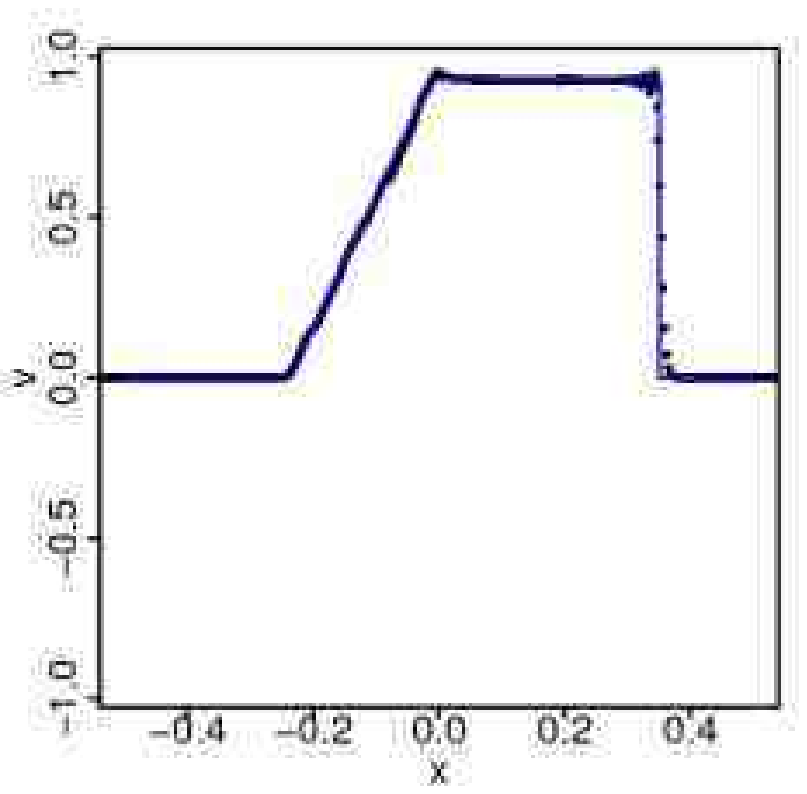}\hspace{\myfix}
\includegraphics[height=0.23\textwidth]{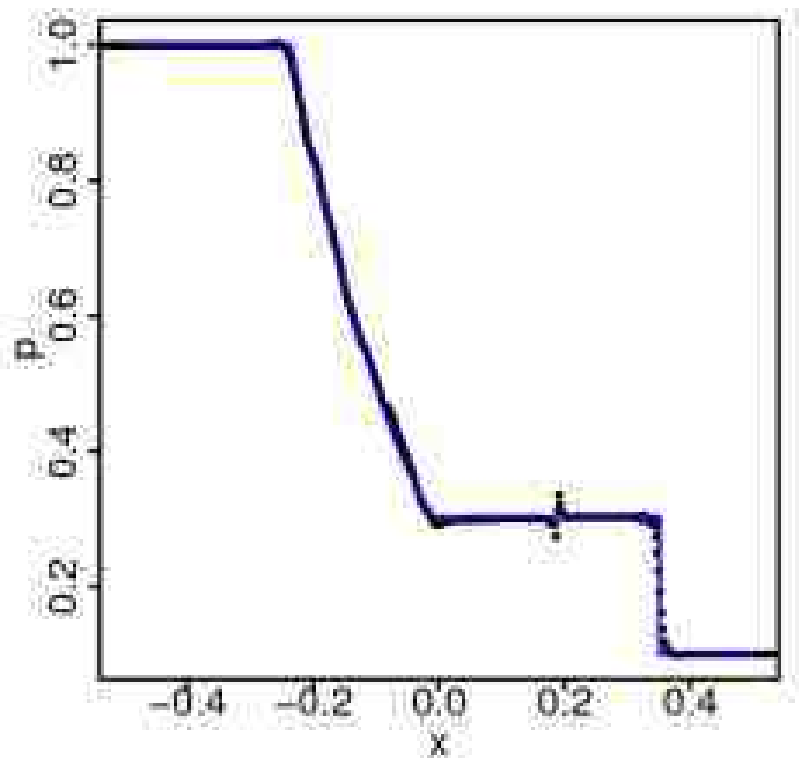}\hspace{\myfix}
\includegraphics[height=0.23\textwidth]{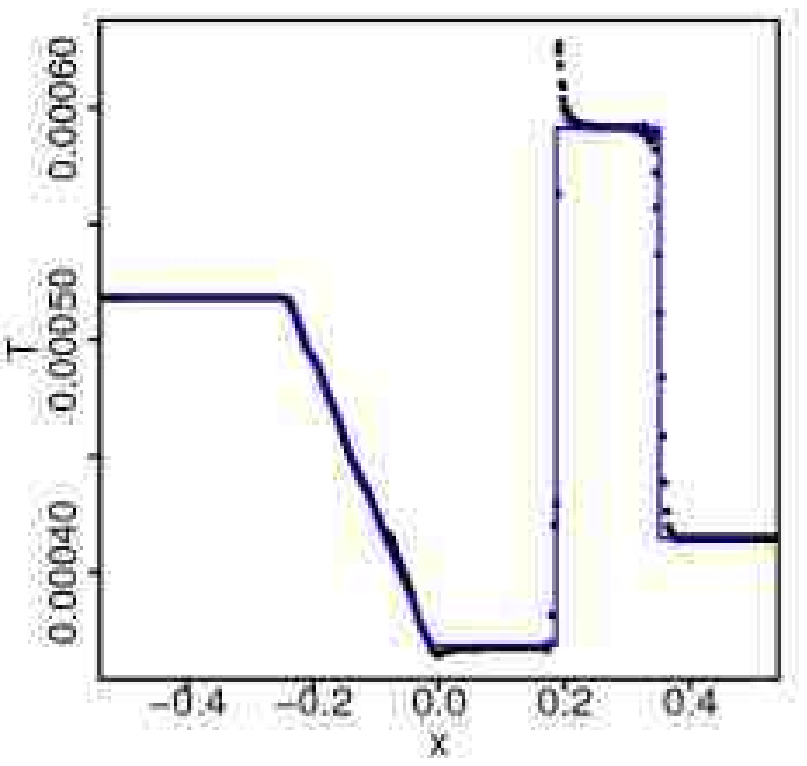}\\
\MSPH-HOCT4-442\\
\includegraphics[height=0.23\textwidth]{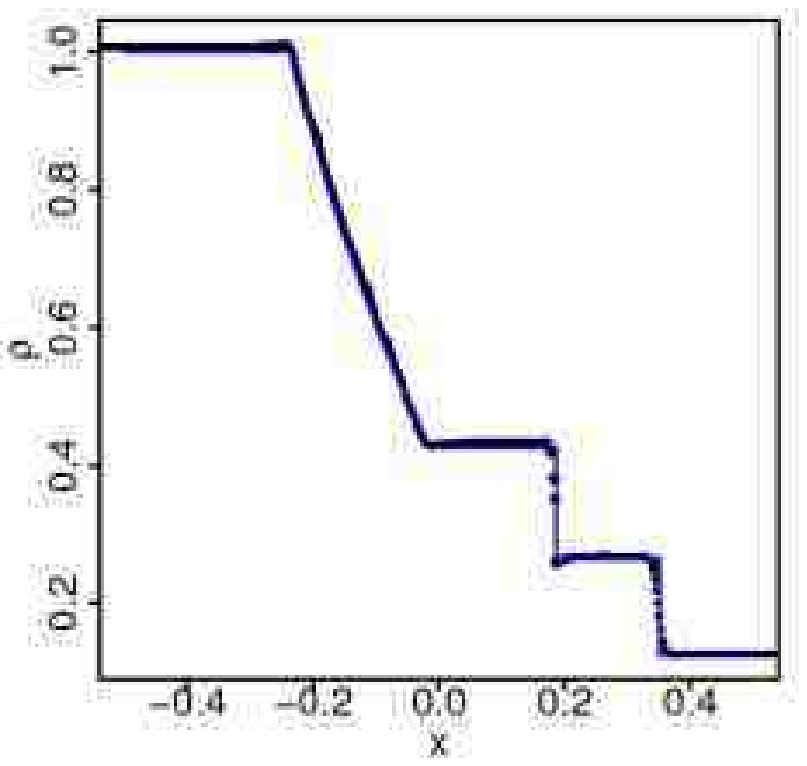}\hspace{\myfix}
\includegraphics[height=0.23\textwidth]{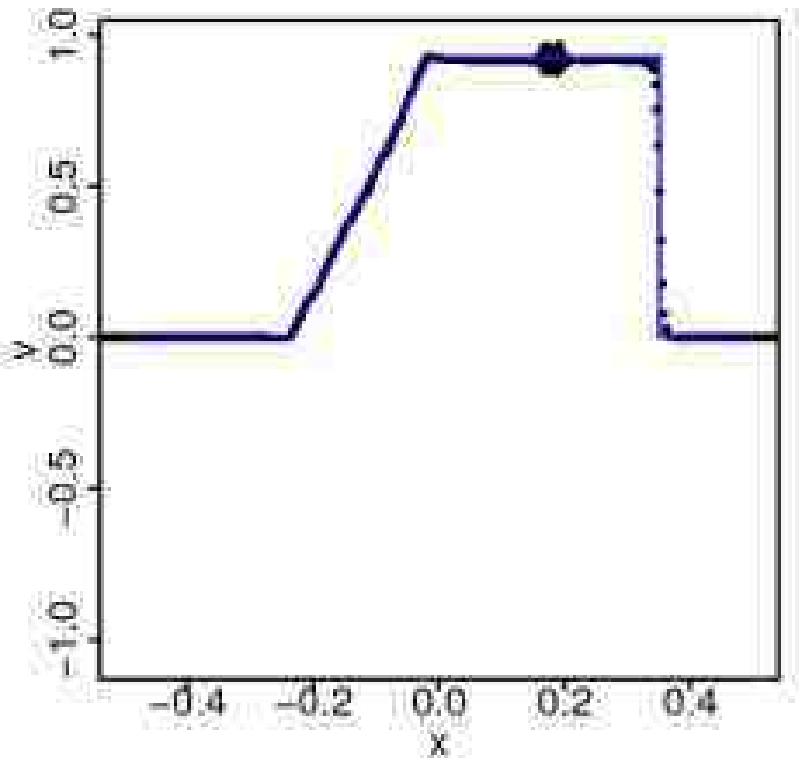}\hspace{\myfix}
\includegraphics[height=0.23\textwidth]{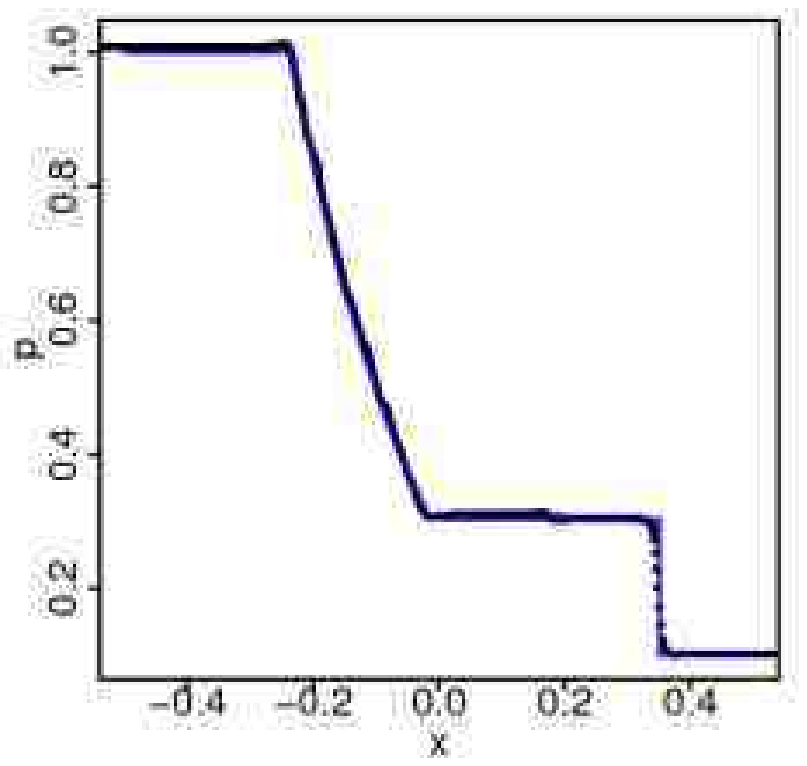}\hspace{\myfix}
\includegraphics[height=0.23\textwidth]{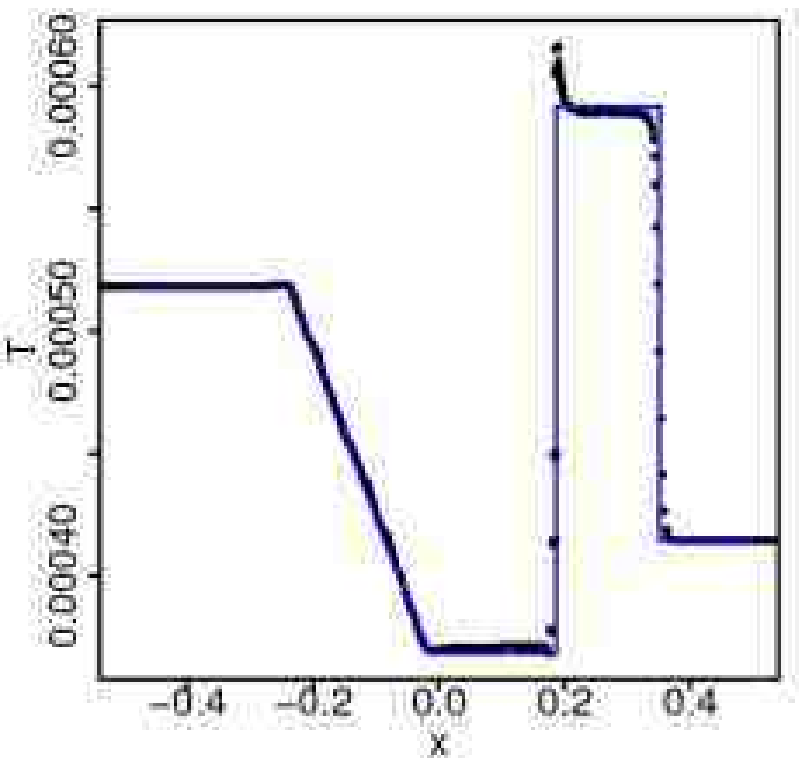}\\
\caption{A Sod shock tube test in SPH (top) and \MSPH\ (bottom). From left to right the panels show the variation in density, velocity, pressure and temperature across the shock, respectively. The blue lines give the analytic solution. This test was performed in 3D.}
\label{fig:sod}
\end{figure*}
\end{center}

\begin{flushleft} 
\begin{figure*}
SPH-CS-32; 31,686 particles in the blob\\
\includegraphics[height=0.24\textwidth]{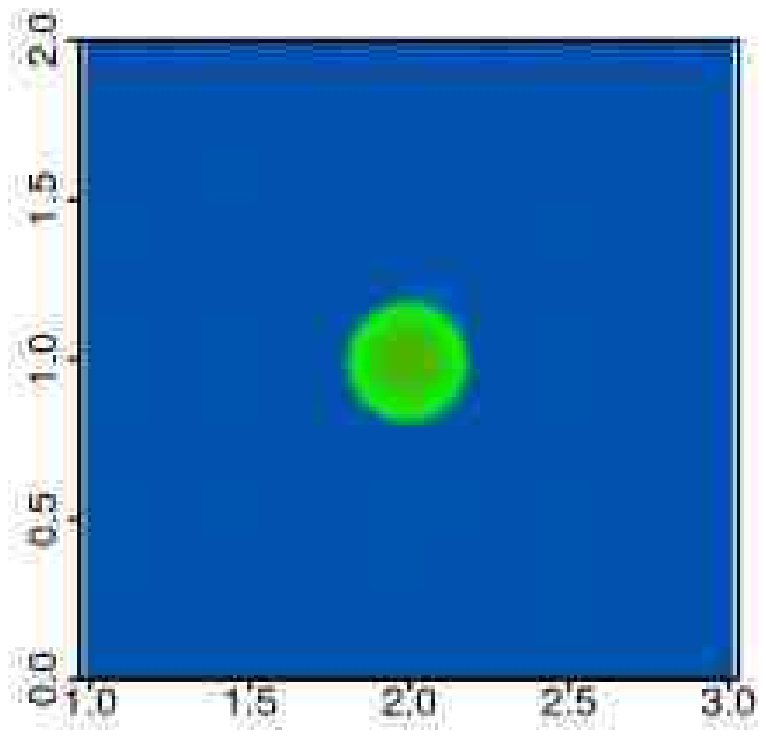}\hspace{\myfix}
\includegraphics[height=0.24\textwidth]{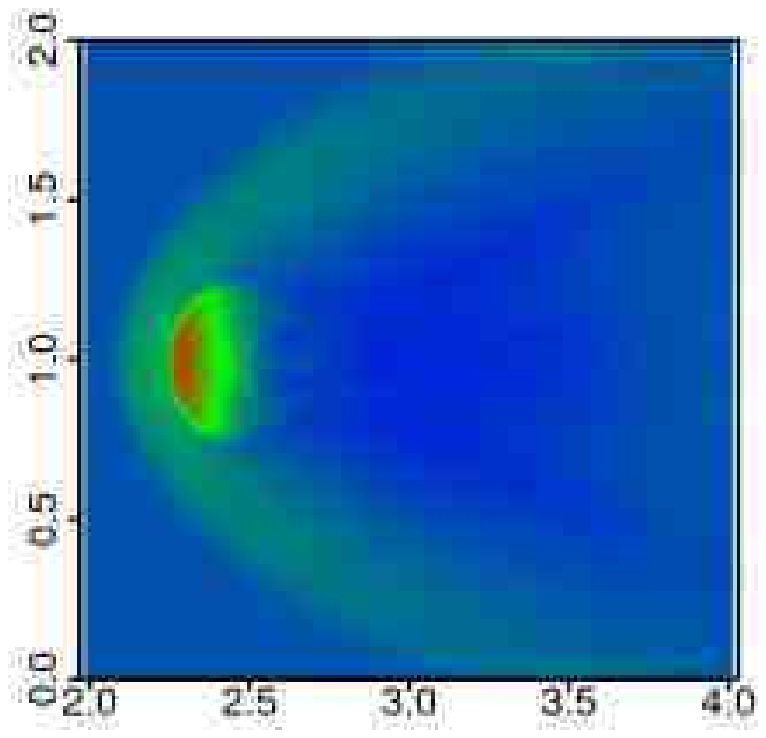}\hspace{\myfix}
\includegraphics[height=0.24\textwidth]{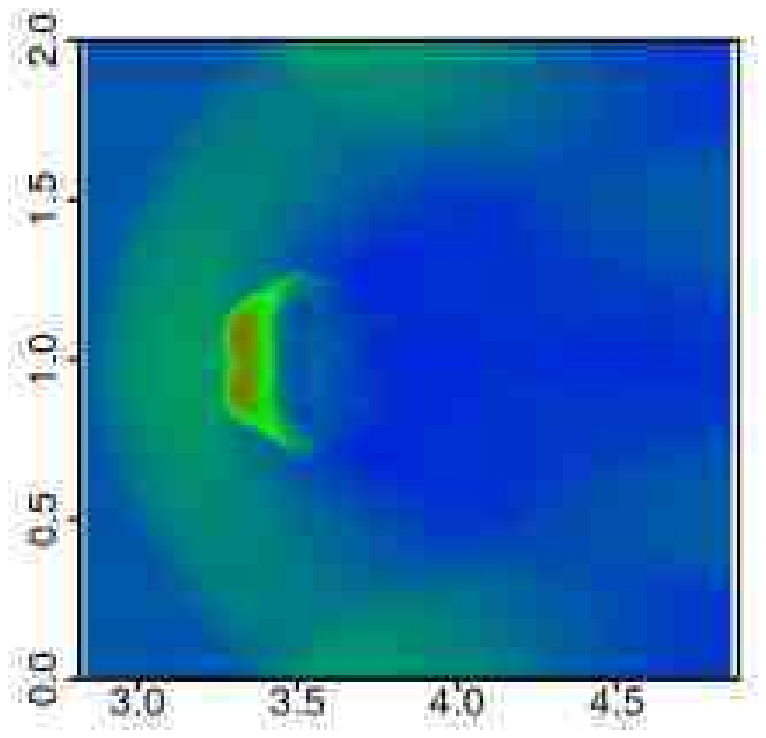}\hspace{\myfix}
\includegraphics[height=0.24\textwidth]{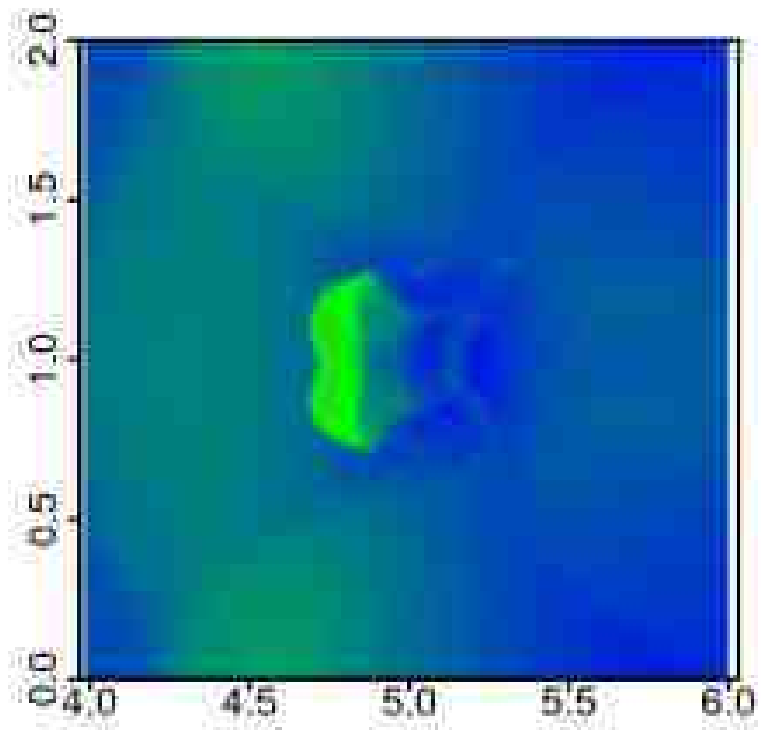}\hspace{\myfix}
\SPHS-HOCT4-442; 126,744 particles in the blob\\
\hspace{0.165\textwidth}\hspace{\myfix}
\includegraphics[height=0.24\textwidth]{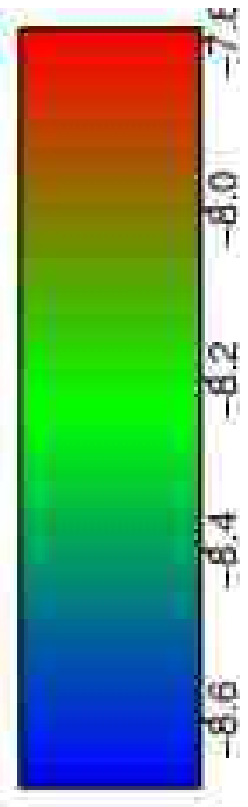}\hspace{\myfix}
\includegraphics[height=0.24\textwidth]{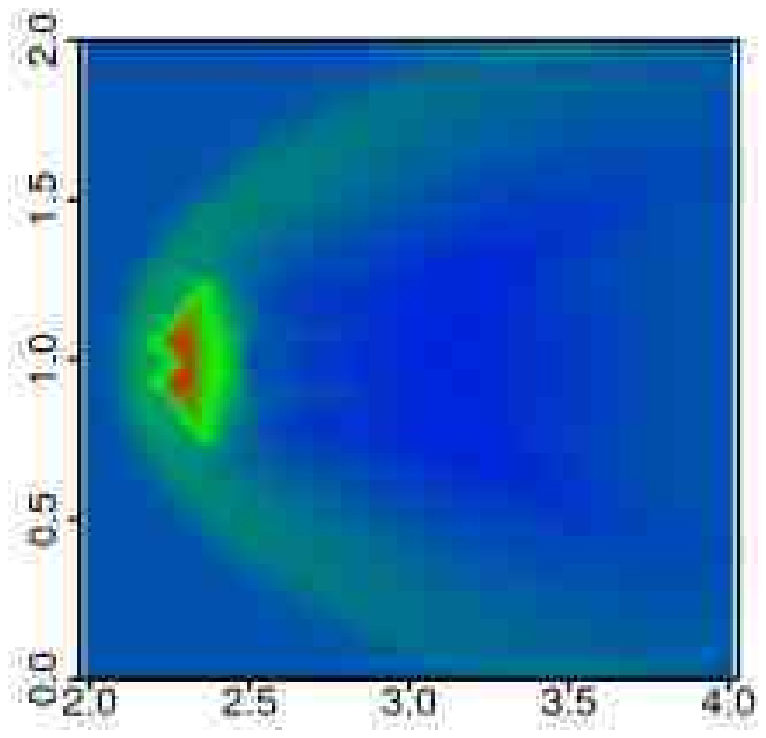}\hspace{\myfix}
\includegraphics[height=0.24\textwidth]{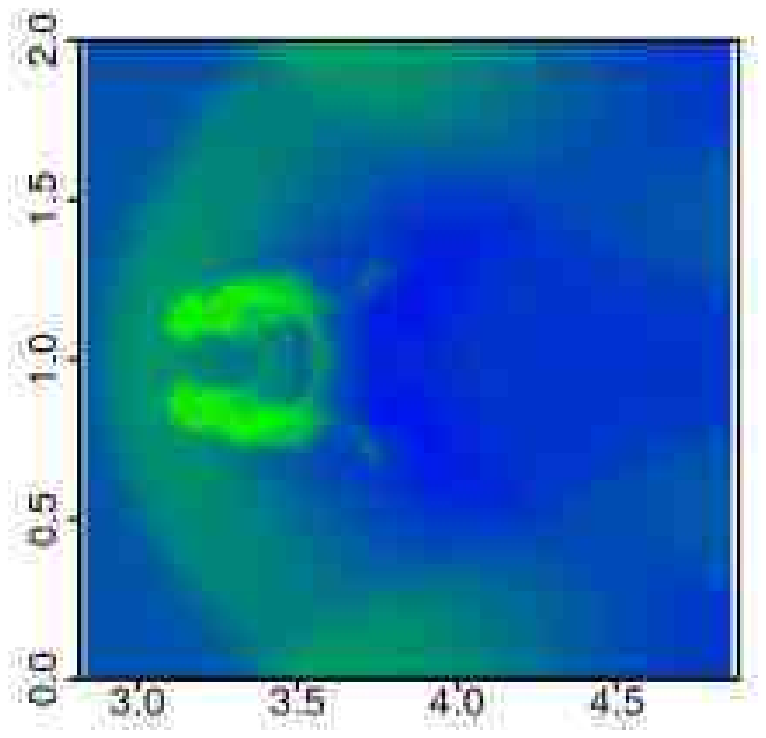}\hspace{\myfix}
\includegraphics[height=0.24\textwidth]{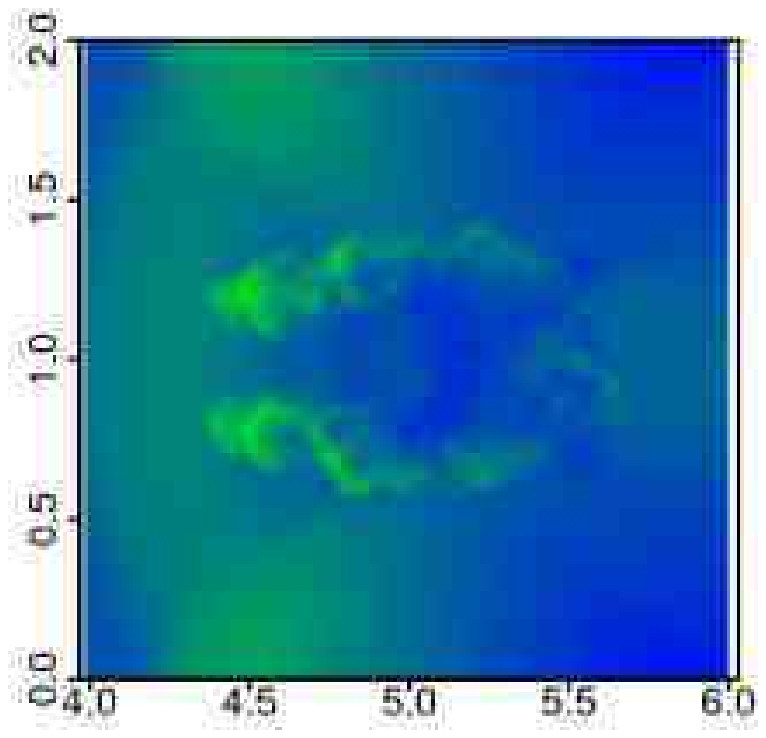}\hspace{\myfix}
\MSPH-HOCT4-442; 126,744 particles in the blob\\
\hspace{0.24\textwidth}\hspace{\myfix}
\includegraphics[height=0.24\textwidth]{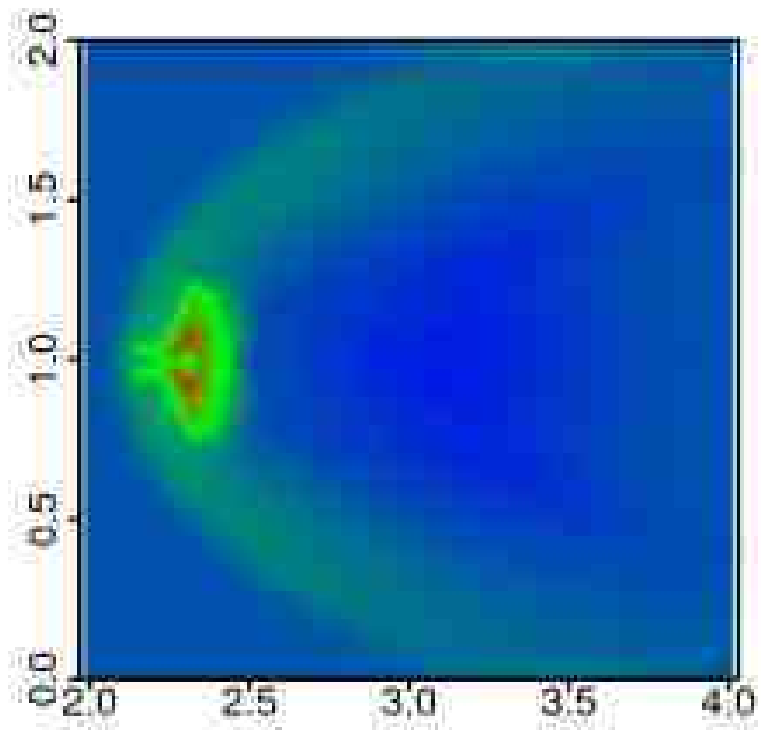}\hspace{\myfix}
\includegraphics[height=0.24\textwidth]{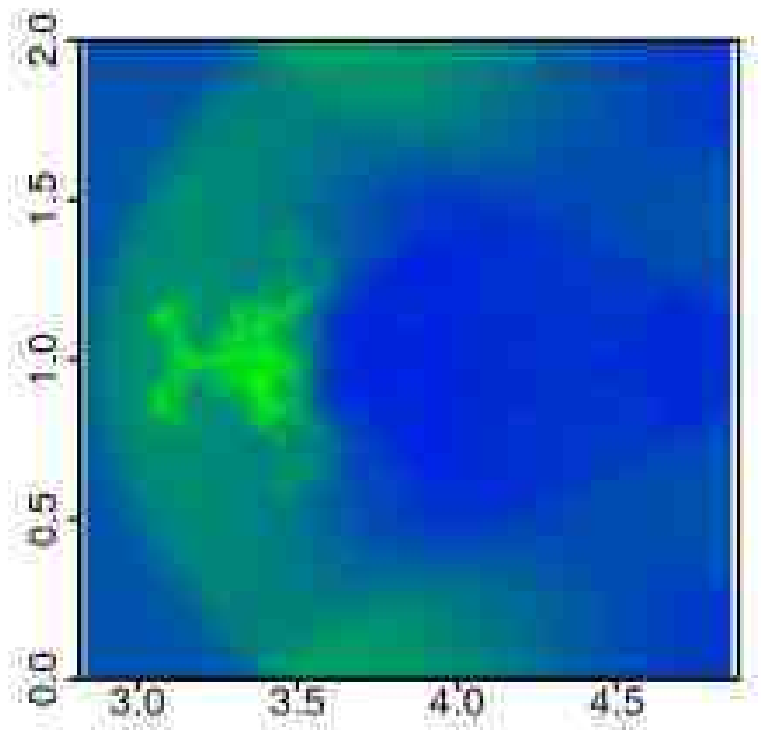}\hspace{\myfix}
\includegraphics[height=0.24\textwidth]{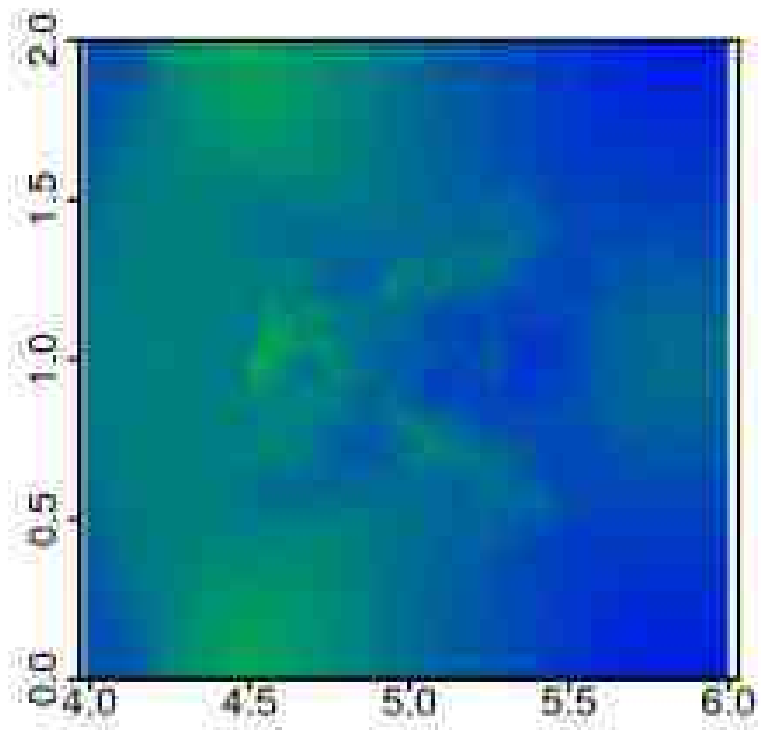}\hspace{\myfix}
{\tt FLASH}; $128\times128\times384$, 8785 cells in the blob, no refinement\\
\hspace{0.24\textwidth}\hspace{\myfix}
\includegraphics[height=0.24\textwidth]{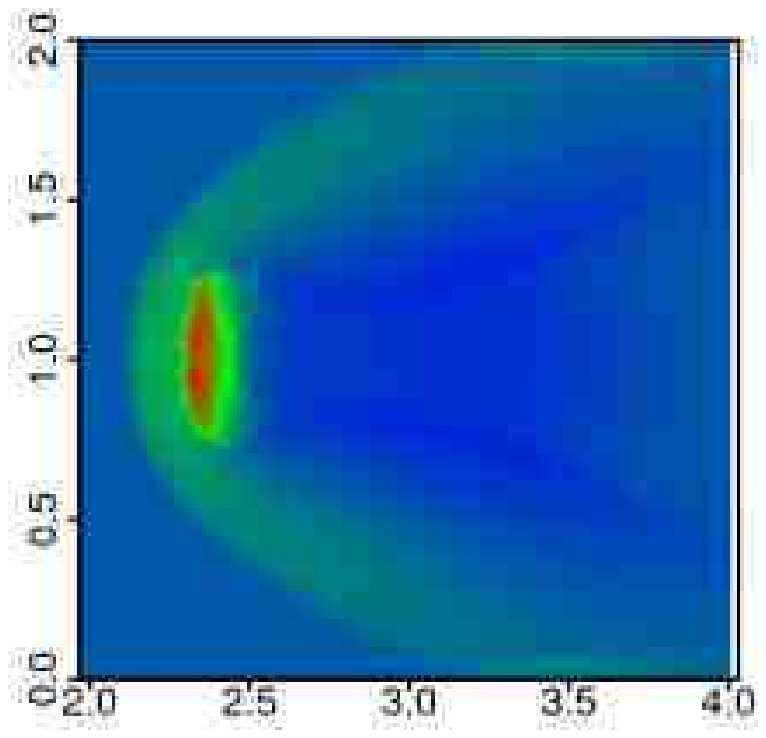}\hspace{\myfix}
\includegraphics[height=0.24\textwidth]{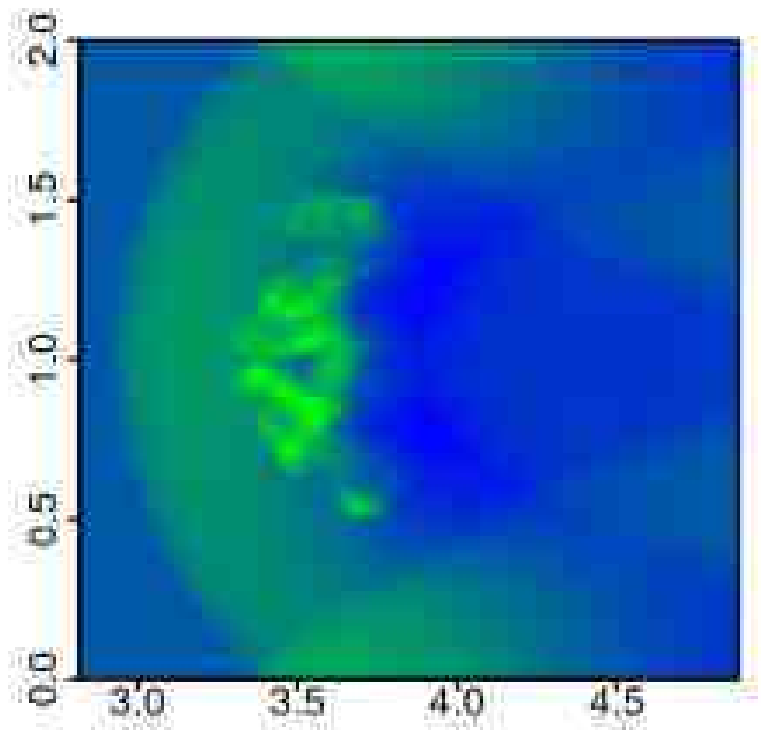}\hspace{\myfix}
\includegraphics[height=0.24\textwidth]{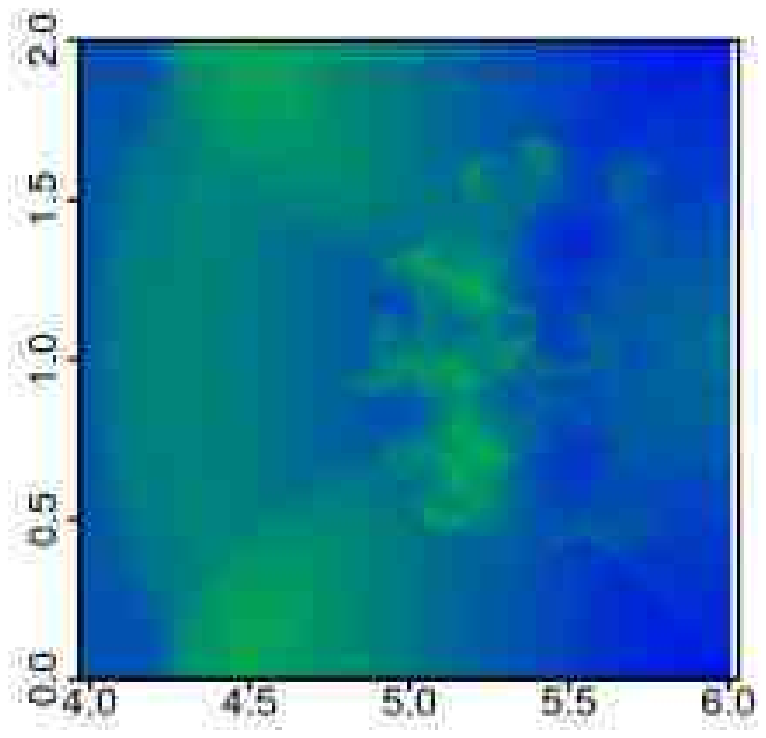}\hspace{\myfix}
\caption{The blob test in SPH, \SPHS, \MSPH\ and the Eulerian code {\tt
    FLASH}. From left to right the plots show density contours at times
  $\tau_\mathrm{KH}=0,1,2$ and 3. The contour bar gives
  logarithmic density in cgi units.}
\label{fig:blob}
\end{figure*}
\end{flushleft}

\begin{center} 
\begin{figure*}
\includegraphics[height=0.33\textwidth]{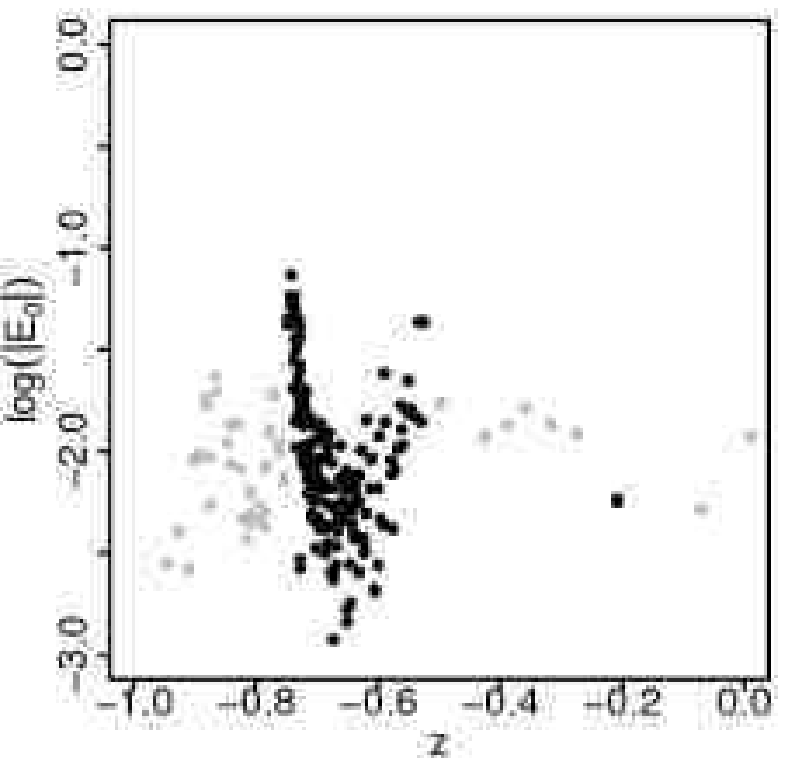}\hspace{\myfix}
\includegraphics[height=0.33\textwidth]{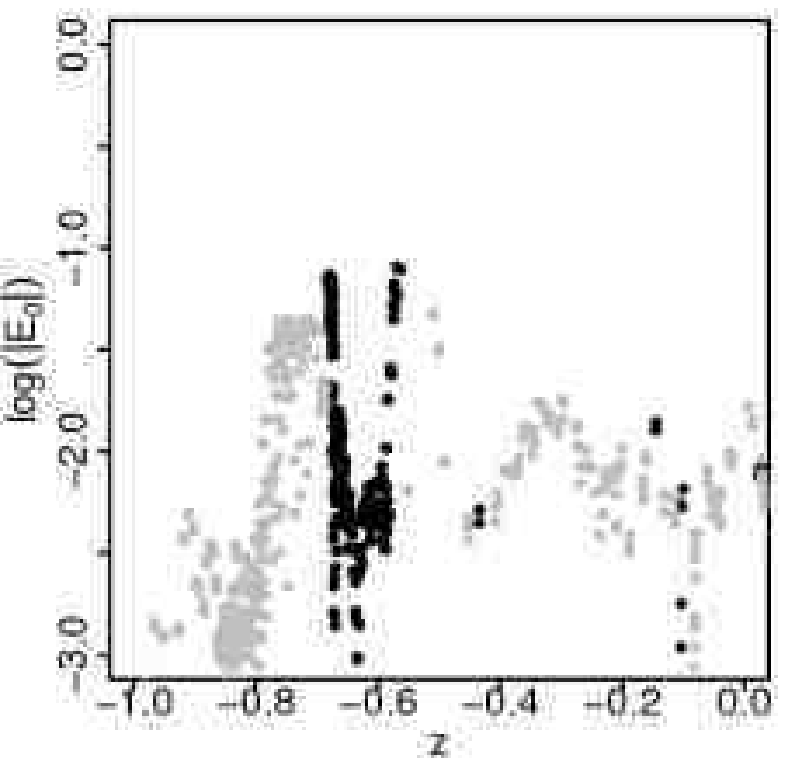}\hspace{\myfix}
\includegraphics[height=0.33\textwidth]{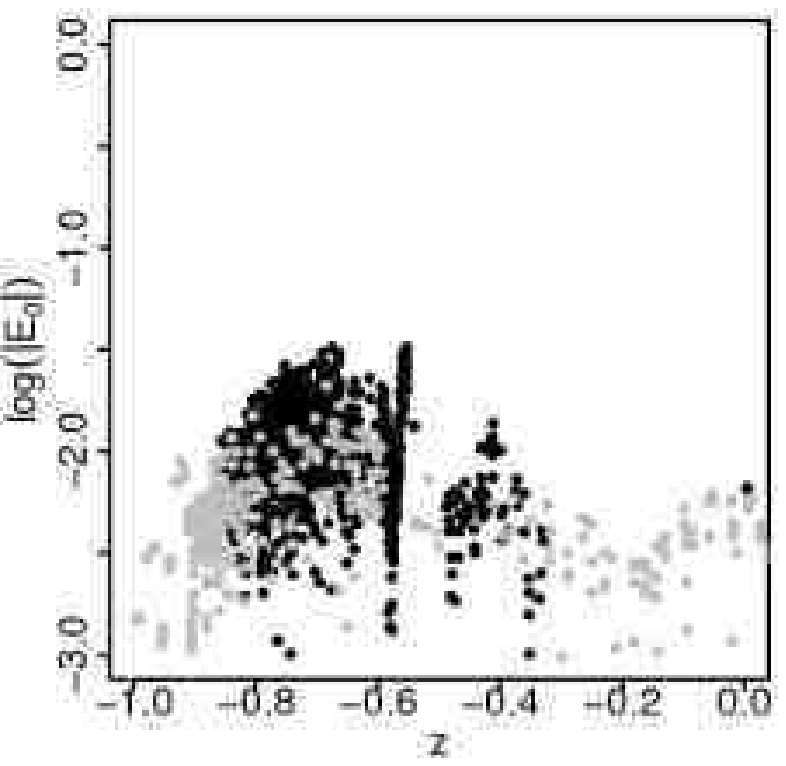}\hspace{\myfix}\\
\includegraphics[height=0.33\textwidth]{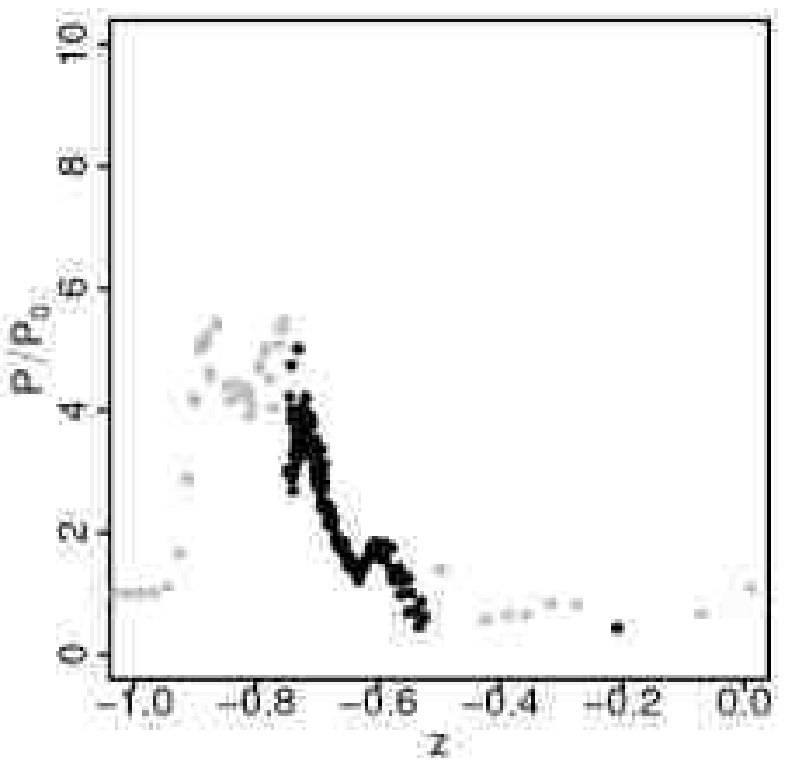}\hspace{\myfix}
\includegraphics[height=0.33\textwidth]{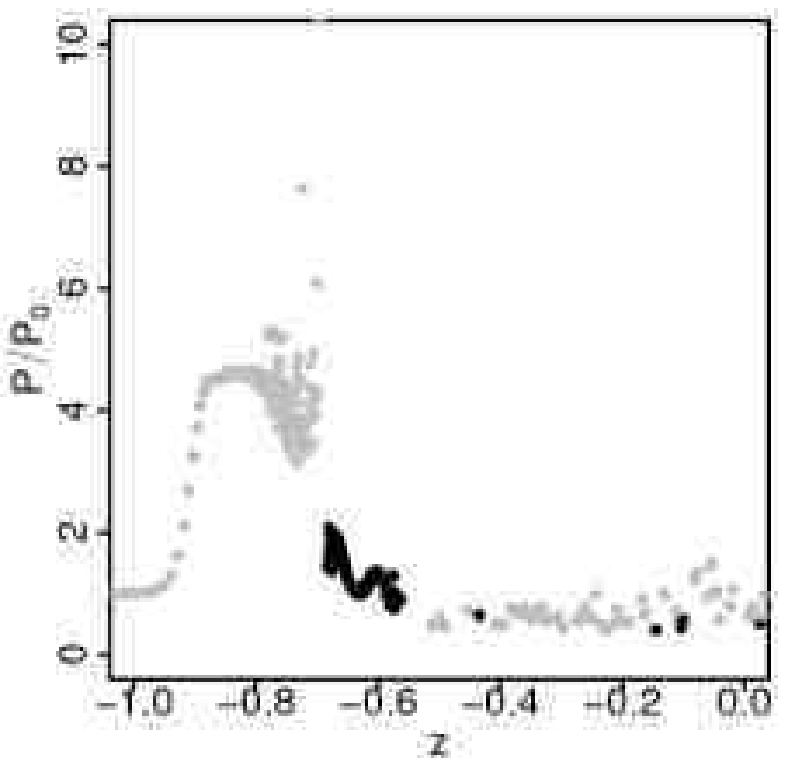}\hspace{\myfix}
\includegraphics[height=0.33\textwidth]{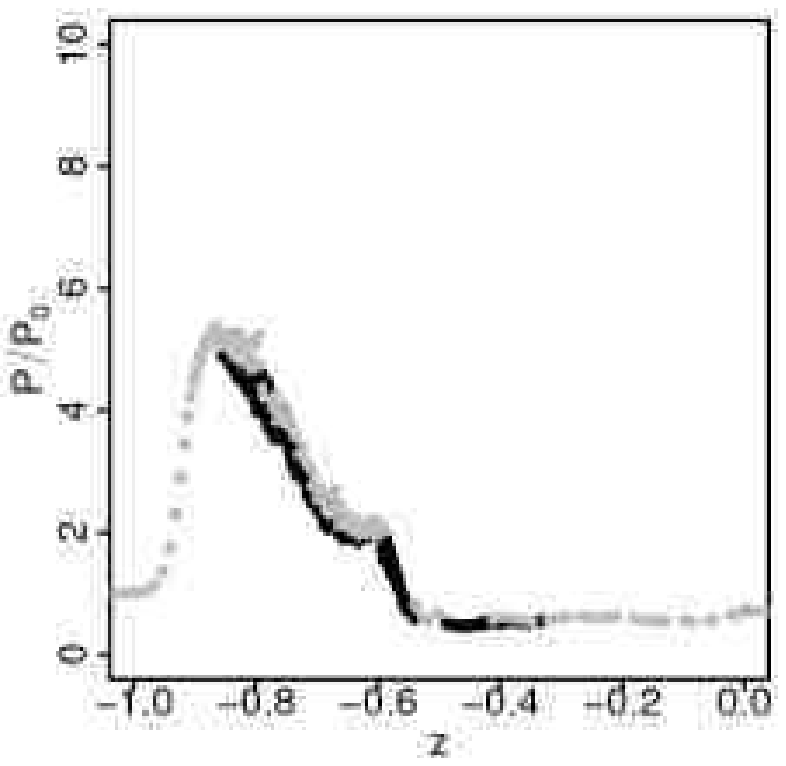}\hspace{\myfix}\\
\caption{The magnitude of the $|\uE|$ error in a slice of
  width $dx=1$ about the z-axis, as a function of
  $y$ (top); and the pressure in a slice of width $dx$ about the x-axis
  and $dx$ about the z-axis, as a function of $y$ (bottom) for the blob test
  in SPH-32 (left), \SPHS-HOCT4-442 (middle) and \MSPH-HOCT4-442 (right) at time
  $\tau_\mathrm{KH}=1$.}
\label{fig:bloberror}
\end{figure*}
\end{center}
\vspace{-5mm}

The clumping instability is a problem because it means that increasing
the neighbour number will not give improved sampling of the kernel,
and the $\uE$ error will remain large. However, the situation is
dramatically improved if we add a constant central core to the kernel
gradient $\frac{\partial W}{\partial r}$. This gives a constant force
term at the centre of the Kernel that physically prevents
clumping. We choose a kernel that is maximally similar to the CS kernel,
while obeying $\frac{\partial W}{\partial r} = \mathrm{const.}$
$\forall r < \alpha$, where $\alpha$ is the core size. This leads us
to the Core Triangle (CT) kernel:

\begin{equation}
W = \frac{N}{h^3}\left\{\begin{array}{lr}
\left(-12\alpha + 18\alpha^2\right)x + \beta & 0 < x \le \alpha \\
 1 - 6x^2 + 6x^3 & \alpha < x \le \frac{1}{2}\\
2(1-x)^3 & \frac{1}{2} < x \le 1\\ 
0 & \mathrm{otherwise} \end{array}\right.
\label{eqn:wkern}
\end{equation}
where $\beta = 1 + 6\alpha^2 - 12\alpha^3$, $N
= 8/[\pi \left(6.4\alpha^5 - 16\alpha^6 + 
  1\right)]$, and the core size is fixed at $\alpha = 1/3$ by the
requirement that $\frac{\partial^2 W}{\partial r^2}$ be
continuous.

Figure \ref{fig:stabct} shows stability plots for the CT kernel. The
CT kernel has greatly improved stability for the longitudinal waves (top row)
compared to the CS kernel and should give significantly improved
performance for large neighbour number. We demonstrate this in
\S\ref{sec:kh}.

Note that for all of the kernels we use in this paper, we consistently
apply the kernel for the density estimate and its gradient for the
energy and momentum equations. For small neighbour number, the central
triangle in the CT kernel will degrade the quality of the density
estimate. However, 
in this paper we typically use large neighbour numbers ($>100$). In
this case, very few particles sample the inner regions of the kernel
and the bias introduced in the density is negligible. (The quality of
the density estimate in \MSPH\ can be seen in the Sod shock tube test in
\S\ref{sec:sod}.) We found in tests that retaining the CS kernel just
for the density estimate gives near-identical results. 

\subsubsection{The banding instability}\label{sec:bandinginst}

The clumping instability is a result of unstable longitudinal
waves. A related instability -- the banding instability -- is a result
of unstable transverse waves. For both the CT and CS kernels, there
are broad bands of instability to transverse waves (see Figures
\ref{fig:stab} and \ref{fig:stabct}). If the neighbour number is
carefully chosen to lie in a stable region, banding will not occur. However, banding can still be
excited at boundaries if $h$ changes there, moving into an unstable
region.

For both the CS and CT kernels, it is difficult to find a
suitable neighbour number for which the kernel is stable to all
transverse and longitudinal modes. This suggests hunting for an even
more stable kernel. A full search is beyond the scope of this
paper. Here, we present a simple class 
of kernels that improve stability by moving to higher
order \citep{1996PhDMorris}. Following \citet{2005astro.ph..7472P},
we generalise our CT kernel to order $n_k$ to obtain the following class
of kernels that we call the High Order Core-Triangle (HOCT) kernels:

\begin{equation}
W = \frac{N}{h^3}\left\{\begin{array}{lr}
Px + Q & 0 < x \le \alpha \\
 (1-x)^{n_k} + A(\gamma-x)^{n_k} + & \alpha < x \le \beta\\
B(\beta-x)^{n_k} &\\
(1-x)^{n_k} + A(\gamma-x)^{n_k} & \beta < x \le \gamma\\ 
(1-x)^{n_k}  & \gamma < x \le 1\\ 
0 & \mathrm{otherwise} \end{array}\right.
\label{eqn:hoctkern}
\end{equation}
where:
\begin{equation}
A = \frac{1-\beta^2}{\gamma^{n_k-3}(\gamma^2-\beta^2)}
\end{equation}
\begin{equation}
B = -\frac{1 + A\gamma^{n_k-1}}{\beta^{n_k-1}}
\end{equation}
\begin{equation}
P = -n_k(1-\alpha)^{n_k-1} - n_k A (\gamma-\alpha)^{n_k-1} - n_k
B(\beta-\alpha)^{n_k-1} 
\end{equation}
\begin{equation}
Q = (1-\alpha)^{n_k} + A(\gamma-\alpha)^{n_k} + B(\beta-\alpha)^{n_k} - P\alpha
\end{equation}
and $\alpha$ and $N$ are calculated numerically for a given choice of
$n_k$. Continuity requires that $\alpha$ solves the equation: 
\begin{equation}
0 = (1-\alpha)^{n_k-2} + A(\gamma-\alpha)^{n_k-2} + B(\beta-\alpha)^{n_k-2}
\end{equation}
where $\beta$ and $\gamma$ are free parameters. In this paper we
choose $\beta = 0.5, \gamma = 0.75$. Other choices, and indeed other
high-order kernels, may give better results than those presented
here. We tabulate
values for $A,B,P,Q,\alpha$ and $N$ as a function $n_k$ in Table
\ref{tab:hoctkern}. Notice that the core size $\alpha$ decreases with
$n_k$. 

Stability plots for the HOCT4 kernel (with $n_k = 4$) are given
in Figure \ref{fig:stabhoct}. Notice the improvement over the CT
kernel, particularly for the transverse waves. There are two bands
where the kernel is fully stable to both longitudinal and transverse
waves on a lattice: 96 neighbours and 442 neighbours, corresponding to
$h=2.86$ and $h = 4.75$, respectively. We use the latter choice since
this also gives very low $\uE$. The CT kernel also has a
stability band for $h \sim 4.75$, but this is narrower than for the
HOCT4 kernel, while the HOCT4 kernel with this many neighbours gives
better spatial resolution. (It is important to realise that the
smoothing length for different kernels takes on a different meaning
in terms of spatial resolution. We suggest a resolution criteria based on the
numerical sound speed versus the true sound speed for longitudinal
waves: $\omega^2 / k^2 / c_s^2$. Spatial scales are well resolved if
$\omega^2 / k^2/c_s^2 \simeq 1$. By this definition, our choice of 442
neighbours ($h=4.75$) for the HOCT4 kernel gives a very similar
spatial resolution to 128 neighbours ($h=3.14$) for the CT kernel.)

Note that our stability analysis only applies for particles arranged on a lattice. Hexagonal
close-packed particles, randomly arranged particles, and indeed
boundaries may have different preferred stability regions. A full
analysis is beyond the scope of this present work. 

The banding instability is not as problematic as the clumping instability for the tests we present in \S\ref{sec:kh} and \S\ref{sec:blob}. Unlike the clumping instability, it does not seem to (directly) play a major role in preventing mixing from occurring in SPH (see \S\ref{sec:bandingresults}). 

\begin{table}
\center
\caption{Parameters for a selection of High Order Core-Triangle (HOCT)
kernels. See equation \ref{eqn:hoctkern} for details and definitions.}
\begin{tabular}{ccccccccc}
\hline
$n_k$ & $A$ & $B$ & $P$ & $Q$ & $\alpha$ & $N$\\
\hline 
3 & 2.4 & -9.4 & -1.81 & 1.028 & 0.317 & 3.71\\
4 & 3.2 & -18.8 & -2.15 & 0.98 & 0.214 & 6.52\\
5 & 4.27 & -37.6 & -2.56 & 0.962 & 0.161 & 10.4\\
8 & 10.1 & -300.8 & -3.86 & 0.942 & 0.0927 & 30.75\\
\hline
\label{tab:hoctkern}
\end{tabular}
\end{table}

\subsection{The local mixing instability \& RT densities}\label{sec:localmix}

Our error analysis in \S\ref{sec:error} missed one very important error term. This is because the Taylor expansion assumed that both the pressures and velocities in the flow are {\it smooth}. Unfortunately, in SPH at sharp boundary this is not the case. The reason for this is easiest to understand using the entropy form of SPH, as follows (similar arguments apply also for the energy form). 

Imagine a density step of ratio $R_\rho =
\rho_1/\rho_2$ initially in pressure equilibrium, such that the
entropy function (equation \ref{eqn:entropystate}) is given by
$A_1/A_2 = 1/R_\rho^\gamma$. Now imagine that we perturb the boundary
very slightly by pushing a low density particle towards it. The
particle's entropy is conserved, but its density increases very
rapidly proportional to $R_\rho$. This leads to an increase in
pressure: $P_1 = P + \kappa_1 R$, where $\kappa_1$ is some constant that
depends on the perturbation size and the kernel. On the other side of
the boundary, if we push a high density particle towards the low
density region, however, its density will rapidly decrease giving a
decrease in pressure: $P_2 = P - \kappa_2 R_\rho$. This drives us
towards a pressure discontinuity at the boundary which drives an
associated error in the momentum equation. It can be thought of as a
fundamental result of particles trying to mix on the kernel scale, but
being unable to as a result of entropy conservation. We call this the local mixing instability (LMI). 

Although not phrased in terms of the LMI, the LMI is a recognised
problem in the literature and there are essentially two classes of
solution. We can generate entropy at the boundary to give smooth
entropies and therefore smooth pressures, as in \citep{2008MNRAS.387..427W} and
\citet{2007arXiv0709.2772P}; or we can try to obtain sharper densities
that are consistent with the discrete entropies. This is the approach
adopted by \citet{2001MNRAS.323..743R}, and the approach we take in
this paper. The key advantage of sharpening the densities is that we
do not need to specify a subgrid mixing model.

The sharper densities we require are exactly what we get from the
density estimate given in equation \ref{eqn:sphcontrtent}, and originally
proposed by \citet{2001MNRAS.323..743R} -- the `RT' density
estimate. Consider the perturbation discussed above, but now using
the RT density estimate\footnote{We use here the entropy form given in equation \ref{eqn:sphcontrtent} since we use the entropy form of SPH in this analysis. If instead, we use the energy form of SPH then we should use instead equation \ref{eqn:sphcontrteng}.}. A low density particle which has half of its kernel in the high density phase (an extreme example), will have a density: 

\begin{equation}
\rho_\mathrm{low} = \sum_j^{N_l} m_j \wijtild +
\sum_j^{N_r}\left(\frac{A_j}{A_i}\right)^{\frac{1}{\gamma}} m_j \wijtild
\end{equation}
where $N_l$ is the number of particles in the low density region, 
$N_r$ is the number in the high density region, and we have used the
fact that the ratio $\left(A_j/A_i\right)^{\frac{1}{\gamma}} =
1$ for the low density region.

If the simulation is adiabatic and started in pressure equilibrium,
then for the high density region
$\left(A_j/A_i\right)^{\frac{1}{\gamma}} = 1/R_\rho$, and since
the high density particles sample the kernel $R_\rho$ times more often
than the low density particles, we recover: 

\begin{equation}
\rho_\mathrm{low} = \sum_j^{N} m_j \wijtild 
\end{equation}
which is identical to a particle in the low density region. A similar
derivation applies to a high density particle at the boundary. Thus,
the RT density estimate ensures that densities remain sharp. It is straightforward to show that it also ensures the pressures are single valued throughout the flow. Substituting the RT density estimator (equation \ref{eqn:sphcontrtent}) into equation \ref{eqn:entropystate}, we obtain: 

\begin{equation}
P_i = \left[\sum_j^N m_j A_j^{\frac{1}{\gamma}} \wijtild\right]^\gamma
\end{equation}

Notice that the entropy function $A_j$ now appears {\it inside} the sum, whereas in standard SPH it would appear as $A_i$ outside of the sum. This difference ensures that the $P_i$ will be everywhere single valued throughout the flow -- even at boundaries. A similar derivation can be made for the energy form of SPH, in which case we should use the density estimate given in equation \ref{eqn:sphcontrteng}.

The RT density estimate is robust to particle mixing on the kernel scale and should lead to a dramatically reduced LMI. We demonstrate this in \S\ref{sec:kh}. Furthermore, the RT density estimate ensures that our error analysis in \S\ref{sec:error} is valid by construction since it ensures smooth pressures (recall that we assumed that both the pressures and the velocities were smooth, but not the densities). And, since the RT density estimate leads to sharper densities, it gives improved volume estimates for the particles. This suggests that we can expect the RT density estimate to reduce $\uE$ at boundaries. We demonstrate this also in \S\ref{sec:kh}. 

Note that the RT density estimate is chosen to ensure single valued pressures throughout the flow. However, when extracting results from a simulation, it is the positions of the particles themselves that describe the state of the fluid. This suggests using the density estimate in equation \ref{eqn:sphcont} for calculating the observable flow density, rather than the RT density estimate. This is the approach we adopt in this paper, though the difference is negligible. 

\section{Implementation}\label{sec:implement}

We implemented \MSPH\ in the {\tt GASOLINE} code \citep{2004NewA....9..137W},
a parallel implementation of TreeSPH that uses a fixed
number $N$ of smoothing neighbours\footnote{Allowing for varying
  neighbour number is needlessly dissipative 
(\bcite{nelsonpapagradh94}; \bcite{attwoodadap07}).}, and a standard
prescription for the artificial viscosity as in \citet{monaghanvisc83} 
with $\alpha = 1$, $\beta = 2$, controlled with a Balsara switch
\citep{balsaraphd89}. We used variable timesteps controlled by the
Courant time with a Courant factor of $0.4$.  

The improved stability and error properties of \MSPH\ motivate a full re-examination of the standard SPH artificial viscosity. This is beyond the scope of this present work. However, we note that the
improved stability in \MSPH\ means that particles better follow
characteristics of the flow, while the gradients in the Balsara switch
will be less noisy. Both of these effects should act to decrease the 
viscosity in regions of steady flow. (Note that all numerical schemes carry numerical viscosity, whether
it is manifested through limited resolution or artificial shock-capturing viscosity. Indeed, these viscous terms are vital for successfully modelling shocks.) In \S\ref{sec:kh}, \S\ref{sec:sod} and
\S\ref{sec:blob}, we show that our \MSPH\ results agree very well with
analytic expectations, and with the results from Eulerian codes. This
suggests that the viscosity prescription in \MSPH\ is not a
significant source of error. Certainly it is not responsible for SPH's
inability to model mixing processes.

\section{The Kelvin-Helmholtz instability}\label{sec:kh}

In this section, we use a 1:2 and 1:8 density ratio shearing fluid simulation
to test mixing in \MSPH. We use the naming convention XSPH-K-N, where X
denotes the variety of SPH, K the choice of kernel, and N the neighbour number (see Table \ref{tab:sims}).

\begin{table}
\begin{center}
\begin{tabular}{ccccc}

Flavour of SPH & $\phir$ & $\phiv$ & $\phiu$ & Kernel \\
\hline
SPH & 1 & 1 & 1 & CS \\
\SPHS & 1 & $\rho$ & 1 & CS, CT, HOCT4 \\
\MSPH & 1/u & $\rho$ & 1/u & HOCT4 

\end{tabular}
\caption{The different flavours of SPH we explore in this work. The free functions $\phir, \phiv$ and $\phiu$ are defined in equations \ref{eqn:cont}, \ref{eqn:moment} and \ref{eqn:energy}. The Cubic Spline (CS) kernel is given by equation \ref{eqn:cubicspline}, the Core-Triangle (CT) kernel is given by equation \ref{eqn:wkern}, and the fourth order High-Oder Core Triangle (HOCT4) kernel is given by equation \ref{eqn:hoctkern} with $n_k = 4$.}
\label{tab:sims}
\end{center}
\end{table}

\subsection{Numerical set-up}

A Kelvin-Helmholtz instability (KHI) occurs when two shearing fluids
are subjected to an infinitesimal perturbation at the boundary
layer. The result of the perturbation is a linearly growing phase in
which the layers start to interpenetrate each other, progressively
developing into a vortex in the non-linear phase that mixes
the two fluid layers. The growth-rate of the instability is in general
a complicated function of the shear velocity, fluid densities,
compressibility, interface thickness, gravity, viscosity, surface
tension, magnetic field strength etc. In this test, we are only
interested in the behaviour of inviscid, incompressible (i.e. with bulk motions very much less than the sound speed) perfect fluids neglecting gravity. In this case, the linear growth rate of the KHI is
(\bcite{1961hhs..book.....C}):
\begin{equation}
w=k\frac{(\rho_1\rho_2)^{1/2}v}{(\rho_1+\rho_2)},
\end{equation} 
where $k=2\pi/\lambda$ is the wavenumber of the instability, $\rho_1$
and $\rho_2$ are the densities of the respective layers and $v=v_1-v_2$ is
the relative shear velocity. The characteristic growth time for the
KHI is then: 
\begin{equation}
\label{eq:KHI}
\tau_{\rm
  KH}\equiv\frac{2\pi}{w}=\frac{(\rho_1+\rho_2)\lambda}{(\rho_1\rho_2)^{1/2}v}.
\end{equation}
This is a particularly challenging test for SPH/\MSPH\ because the
velocity due to particle noise can approach the sound speed which can
wash out the physical velocity perturbation relevant for this test.

For the simulations, we set up the problem in three dimensions using a
periodic thin slab defined by $x\in \{-0.5,0.5\}$, $y\in \{-0.5,0.5\}$
and $z\in \{-1/64,1/64\}$. The domain satisfied: 
\begin{equation}
\rho,T,v_x=\left\{
 \begin{array}{rl} 
 	\rho_1,T_1,v_1  & |y|< 0.25\\
 	\rho_2,T_2,v_2  & |y|> 0.25
 \end{array} \right.
\end{equation}
The density and temperature ratio were $R_\rho=\rho_{\rm 1}/\rho_{\rm
  2}=T_{\rm 2}/T_{\rm 1}=c_{\rm 2}^2/c_{\rm 1}^2$, ensuring that the
whole system was pressure equilibrium. The two layers were given
constant and opposing shearing velocities, with the low density layer
moving at a Mach number $\mathcal{M}_{\rm{2}}=-v_{\rm 2}/c_{\rm 2}
\approx 0.11$ and the dense layer moving at
$\mathcal{M}_{\rm{1}}=\mathcal{M}_{\rm{2}}\sqrt{R_\rho}$. The density
ratios considered in this work are small which assures a subsonic
regime where the growth of instabilities can be treated using equation
\ref{eq:KHI} \citep{1997ApJ...483..262V}.
 
To trigger instabilities, velocity perturbations were imposed on the
two boundaries of the form:
\begin{eqnarray}
v_y & =& \delta v_{\rm y}[\sin(2\pi
(x+\lambda/2)/\lambda)\exp(-(10(y-0.25))^2) \nonumber \\
& & - \sin(2\pi x/\lambda)\exp(-(10(y+0.25))^2)]
\end{eqnarray}
where the perturbation velocity $\delta v_{\rm y}/v=1/8$ and
$\lambda=0.5$ is the wavelength of the mode.

Equal mass particles were placed in lattice configurations to satisfy
the setup described above. To satisfy pressure equilibrium everywhere,
in \SPHS\ the temperatures were adjusted at boundaries to be coherent with the
smoothed density step measured by equation \ref{eqn:sphcont}. This was
not done for the \MSPH\ simulations since these sharpen the densities
using the discrete initial temperatures. 

The low density region $\rho_2$ was set up using 256 particles
in the $x$-direction and the appropriate number of particles in the
other dimensions to satisfy a fixed inter-particle distance. The high
density region $\rho_1$ was created in the same way with 320
particles in the $x$-direction. We adopted a periodic simulation
domain. 

The {\tt RAMSES} simulation used the same numerical set-up as described
above, but in 2D rather than in a thin slab. We performed the
$R_\rho=2$ simulation using the LLF Riemann solver \citep{toro} on a
$256\times256$ fixed Cartesian grid. The LLF solver is rather
diffusive and is used in order to suppress the growth of undesirable
small scale KHIs arising from grid irregularities. 

We note that all numerical schemes carry numerical viscosity, whether
it is manifested through limited resolution or artificial
shock-capturing viscosity. A detailed study of this effect on the KHI
and the relation to physical viscosity is beyond the scope of this
paper.

\subsection{Results}

Figure \ref{fig:clump} shows our results for the KHI test
(density ratio $R_\rho = 2$) at $\tau_\mathrm{KH}=1$ modelled with
SPH, \SPHS\ and \MSPH, using three different kernels: CS, CT and HOCT4,
and different neighbour numbers as marked on each plot (see also Table \ref{tab:sims}). From left to right, the panels show, in a slice
of width $dx=1$ about the z-axis: density
contours of the simulation box, a zoom in on the particle distribution
around on of the rolls, the magnitude of the $|\uE|$ error
(equation \ref{eqn:e0int}) as a function of $y$, and the pressure as
a function of $y$ in a slice of width $dx=1$ about the x-axis. 

\subsubsection{The clumping instability}

Using the standard CS kernel, SPH-CS-128 (top row, Figure \ref{fig:clump})
and \SPHS-CS-128 (second row) give poor results that
improve very slowly with increasing neighbour number. This can be seen
both in the lack of strong evolution on the boundary, and in the large
$|\uE|$ error, even for 128 neighbours. \SPHS-CS-128 gives slightly better
results than SPH-CS-128, showing the first beginnings of a KHI roll, but both
are in poor agreement with the {\tt RAMSES} results (bottom row).

The reason for the poor performance in both SPH-CS-128 and \SPHS-CS-128
is the clumping instability (\S \ref{sec:clumpinginst}). Particles gather together on the kernel
scale, giving poor kernel sampling, and poor 
associated error. This can be seen in the particle distribution for
SPH-CS-128 and \SPHS-CS-128 (second row, Figure \ref{fig:clump}) which
show visible 
holes and over-densities in the particle distribution. Using instead
the CT kernel introduced in \S\ref{sec:clumpinginst}, the results improve
dramatically (third row, Figure \ref{fig:clump}). Now the
errors reduce for increasing neighbour number (see Appendix \ref{sec:partnumber}). With 128 neighbours, we successfully resolve a KH roll up to $\tau_\mathrm{KH}=1$ with the
correct growth time.

It has been noted previously in the literature that putting a small
core inside a cubic spline kernel suppresses the clumping instability
(\bcite{1992MNRAS.257...11T}; \bcite{1994MmSAI..65.1013H}), though its
importance for modelling 
multiphase flow was not realised. Alternative fixes include adding an
negative pressure term \citep{2000Monaghan}, which in tests we find
works also. However, we prefer changing the kernel to introducing new
forces since we may then still estimate our errors through $|\uE|$.

\subsubsection{The banding instability}\label{sec:bandingresults}

In addition to the clumping instability, there is also an instability
to transverse waves -- the banding instability
(\S\ref{sec:bandinginst}). For the KHI tests we present here, the banding instability occurs only on the boundary and appears to be relatively benign. This is shown in Figure \ref{fig:banding}, that shows a zoom in on the boundary at $\tau_\mathrm{KH}=1$ for
\SPHS-CT-128, \SPHS-HOCT4-442 and \MSPH-HOCT4-442. The \SPHS-CT-128
simulation has a kernel and neighbour number combination that are
unstable to transverse waves (see Figure \ref{fig:stabct}), and
banding is clearly visible on the boundary. However, \SPHS-HOCT4-442
should be stable to transverse waves, yet the banding
persists. Only in our full scheme, \MSPH-HOCT4-442, is the banding is
gone.

To understand the above results, we ran an additional test that we
omit for brevity -- \SPHS-HOCT4-96. This simulation showed little
boundary evolution because the low neighbour number and associated
large $|\uE|$ significantly damped the KHI. However, interestingly,
there was no banding observed on the boundary 
(recall that 96 neighbours for the HOCT4 kernel should be stable to
both transverse and longitudinal wave perturbations).

Taken together, our results suggest that the observed banding at the
boundary is a result of a transverse wave instability driven by 
the local mixing instability (LMI; \S\ref{sec:localmix}). Where there
is little evolution at the boundary {\it and} the kernel is chosen to
be stable to transverse waves, the banding disappears, as was the case
for our extra \SPHS-HOCT4-96 simulation. Where there is strong
evolution at the boundary, as was the case for \SPHS-HOCT4-442, the LMI
drives banding irrespective of the choice of kernel. Only in our full
scheme, \MSPH-HOCT4-442, where the LMI is cured and the kernel is
stable to transverse waves, is the banding cured.

\subsubsection{The $\uE$ error}

Away from boundaries, the $\uE$ error in \SPHS\ decreases with the
neighbour number, as expected for smooth flow (see Appendix
\ref{sec:partnumber}). However, on the boundary the $|\uE|$ error
grows by  2-3 orders of magnitude. Increasing the neighbour number
does result in better long-term evolution, but the results improve
very slowly. This is shown in Figure \ref{fig:longterm}. Notice that
\SPHS-HOCT4-442 resolves two wraps of the KH roll at
$\tau_\mathrm{KH}=2$, whereas \SPHS-CT-128 only manages one. However,
even in \SPHS-HOCT4-442, the long-term evolution eventually
degrades. By $\tau_\mathrm{KH}=3$, the results are `gloopy', rather
similar to simulations that explicitly model fluid surface tension
(see e.g. \bcite{1066457}).  

The poor $\uE$ on the boundary is the result of a poor volume estimate for each particle $m_j/\rho_j$ (see \S\ref{sec:error}). However $\uE$ is not solely responsible for the gloopy behaviour. There is a second problem -- similar to a numerical surface tension term -- that needs to be solved extra to minimising $\uE$. This is the local mixing instability (LMI) error (\S\ref{sec:localmix}).

\subsubsection{The local mixing instability error}\label{sec:localmixresults}

The right panels of Figure \ref{fig:clump} show the pressure as a
function of $y$ in a slice of width $dx=1$ about the z-axis and width
$dx=1$ about the x-axis. In SPH and \SPHS\, there is a clear pressure
discontinuity on the boundary. This is caused by the local mixing instability (LMI)
discussed in \S\ref{sec:localmix}. 

Notice that the pressure blip is larger in \SPHS\ than in SPH, yet the KHI roll progresses further in \SPHS\ than in SPH. This apparent paradox is the result of the improved performance in \SPHS. As the KHI roll progresses in \SPHS, particles are pushed closer to the boundary making the LMI worse and increasing the pressure blip. In SPH, there is a larger gap at the boundary due to the larger surface tension error. This leads to less evolution and a smaller associated pressure blip. We will see a similar effect occurring in the blob test in \S\ref{sec:blob}. 

As discussed in \S\ref{sec:localmix}, the LMI should be cured by the RT
density estimate (equation \ref{eqn:sphcontrteng}). This is shown in Figure
\ref{fig:clump}, third row which shows the results for our full \MSPH\
scheme. With the RT densities, the pressure at the boundary has a much
smaller blip, while $\uE$ is reduced by over an order of magnitude. This
latter effect occurs since the RT densities also give an improved volume estimate for each particle (see \S\ref{sec:error} and \S\ref{sec:localmix}). The long term evolution is now in excellent agreement with the {\tt RAMSES} results (compare Figure \ref{fig:clump} third and bottom rows).

Although \MSPH\ gives significantly improved results as compared with SPH, the scheme is numerically expensive. Simulations with larger density gradients require very high resolution. This is shown in Figure \ref{fig:longterm}, second from bottom row. This shows the long term evolution of a KHI test
with density ratio $R_\rho = 8$ in \MSPH. The solution should be similar to the $R_\rho = 2$ simulation, but it is not. The `gloopy' behaviour indicative of large surface tension errors has returned. Further increasing the neighbour numbers would reduce this problem, but at increased numerical cost. We will address this issue in future work (Hayfield \& Read in prep.). 

\section{The Sod shock tube}\label{sec:sod}

Before we embark on the blob test in \S\ref{sec:blob}, it is worth
checking that our new \MSPH\ scheme can still correctly resolve
shocks. To test this, we use a standard Sod shock tube test
\citep{citeulike:2487799}. 

The Sod shock tube consists of a 1D tube on the interval
$[-0.5,0.5]$ with a discontinuous change in properties at $x = 0$ designed
to generate a shock. The left state is described by $\rho_l = 1.0$,
$P_l = 1.0$, $v_l = 0$, and the right state by $\rho_r = 0.125$, $P_r
= 0.1$, $v_r = 0$, where $\rho, P$ and $v$ are the density, pressure
and velocity along the $x$ axis. We use an adiabatic equation of state
with $\gamma = 1.4$. The subsequent evolution of the
problem has a self-similar analytic solution that has a 
number distinct features which quite generally test a code's
conservation properties, artificial viscosity, ability to handle
nonlinear waves, and shock resolution. 

Figure \ref{fig:sod} shows the results for the Sod shock tube
test at time $t = 0.2$ in SPH (top) \MSPH\ (bottom). Since we are primarily concerned with the
3D performance of the code, the test was performed in 3D on the union
of a $24\times 24\times 300$ lattice on the left, with a $12\times 12 \times 150$ lattice on
the right, giving a 1D resolution of 450 points. We use 442 neighbours for this test in both SPH and OSPH to ensure that any difference is not simply due to improved kernel sampling in OSPH.

For SPH, the only strong disagreement with the analytic solution is in the pressures that have a blip at $x = 0.2$, and the temperatures that overshoot at $x=0.2$. The former feature is due to the LMI (see \S\ref{sec:localmix} and \S\ref{sec:localmixresults}). The latter feature is seen in all SPH Sod shock tube tests and results from the well-known `wall heating' effect \citep{Noh198778}. This is an error due to the artificial viscosity prescription and is beyond the scope of this present work. 

For OSPH, the results are even better than for SPH. The pressure blip is now gone, while the temperature overshoot at $x = 0.2$ is reduced. Only the velocities appear to be worse, with some remaining dispersion at $x = 0.2$. This owes to the jump in density at this point, and the associated jump in $|\uE|$. This gives a force error at the discontinuity which introduces some dispersion into the velocities. In SPH this cannot occur since the LMI causes a pressure blip at the boundary that prevents mixing. We will discuss this issue further in a forthcoming paper (Hayfield \& Read in prep.). 

\section{The blob test}\label{sec:blob}

The KHI test presented in \S\ref{sec:kh} is a worst-case scenario for
\MSPH, since it has a pure adiabatic sharp boundary. For many practical
situations, boundaries will be less sharp, while physical entropy
generation due to shocks and/or cooling will suppress the LMI. We give
a practical example of this in this section using the blob test
described in \citet{2006astro.ph.10051A}. A spherical cloud of gas of
radius $R_{\rm cl}$ is placed in a wind 
tunnel with periodic boundary conditions. The ambient medium is ten 
times hotter and ten times less dense than the cloud so that it is in
pressure equilibrium with the latter. We refer to the initial
density contrast between the cloud and the medium as $R_{\rho,{\rm
    ini}}$. The wind velocity ($v_{\rm wind}=c_s\mathcal{M}$) has an
associated Mach number $\mathcal{M}=2.7$. This leads to the formation
of a bow shock after which the post-shock subsonic flow interacts with
the cloud and turns supersonic as it flows past it. 

The blob test is useful for investigating how different hydrodynamics
codes model astrophysical processes important for multiphase systems,
such as shocks, ram-pressure stripping and fragmentation 
through KH and Rayleigh-Taylor (RT) instabilities. As $\tau_{\rm
  KH}<\tau_{\rm RT}$, an approximate timescale of the cloud
destruction is that of the full growth of the largest KH mode,
i.e. the wavelength of the cloud's radius. This can be obtained by
considering the post-shock flow on the cloud and its time-dependence
as the shock weakens and the cloud is accelerated. A full analysis of
this test is presented in \citet{2006astro.ph.10051A} and gives
$\tau_{\rm KH}\approx1.6\tau_{\rm cr}$ where $\tau_{\rm
  cr}=2R_{\rm{cl}}R_{\rho,{\rm ini}}^{1/2}/v$ is the crushing time and
the velocity $v$ refers to the streaming velocity in the reference
frame of the cloud. After this time, the cloud is expected to show a
more complicated non-linear behaviour leading to disruption. The
original blob test was initialised in a glass-like configuration
obtained using simulated annealing using a standard SPH code. Since we
now use \MSPH\ rather than SPH, we must set up new ICs for the blob. Our
new IC set up is described in detail in Appendix
\ref{sec:blobsetup}. Unlike the previous blob test, where perturbations
were seeded by random noise in the particle distribution, here we
deliberately seed an inward growing mode on the front surface of the
blob. This makes comparison between \MSPH\ and {\tt FLASH} simpler, since then
the morphology of the blob is less dependent on small scale numerical
noise.

The results are presented in Figure \ref{fig:blob}, where we compare
SPH, \SPHS\ and \MSPH\ with increasing resolution with similar results
from the Eulerian code {\tt FLASH} \citep{2000ApJS..131..273F}. The SPH
results (top panels) are similar to those presented in
\citet{2006astro.ph.10051A}. The blob is squashed by the shock, but
does not break up. There are no visible surface KHI or Rayleigh-Taylor
instabilities. \SPHS\ (second row) gives significantly improved
results. The central depression is now resolved and the blob is mostly
destroyed by $\tau_{\rm KH} = 3$. However, the density remains clumpy
as compared to the {\tt FLASH} simulation (bottom row). Our full \MSPH\
scheme (third row) gives excellent agreement with the {\tt FLASH}
results. There are 
clear surface KHI and RT instabilities and the blob breaks up fully
by $\tau_{\rm KH} = 3$. The precise details of the break up in {\tt
  FLASH} and \MSPH\ are different. However, these differences are smaller
than those we observed between {\tt FLASH} simulations of
varying resolution. They are caused by the non-linear break up of the
blob that is affected by resolution-scale perturbations.

Figure \ref{fig:bloberror} shows $\uE$ and the pressure blips for the
blob test in SPH (left), \SPHS\ (middle) and \MSPH\ (right) at $\tau_{\rm
  KH} = 1$. In SPH and \SPHS, the two fluid phases (marked by the black
and grey solid circles) remain well separated at all times. In both
cases the pressure distribution shows discontinuities. By contrast,
in \MSPH\ the fluids are already mixed at $\tau_{\rm KH} = 1$, while the
pressures are smooth and single-valued throughout the flow. 

There is a more modest improvement in $\uE$ between SPH, \SPHS\ and \MSPH\
than that seen in the KHI tests presented in \S\ref{sec:kh}. \MSPH\
gives a $\uE$ smaller by a factor $\sim 5$ compared to the SPH and
\SPHS\ simulations, whereas in the KHI tests, there was an improvement
of over an order of magnitude. There are 
two reasons for this. Firstly, since the SPH simulation shows little
evolution at the boundary, the initial $\uE$ is relatively well
conserved. By contrast, \SPHS\ shows significant boundary evolution due
to improved mixing. This can actually worsen $\uE$ at the
boundary since the particles are still unable to properly
interpenetrate as a result of the LMI. Figure \ref{fig:blob} clearly
shows, however, that \SPHS\ gives improved mixing. Secondly, entropy
generation at the shock softens the density step around the blob,
leading to an improved $\uE$ even for the SPH simulation.

\section{Conclusions}\label{sec:conclusions}

Standard formulations of smoothed particle hydrodynamics (SPH) cannot
resolve fluid mixing and instabilities at flow boundaries. We have
used an error and stability analysis of the generalised SPH
equations of motion to show that mixing fails for two distinct
reasons. The first is a leading order error in the momentum
equation. This should 
decrease with increasing neighbour number, but does not because
numerical instabilities cause the kernel 
to be irregularly sampled. We identified two important instabilities:
the clumping instability and the banding instability, and we showed
that both are cured by a suitable choice of kernel. The second problem
is the local mixing instability (LMI). This occurs as particles attempt 
to mix on the kernel scale, but are unable to due to entropy
conservation. The result is a pressure discontinuity at boundaries that
pushes fluids of different entropy apart. We cured the LMI by using a weighted density estimate proposed by \citet{2001MNRAS.323..743R}. We showed that this both reduces errors in the continuity equation and allows individual particles to mix at constant pressure. 

We demonstrated mixing in our new \Multiphase\
Smoothed Particle Hydrodynamics (\MSPH) scheme using a Kelvin Helmholtz
instability (KHI) test with density contrast 1:2, and the `blob test'
-- a 1:10 density ratio gas sphere in a wind tunnel -- finding excellent
agreement between \MSPH\ and Eulerian codes. 

\MSPH\ is a multiphase Lagrangian method that conserves momentum, mass and
entropy, and demonstrates that it is possible to model multiphase
fluid flow using SPH. However, \MSPH\ remains a low-order method, requiring large neighbour number to keep the $\uE$ error small. We will address this problem in a forthcoming paper, where we
use the lessons learnt in this present work to move to higher order
particle methods (Hayfield \& Read in prep.).

\section{Acknowledgements}
We would like to thank the referee Paul Clark for useful comments and suggestions that improved the clarity of this work. We would also like to thank James Wadsley, Walter Dehnen, Joachim Stadel, Romain Teyssier, Volker Springel, Peter Thomas, Peter Englmaier, and Melvyn Davis for useful discussions and comments. J.R. would like to acknowledge support from a Forschungskredit grant from the University of Z\"urich and an ESF Astrosim grant. The {\tt FLASH} software used for one of the simulations in this paper was developed by the DOE supported ASCI/Alliances Center for Astrophysical Thermonuclear Flashes at the University of Chicago.

\appendix
\section{The effect of increasing neighbour number in \SPHS}\label{sec:partnumber}

\begin{center}
\begin{figure*}
\SPHS-CT-32\\
\includegraphics[height=0.23\textwidth]{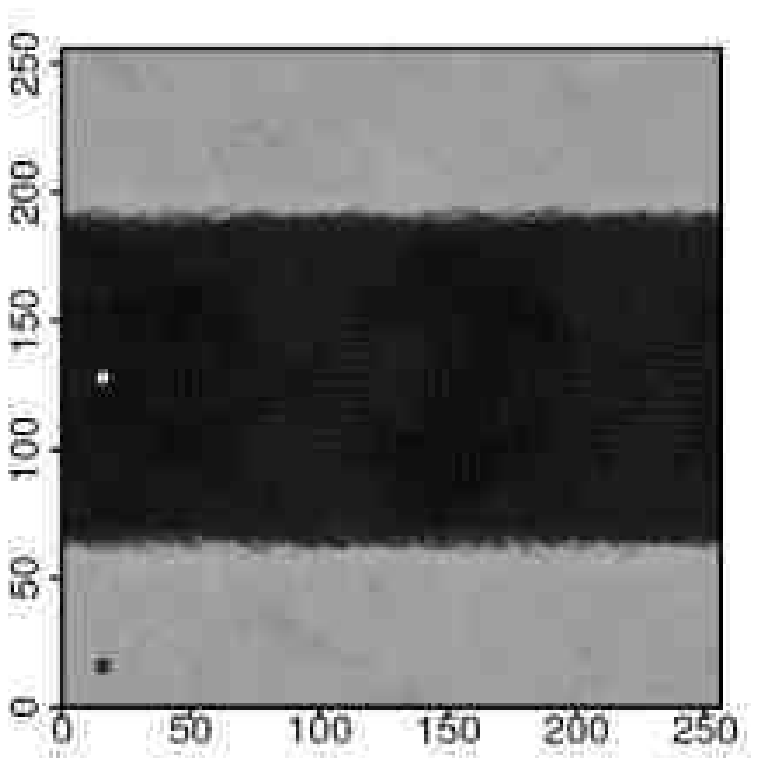}\hspace{\myfix}
\includegraphics[height=0.23\textwidth]{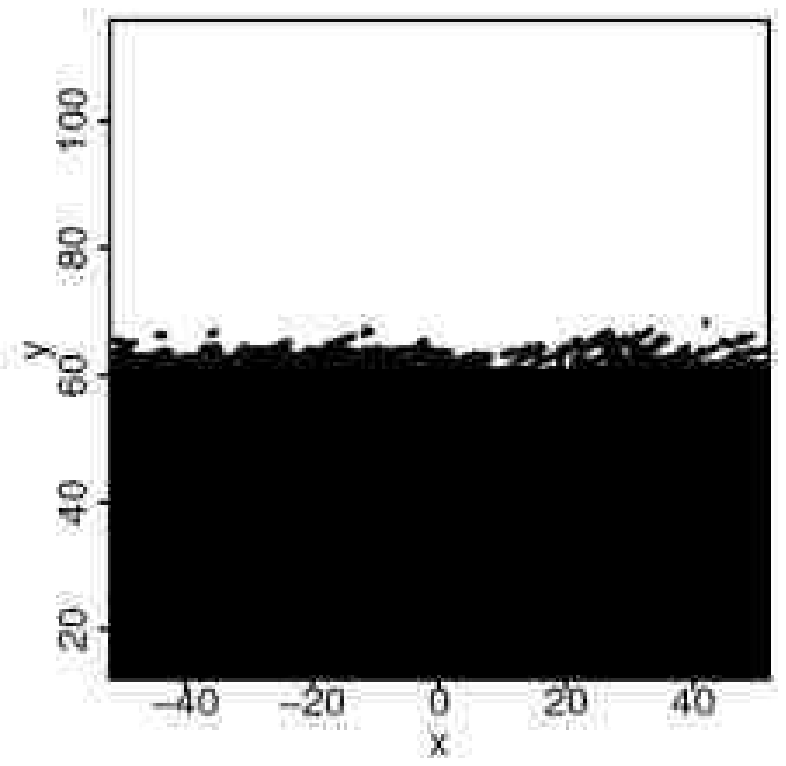}\hspace{\myfix}
\includegraphics[height=0.23\textwidth]{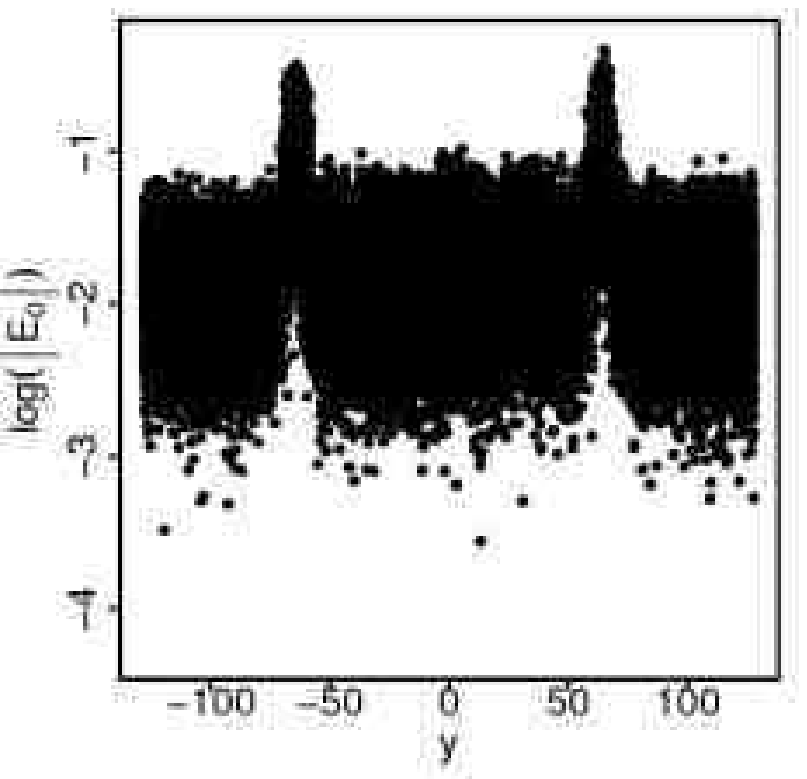}\hspace{\myfix}
\includegraphics[height=0.23\textwidth]{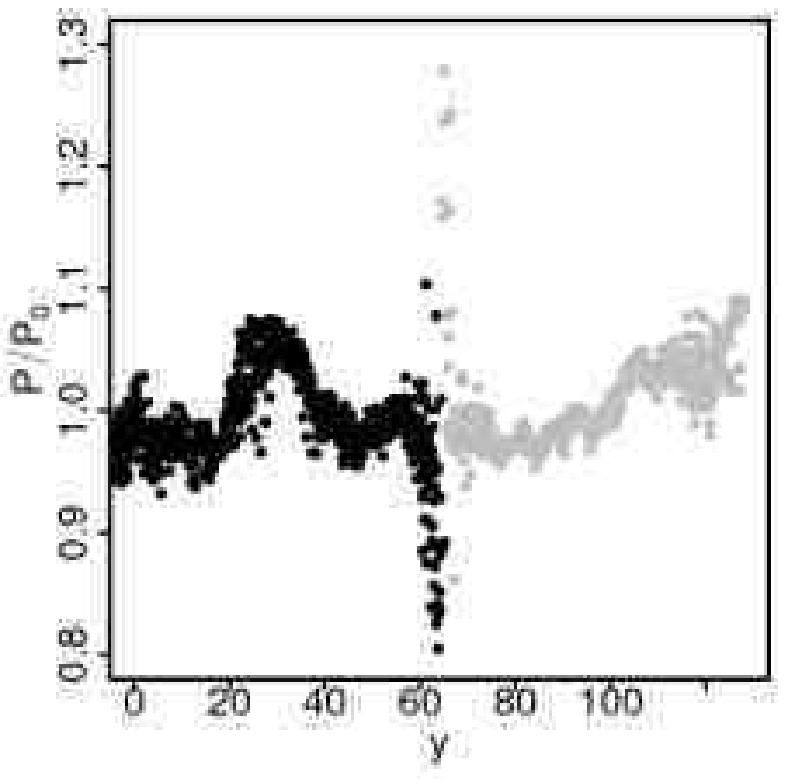}\\
\SPHS-CT-64\\
\includegraphics[height=0.23\textwidth]{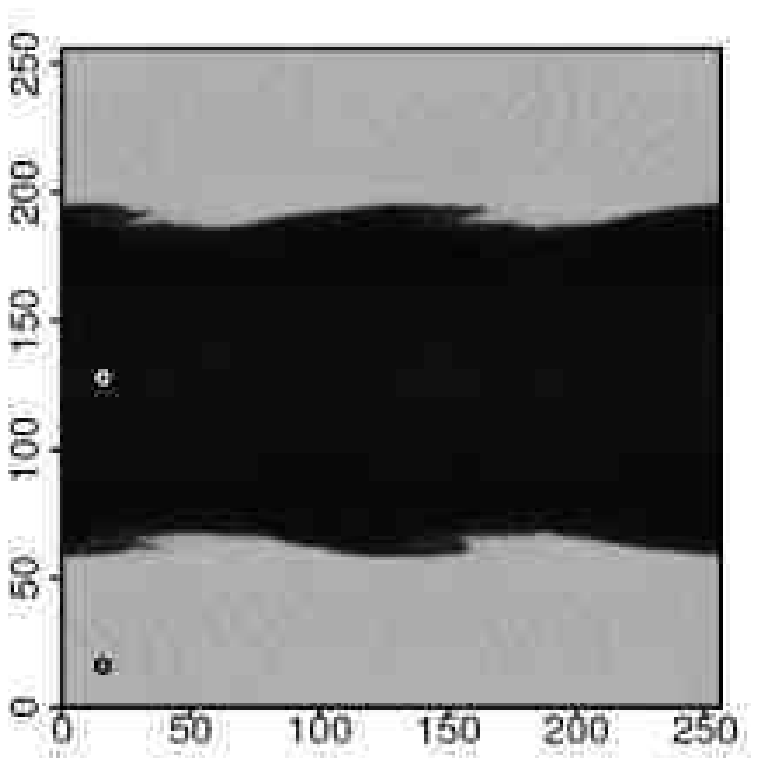}\hspace{\myfix}
\includegraphics[height=0.23\textwidth]{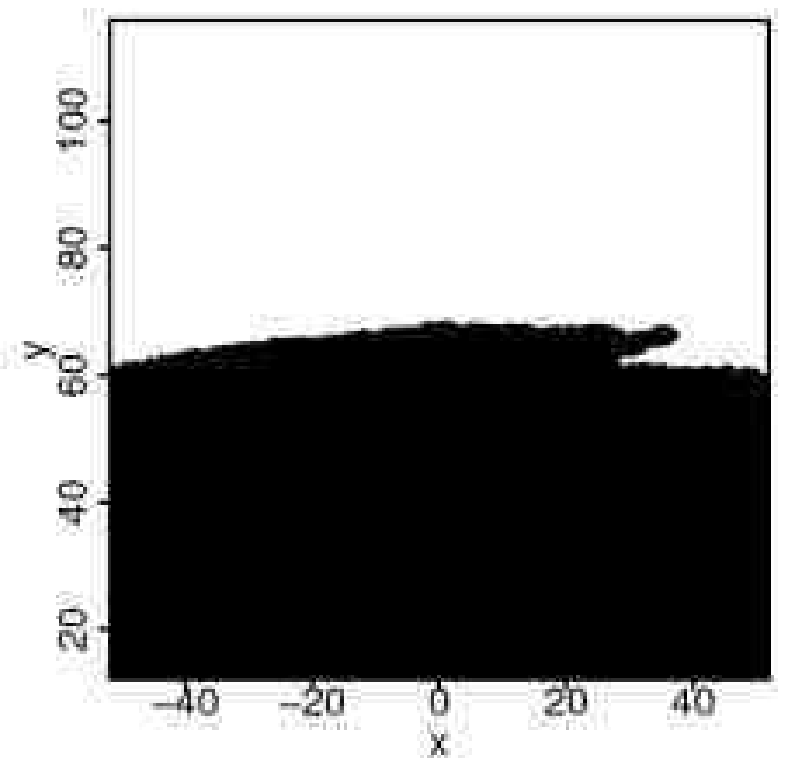}\hspace{\myfix}
\includegraphics[height=0.23\textwidth]{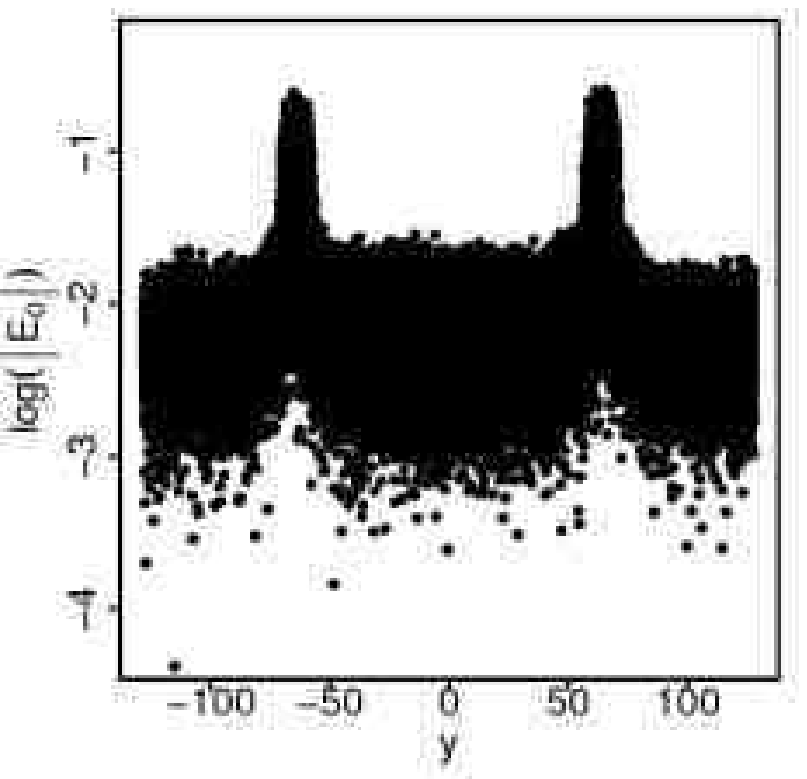}\hspace{\myfix}
\includegraphics[height=0.23\textwidth]{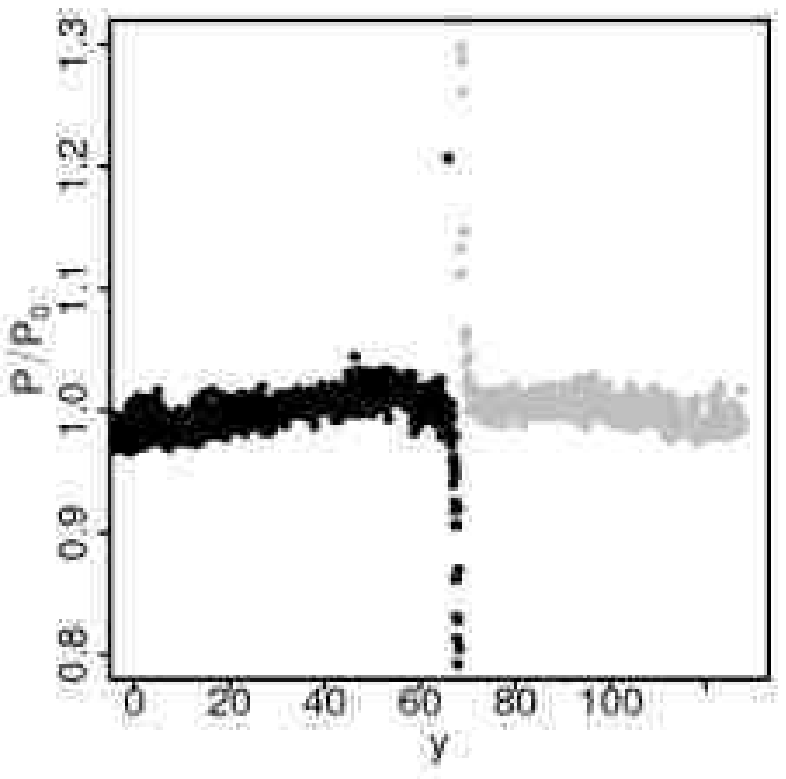}\\
\caption{As Figure \ref{fig:clump}, but showing the effect of
  increasing particle number for \SPHS-CT.}
\label{fig:partnumber}
\end{figure*}
\end{center}

In this appendix we show the effect of increasing neighbour number for
\SPHS-CT (i.e. without the clumping instability). The results for the
same KH instability test shown in Figure \ref{fig:clump} are shown in
Figure \ref{fig:partnumber} for 32 and 64 
neighbours. Notice that with increasing particle number, the error
vector $\uE$ is reduced, the pressure blip at the boundary is reduced,
and the results improve. 

\section{The blob test setup}\label{sec:blobsetup}
The hydrodynamical properties of the blob test are described in
\S\ref{sec:blob}. We use a periodic simulation box of size, in
units of the cloud radius $R_{\rm cl}$, $\{L_x,L_y,L_z\}=\{10,10,30\}$
and we centre the cloud at $\{x,y,z\}=\{5,5,5\}$. The equal mass SPH
particles constituting the ambient medium and the cloud are arranged
in lattice configurations to achieve the relevant density contrast
$R_{\rho,{\rm ini}}=10$. The particle temperatures ($T\sim P/\rho$)
are then assigned to achieve pressure equilibrium where the local
density measurement of equation \ref{eqn:sphcont} is used for
consistency. The wind velocity ($v_{\rm wind}=c_s\mathcal{M}$) has an
associated Mach number $\mathcal{M}=2.7$, where the sound speed is
$c_{\rm s}=\sqrt{\gamma P/\rho}$ using an adiabatic index $\gamma=5/3$. 

In the original blob test described in \citet{2006astro.ph.10051A},
SPH particle noise was used to trigger instabilities. This procedure
is not applicable when using a noise free lattice
configuration. Hence, we use spherical harmonics to apply large scale
perturbations to the surface layer of the cloud. The full
perturbation, in spherical coordinates centred on the cloud, can be
expressed as $v_{\rm pert}(r,\theta,\phi)=\delta vR(r){\rm
  Re}[Y(\theta,\phi)_l^m)]/C$, where the radial part,
$R(r)=\exp{(2(r-r_{\rm cl})/r_{\rm cl})}$, is defined for $r\leq
r_{\rm cl}$ and the spherical harmonic is, adopting $l=5$, $m=3$:
\begin{equation}
\label{eq:harmonic}
Y_5^3=\frac{-1}{32}\sqrt{\frac{385}{\pi}}e^{3i\phi}\sin^3{\theta}
(9\cos^2{\theta}-1).
\end{equation}
The constant $C$ simply normalises the real part of the harmonic to
reach a maximum value of 1. We chose a subsonic perturbation
$\delta v=-0.06\,v_{\rm wind}$. 

\bibliographystyle{mn2e}
\bibliography{../../../BibTeX/refs}

\begin{thebibliography}{}

\bibitem[\protect\citeauthoryear{{Agertz}, {Moore}, {Stadel}, {Potter},
  {Miniati}, {Read}, {Mayer}, {Gawryszczak}, {Kravtsov}, {Nordlund}, {Pearce},
  {Quilis}, {Rudd}, {Springel}, {Stone}, {Tasker}, {Teyssier}, {Wadsley} \&
  {Walder}}{{Agertz} et~al.}{2007}]{2006astro.ph.10051A}
{Agertz} O.,  {Moore} B.,  {Stadel} J.,  {Potter} D.,  {Miniati} F.,  {Read}
  J.,  {Mayer} L.,  {Gawryszczak} A.,  {Kravtsov} A.,  {Nordlund} {\AA}.,
  {Pearce} F.,  {Quilis} V.,  {Rudd} D.,  {Springel} V.,  {Stone} J.,  {Tasker}
  E.,  {Teyssier} R.,  {Wadsley} J.,    {Walder} R.,  2007, \mnras, 380, 963

\bibitem[\protect\citeauthoryear{{Attwood}, {Goodwin} \& {Whitworth}}{{Attwood}
  et~al.}{2007}]{attwoodadap07}
{Attwood} R.~E.,  {Goodwin} S.~P.,    {Whitworth} A.~P.,  2007, \aap, 464, 447

\bibitem[\protect\citeauthoryear{{Balsara}}{{Balsara}}{1989}]{balsaraphd89}
{Balsara} D.~S.,  1989, PhD thesis, , Univ.~Illinois at Urbana-Champaign,
  (1989)

\bibitem[\protect\citeauthoryear{{Bennett}}{{Bennett}}{2006}]{bennett}
{Bennett} A.,  2006, Lagrangian fluid dynamics.
Cambridge Monographs on Mechanics

\bibitem[\protect\citeauthoryear{{Chandrasekhar}}{{Chandrasekhar}}{1961}]{1961%
hhs..book.....C}
{Chandrasekhar} S.,  1961, {Hydrodynamic and hydromagnetic stability}.
International Series of Monographs on Physics, Oxford: Clarendon, 1961

\bibitem[\protect\citeauthoryear{{Dehnen}}{{Dehnen}}{2000}]{2000ApJ...536L..39%
D}
{Dehnen} W.,  2000, \apjl, 536, L39

\bibitem[\protect\citeauthoryear{Dilts}{Dilts}{1999}]{1999Dilts}
Dilts G.,  1999, International journal for numerical methods in engineering,
  44, 1115

\bibitem[\protect\citeauthoryear{{Fryxell}, {Olson}, {Ricker}, {Timmes},
  {Zingale}, {Lamb}, {MacNeice}, {Rosner}, {Truran} \& {Tufo}}{{Fryxell}
  et~al.}{2000}]{2000ApJS..131..273F}
{Fryxell} B.,  {Olson} K.,  {Ricker} P.,  {Timmes} F.~X.,  {Zingale} M.,
  {Lamb} D.~Q.,  {MacNeice} P.,  {Rosner} R.,  {Truran} J.~W.,    {Tufo} H.,
  2000, \apjs, 131, 273

\bibitem[\protect\citeauthoryear{{Gingold} \& {Monaghan}}{{Gingold} \&
  {Monaghan}}{1977}]{1977MNRAS.181..375G}
{Gingold} R.~A.,  {Monaghan} J.~J.,  1977, \mnras, 181, 375

\bibitem[\protect\citeauthoryear{{Gingold} \& {Monaghan}}{{Gingold} \&
  {Monaghan}}{1983}]{monaghanvisc83}
{Gingold} R.~A.,  {Monaghan} J.~J.,  1983, \mnras, 204, 715

\bibitem[\protect\citeauthoryear{{Goodman} \& {Hernquist}}{{Goodman} \&
  {Hernquist}}{1991}]{1991ApJ...378..637G}
{Goodman} J.,  {Hernquist} L.,  1991, \apj, 378, 637

\bibitem[\protect\citeauthoryear{{Greengard} \& {Rokhlin}}{{Greengard} \&
  {Rokhlin}}{1987}]{1987JCoPh..73..325G}
{Greengard} L.,  {Rokhlin} V.,  1987, Journal of Computational Physics, 73, 325

\bibitem[\protect\citeauthoryear{{Herant}}{{Herant}}{1994}]{1994MmSAI..65.1013%
H}
{Herant} M.,  1994, Memorie della Societa Astronomica Italiana, 65, 1013

\bibitem[\protect\citeauthoryear{{Hernquist} \& {Katz}}{{Hernquist} \&
  {Katz}}{1989}]{1989ApJS...70..419H}
{Hernquist} L.,  {Katz} N.,  1989, \apjs, 70, 419

\bibitem[\protect\citeauthoryear{Herrmann}{Herrmann}{2005}]{1066457}
Herrmann M.,  2005, J. Comput. Phys., 203, 539

\bibitem[\protect\citeauthoryear{{Hieber} \& {Koumoutsakos}}{{Hieber} \&
  {Koumoutsakos}}{2008}]{2008JCoPh.227.9195H}
{Hieber} S.~E.,  {Koumoutsakos} P.,  2008, Journal of Computational Physics,
  227, 9195

\bibitem[\protect\citeauthoryear{{Libersky}, {Petschek}, {Carney}, {Hipp} \&
  {Allahdadi}}{{Libersky} et~al.}{1993}]{1993JCoPh.109...67L}
{Libersky} L.~D.,  {Petschek} A.~G.,  {Carney} T.~C.,  {Hipp} J.~R.,
  {Allahdadi} F.~A.,  1993, Journal of Computational Physics, 109, 67

\bibitem[\protect\citeauthoryear{{Lucy}}{{Lucy}}{1977}]{1977AJ.....82.1013L}
{Lucy} L.~B.,  1977, \aj, 82, 1013

\bibitem[\protect\citeauthoryear{{Maron} \& {Howes}}{{Maron} \&
  {Howes}}{2003}]{2003ApJ...595..564M}
{Maron} J.~L.,  {Howes} G.~G.,  2003, \apj, 595, 564

\bibitem[\protect\citeauthoryear{{Marri} \& {White}}{{Marri} \&
  {White}}{2003}]{2003MNRAS.345..561M}
{Marri} S.,  {White} S.~D.~M.,  2003, \mnras, 345, 561

\bibitem[\protect\citeauthoryear{{Monaghan}}{{Monaghan}}{1992}]{1992ARA&A..30.%
.543M}
{Monaghan} J.~J.,  1992, \araa, 30, 543

\bibitem[\protect\citeauthoryear{{Monaghan}}{{Monaghan}}{2000}]{2000Monaghan}
{Monaghan} J.~J.,  2000, Journal of Computational Physics, 159, 290

\bibitem[\protect\citeauthoryear{{Morris}}{{Morris}}{1996a}]{1996PASA...13...9%
7M}
{Morris} J.~P.,  1996a, Publications of the Astronomical Society of Australia,
  13, 97

\bibitem[\protect\citeauthoryear{{Morris}}{{Morris}}{1996b}]{1996PhDMorris}
{Morris} J.~P.,  1996b, Ph.D.~Thesis, Department of Mathematics, Monash
  University

\bibitem[\protect\citeauthoryear{{Nelson} \& {Papaloizou}}{{Nelson} \&
  {Papaloizou}}{1994}]{nelsonpapagradh94}
{Nelson} R.~P.,  {Papaloizou} J.~C.~B.,  1994, \mnras, 270, 1

\bibitem[\protect\citeauthoryear{Noh}{Noh}{1987}]{Noh198778}
Noh W.,  1987, Journal of Computational Physics, 72, 78

\bibitem[\protect\citeauthoryear{{Oger}, {Doring}, {Alessandrini} \&
  {Ferrant}}{{Oger} et~al.}{2007}]{2007Oger}
{Oger} G.,  {Doring} M.,  {Alessandrini} B.,    {Ferrant} P.,  2007, Journal of
  Computational Physics, 225, 1472

\bibitem[\protect\citeauthoryear{{Price}}{{Price}}{2005}]{2005astro.ph..7472P}
{Price} D.,  2005, ArXiv Astrophysics e-prints

\bibitem[\protect\citeauthoryear{{Price}}{{Price}}{2007}]{2007arXiv0709.2772P}
{Price} D.~J.,  2007, ArXiv e-prints, 709

\bibitem[\protect\citeauthoryear{{Ritchie} \& {Thomas}}{{Ritchie} \&
  {Thomas}}{2001}]{2001MNRAS.323..743R}
{Ritchie} B.~W.,  {Thomas} P.~A.,  2001, \mnras, 323, 743

\bibitem[\protect\citeauthoryear{{Rosswog}}{{Rosswog}}{2009}]{2009NewAR..53...%
78R}
{Rosswog} S.,  2009, New Astronomy Review, 53, 78

\bibitem[\protect\citeauthoryear{{Schuessler} \& {Schmitt}}{{Schuessler} \&
  {Schmitt}}{1981}]{1981A&A....97..373S}
{Schuessler} I.,  {Schmitt} D.,  1981, \aap, 97, 373

\bibitem[\protect\citeauthoryear{Sod}{Sod}{1978}]{citeulike:2487799}
Sod G.~A.,  1978, Journal of Computational Physics, 27, 1

\bibitem[\protect\citeauthoryear{{Springel}}{{Springel}}{2009}]{2009arXiv0901.%
4107S}
{Springel} V.,  2009, ArXiv e-prints

\bibitem[\protect\citeauthoryear{{Springel} \& {Hernquist}}{{Springel} \&
  {Hernquist}}{2002}]{2002MNRAS.333..649S}
{Springel} V.,  {Hernquist} L.,  2002, \mnras, 333, 649

\bibitem[\protect\citeauthoryear{{Teyssier}}{{Teyssier}}{2002}]{2002A&A...385.%
.337T}
{Teyssier} R.,  2002, \aap, 385, 337

\bibitem[\protect\citeauthoryear{{Thomas} \& {Couchman}}{{Thomas} \&
  {Couchman}}{1992}]{1992MNRAS.257...11T}
{Thomas} P.~A.,  {Couchman} H.~M.~P.,  1992, \mnras, 257, 11

\bibitem[\protect\citeauthoryear{{Toro}}{{Toro}}{1999}]{toro}
{Toro} E.~F.,  1999, Riemann solvers and numerical methods for fluid dynamics -
  {A} practical introduction - 2nd edition.
Springer, Berlin

\bibitem[\protect\citeauthoryear{{Vietri}, {Ferrara} \& {Miniati}}{{Vietri}
  et~al.}{1997}]{1997ApJ...483..262V}
{Vietri} M.,  {Ferrara} A.,    {Miniati} F.,  1997, \apj, 483, 262

\bibitem[\protect\citeauthoryear{{Wadsley}, {Stadel} \& {Quinn}}{{Wadsley}
  et~al.}{2004}]{2004NewA....9..137W}
{Wadsley} J.~W.,  {Stadel} J.,    {Quinn} T.,  2004, New Astronomy, 9, 137

\bibitem[\protect\citeauthoryear{{Wadsley}, {Veeravalli} \&
  {Couchman}}{{Wadsley} et~al.}{2008}]{2008MNRAS.387..427W}
{Wadsley} J.~W.,  {Veeravalli} G.,    {Couchman} H.~M.~P.,  2008, \mnras, 387,
  427

\end{thebibliography}

\end{document}